\documentclass[12pt]{article}
\usepackage{youngtab}
\usepackage{amssymb,amsmath}
\usepackage{cite}
\usepackage[linktocpage=true]{hyperref} 
\input xy
\xyoption{all}

\begin{document}

\vspace*{0.5in}

\begin{center}

{\large\bf Notes on nonabelian (0,2) theories and dualities}

\vspace*{0.2in}

Bei Jia, Eric Sharpe, Ruoxu Wu

Department of Physics\\
Robeson Hall, 0435\\
Virginia Tech\\
Blacksburg, VA  24061\\

{\tt beijia@vt.edu}, 
{\tt ersharpe@vt.edu},
{\tt ronwu@vt.edu}

$\,$

\end{center}

In this paper we explore basic aspects of nonabelian (0,2) GLSMs in two
dimensions for unitary gauge groups, 
an arena that until recently has largely been unexplored.  
We begin by discussing general aspects of (0,2) theories, including
checks of dynamical supersymmetry breaking, spectators and weak coupling
limits, and also build some toy models of
(0,2) theories for bundles on Grassmannians, which
gives us an opportunity to relate physical anomalies and trace conditions to
mathematical properties.  
We apply these ideas to study (0,2) theories
on Pfaffians, applying recent perturbative constructions of
Pfaffians of Jockers {\it et al}.  
We discuss how existing dualities in (2,2) nonabelian gauge theories have
a simple mathematical understanding, and make predictions for additional
dualities in (2,2) and (0,2) gauge theories.
Finally, we outline how duality works in open strings
in unitary gauge theories, and also describe why, in general terms,
we expect analogous dualities in (0,2) theories to be comparatively rare.

\begin{flushleft}
January 2014
\end{flushleft}

\newpage

\tableofcontents

\newpage

\section{Introduction}

Over the last few years we have seen significant advances in our understanding
of gauged linear sigma models \cite{edphases}, 
ranging from GLSMs for different geometries (see {\it e.g.}
\cite{ael,adl,qss,mqs,mqss}), new understandings of GLSM
phases \cite{horitong,meron,cdhps,hori2,as2,jklmr1,bdfik,
hkm,me-rflat,horiknapp}, 
through more recent applications of supersymmetric
localization \cite{bc1,dgfl} to new computations of Gromov-Witten 
invariants \cite{jklmr2,bstv,ncgw} and elliptic genera (see {\it e.g.}
\cite{gg0,beht1,beht,murthy,adt}), and new dualities 
(see {\it e.g.} \cite{hpt,ggp,kutasovlin}),
among many other
advances, too numerous to comprehensively list here. 

The purpose of this paper is to work out some basic aspects of some
(0,2) nonabelian GLSMs, which have been studied comparatively rarely.
In this paper we will be primarily 
concerned with weak-coupling limits of GLSMs, with clear relations to
large-radius geometries\footnote{
Not all phases of GLSMs flow to nonlinear sigma models; many phases
are related to various Landau-Ginzburg models.  In this paper,
however, we are primarily interested in phases of GLSMs which do flow
to nonlinear sigma models.
}.   

In two dimensions, gauge fields do not have propagating degrees of freedom,
which simplifies certain analyses.  For many purposes, gauge fields
can be treated as Lagrange multipliers and integrated out.  When what is
left is a weakly coupled nonlinear sigma model, questions about the
GLSM can often usefully be turned into questions about geometry.
One of our interests in this paper lies in applying such ideas to 
two-dimensional dualities.  After all, if one can argue that two
different GLSMs RG flow to the same weakly-coupled nonlinear sigma model,
then in principle one has shown that they have the same IR limit, establishing
a two-dimensional analogue of a Seiberg-like duality.  Such IR matching
implies matching Higgs moduli spaces, chiral rings, and global symmetries,
which in higher dimensions are used as indirect 
tests\footnote{
Such tests should be applied with care; for example, examples were given
in \cite{am1} of different SCFTs with matching chiral rings.
} of
a common RG IR endpoint, rather than as consequences of a known IR matching.
We will use such geometric identifications
in theories flowing
to weakly-coupled nonlinear sigma models to make several predictions for
dualities in two-dimensional (2,2) and (0,2) theories, predictions checked
by {\it e.g.} comparing elliptic genera.

Another of our interests lies in understanding string compactifications,
in this paper including (0,2) versions of Pfaffian constructions,
and when bending GLSMs above to such purposes,
determining whether the lower-energy nonlinear sigma model has a
nontrivial IR fixed point is usually the significant complication.
For example, in a heterotic nonlinear sigma model on a Calabi-Yau, 
if the gauge bundle
is not stable, there is not expected to be a nontrivial RG fixed
point, a nontrivial SCFT associated to that bundle, but checking
stability is extremely complicated, even more so when working in a 
UV GLSM.  In this paper, in discussing
Calabi-Yau examples in which existence of a nontrivial fixed point
is possible, we will use recent advances to compute central charges
as a check for existence of such a fixed point.  In addition, we will
also discuss the possibility of dynamical supersymmetry breaking, which
has recently been discussed in the (0,2) literature.

We begin in section~\ref{sect:genl-features} by describing some basic
aspects of (0,2) theories which are utilized later.  We begin with a general
discussion of dynamical supersymmetry breaking, then turn
to a abstract overview of bundles in GLSMs.
We also discuss the role of spectators in fixing technical issues with
understanding RG flow of Fayet-Iliopoulos parameters.

In section~\ref{sect:02-grass} we discuss some toy (0,2) GLSMs on
Grassmannians, as basic examples and warm-ups for later constructions.
We relate gauge anomaly cancellation to cohomological
conditions on Chern classes in Grassmannians, discuss the details of
several examples, and also work through some dualities in these models,
concluding with an outline of some tests of those dualities and
a discussion of supersymmetry breaking in those toy examples.

In section~\ref{sect:other-cases}
we outline some constructions of nonabelian (0,2) theories corresponding
to complete intersections in Grassmannians and affine Grassmannians, some
dualities that should be obeyed in such constructions, and outline tests
of those dualities and supersymmetry breaking, computed via elliptic genera.
In section~\ref{sect:pfaff}, we 
discuss (0,2) models on Pfaffians.

We then turn to a mathematically-oriented study of dualities in
two-dimensional nonabelian GLSMs.  The heart of our discussion is the
observation that if two weakly-coupled theories are believed to RG flow
to nonlinear sigma models on the same space, then by definition, they have
the same Higgs moduli space, the same chiral ring, and the same global
symmetries, which in four dimensions are typical criteria for identifying
dualities.

There are a few known examples of Seiberg-like dualities in 
two-dimensional (2,2) theories with nonabelian gauge groups.  (For abelian
gauge groups, there are numerous examples of duality, perhaps most
prominently including mirror symmetry, as well as more recent examples
such as the (0,2) gerbe dualities in \cite{hetstx}.)  
The prototype for nonabelian
examples is encapsulated 
mathematically in two presentations of the same Grassmannian:
the Grassmannian $G(k,n)$ of $k$-planes in ${\mathbb C}^n$ is the
same as the Grassmannian $G(n-k,n)$ of $n-k$ planes in ${\mathbb C}^n$,
which becomes a statement relating universality classes of
$U(k)$ gauge theories with $n$ chiral superfields in the fundamental
representation to $U(n-k)$ gauge theories
also with $n$ chiral superfields in the fundamental representation.
We discuss how this generalizes mathematically to dualities in theories
with both fundamentals and antifundamentals in section~\ref{sect:22-dual}, and
describe how physical dualities can be understood as relating different
presentations of nonlinear sigma models on the same space.
Our approach has the advantage that it applies to
generic weakly-coupled (2,2) and (0,2) theories, in which flavor symmetries are
explicitly broken by choices of superpotentials (holomorphic maps), 
so {\it e.g.} 't Hooft
anomaly matching is of little utility.
We also similarly use geometry to make a prediction for a duality 
between Grassmannians $G(2,n)$ and certain Pfaffians, realizing a 
mathematical equivalence.

We then turn to dualities in (0,2) theories.
We begin with discussion of a duality between (0,2) theories describing
a space $X$ with bundle ${\cal E}$, and (0,2) theories describing the
same space but with dual bundle ${\cal E}^*$, in section~\ref{sect:dualbundle}.
This duality has been considered by others, as we discuss, but is
neither well-known nor thoroughly justified; our purpose
is both to advertise its existence and give additional justifications.

In section~\ref{sect:ggp} 
we then turn to dualities in nonabelian (0,2) theories.
Such dualities have only been rarely considered.
One recent example was discussed in \cite{ggp}, involving a triality
between two-dimensional (0,2) theories with unitary gauge groups and
matter in (anti)fundamental representations.  We review it from
a mathematical perspective in this section.

In section~\ref{sect:pax-vs-paxy} 
we return to the study of Pfaffians, and outline how
some of the dualities just discussed can illuminate the relationship
between the PAX and PAXY constructions of GLSMs for Pfaffians.

Finally in section~\ref{open-seiberg}
we formally consider dualities in open and heterotic strings
with more general representations.  We describe how dualities should
work for open strings, and argue that dualities for more general
(0,2) models will often not exist.

In an attempt to make this paper reasonably self-contained, we have
also included several appendices.  These appendices contain
technical aspects of the relation
between GLSMs and cohomology, and an overview of
Schur polynomials (used to compute
relations between cohomology classes on Grassmannians).
They also describe
our conventions for representations of $U(k)$ and summarize pertinent
properties.  Two final appendices give details of elliptic genus computations
whose results are summarized and utilized in the main text.

Overall, this paper discusses several different dualities:
\begin{itemize}
\item A nonabelian/abelian duality, relating the nonabelian GLSM
for $G(2,4)$ to the abelian GLSM for ${\mathbb P}^5[2]$ and its
(0,2) cousins, in section~\ref{sect:g24-duality}.
\item Another geometric duality, relating $G(2,n)$ and Pfaffian
constructions, is discussed in section~\ref{sect:g2n-duals}.
\item Generalizations of the $G(k,n) \leftrightarrow G(n-k,n)$ duality
relating $U(k)$ and $U(n-k)$ gauge groups are discussed in
sections~\ref{sect:22-dual}, \ref{sect:ggp}, \ref{sect:pax-vs-paxy},
and \ref{open-seiberg}.
\item A nongeometric duality, relating (0,2) 
theories on spaces $X$ with bundle ${\cal E}$ to (0,2) theories on 
the same space but with dual bundle, is discussed in 
section~\ref{sect:dualbundle}.
\end{itemize}
The first three have an essentially mathematical understanding;
part of 
our point is to apply known mathematics to understand existing dualities
between weakly-coupled theories and propose new relationships.

While this work was being completed, the work
\cite{kutasovlin} appeared, which discusses dualities in two-dimensional
nonabelian (0,2) theories with fundamentals, antifundamentals, and adjoints.
Adding adjoints complicates the mathematical analysis we shall present,
and so we leave a detailed mathematical study of \cite{kutasovlin}'s results for
future work.

\section{General features of nonabelian (0,2) constructions}
\label{sect:genl-features}

\subsection{Dynamical supersymmetry breaking}
\label{sect:genl-susy-breaking}

As is well-known, the Witten index Tr $(-)^F$ (for (0,2) theories,
Tr $(-)^{F_R}$) is a measure of the
possibility of dynamical supersymmetry breaking:  if it vanishes,
dynamical supersymmetry breaking is unobstructed.

Any operator that commutes with the fermion number operator can
be used to give a refinement of the Witten index, a graded version
with the property that vanishing of each separate graded component
is a necessary condition for supersymmetry breaking.  
An example of such a refinement is the elliptic genus, which was
utilized in \cite{ggp} as such a refined Witten index to check for
supersymmetry breaking.

Let us quickly review the application of elliptic genera to supersymmetry
breaking.  For a heterotic nonlinear sigma model describing a compact
space $X$
with holomorphic vector bundle ${\cal E}$ satisfying
\begin{displaymath}
{\rm ch}_2(TX) \: = \: {\rm ch}_2({\cal E}), \: \: \:
c_1(TX) \: \equiv \: \pm c_1({\cal E}) \mbox{ mod } 2
\end{displaymath}
the elliptic genus\footnote{
This particular elliptic genus is sometimes known as the
``Witten genus.''
} 
\begin{equation}  \label{eq:eg-trace-0}
\mbox{Tr}_{{\rm R},{\rm R}} (-)^{F}  q^{L_0} 
\overline{q}^{\overline{L}_0}
\end{equation} 
is well-defined, and given by \cite{edoldeg1}[equ'n (31)]
\begin{eqnarray} \label{eq:32}
\lefteqn{
(-)^{r/2} q^{+(1/12)(r - n)} 
\int_X \hat{A} (TX)  
\wedge {\rm ch}\Biggl( 
\left( \det {\cal E} \right)^{+1/2}
\wedge_{-1}\left(  {\cal E}^{*} \right)
}
\nonumber \\
& & \hspace*{1.5in} \left. \cdot \bigotimes_{k=1,2,3,\cdots}
S_{q^k}((TX)^{\mathbb C}) 
\bigotimes_{k=1,2,3,\cdots}
 \wedge_{-q^k}\left(( {\cal E})^{\mathbb C}\right)
\right),
\end{eqnarray}
or equivalently
\begin{eqnarray} \label{eq:32a}
\lefteqn{
(-)^{r/2} q^{+(1/12)(r - n)} 
\int_X \hat{A} (TX)  
\wedge {\rm ch}\Biggl( 
\left( \det {\cal E} \right)^{-1/2}
\wedge_{-1}\left(  {\cal E}\right)
}
\nonumber \\
& & \hspace*{1.5in} \left. \cdot \bigotimes_{k=1,2,3,\cdots}
S_{q^k}((TX)^{\mathbb C}) 
\bigotimes_{k=1,2,3,\cdots}
 \wedge_{-q^k}\left(( {\cal E})^{\mathbb C}\right)
\right),
\end{eqnarray}
where $r$ is the rank of ${\cal E}$, $n$ is the dimension of $X$,
and
\begin{eqnarray*}
S_q(TX) & = & 1 \: + \: q TX \: + \:
q^2 \mbox{Sym}^2 (TX) \: + \:
q^3 \mbox{Sym}^3 (TX) \: + \: \cdots,
\\
\wedge_q({\cal E}) & = &
1 \: + \: q {\cal E} \: + \: q^2 \wedge^2({\cal E}) \: + \:
q^3 \wedge^3({\cal E}) \: + \: \cdots,
\end{eqnarray*}
and the ${\mathbb C}$ superscripts indicate complexifications,
{\it i.e.} ${\cal E}^{\mathbb C} \cong {\cal E} \oplus \overline{\cal E}
\cong {\cal E} \oplus {\cal E}^*$.
By using the fact that
\begin{displaymath}
(-)^F \: = \: (-)^{F_R} (-)^{F_L} \: = \: (-)^{F_R} (-)^{(J_L)_0}
\end{displaymath}
we can see explicitly that the genus above is a refinement of the Witten
index for (0,2) supersymmetry.  It has been graded via operators 
($L_0$, $J_L$ mod 2)
that commute with the right-moving
fermion number.  In order for (0,2) supersymmetry to break, a necessary
condition is that every graded component of the index above must vanish.

If we have a nonanomalous symmetry, then in principle we can use it
to further grade or refine the index above.  For example,
in the special case that $X$ is Calabi-Yau and the bundle 
${\cal E}$ has trivial determinant, there is a nonanomalous left $U(1)$
current $J_L$, and for the corresponding nonlinear sigma model
we can define\footnote{
Occasionally some references, including this paper, will consider elliptic
genera with general $y$ and anomalous $J_L$.  This is only possible because
the formal expressions in the literature for elliptic genera of GLSMs
(using Jeffrey-Kirwan residues) do not explicitly require currents in exponents to be nonanomalous.  For example, in checking dualities
we will often compare elliptic genera with general $y$ even if $J_L$ is
anomalous, though when we do we will remark on the relevance of
more general $y$.  See also \cite{murthy1,adt1} for related discussions
in different contexts.  Such formal elliptic genera are unlikely to have
traditional modularity properties, and may not be mathematically
well-defined at all.  We leave a thorough discussion of such elliptic
genera to future work.
}
\begin{equation}   \label{eq:eg-trace}
\mbox{Tr}_{{\rm R},{\rm R}} (-)^{F} y^{(J_L)_0} q^{L_0} 
\overline{q}^{\overline{L}_0},
\end{equation} 
which is given by
\cite{km,lgeg}\footnote{
The conventions used here differ slightly from those of \cite{lgeg}.
To convert, $z$ should be identified with $-y^{-1}$.
} 
\begin{eqnarray} \label{eq:32-y}
\lefteqn{
(-)^{r/2} q^{+(1/12)(r - n)} y^{+r/2}
\int_X \hat{A} (TX)  
\wedge {\rm ch}\Biggl( 
\left( \det {\cal E} \right)^{+1/2}
\wedge_{-1}\left( y^{-1} {\cal E}^{*} \right)
}
\nonumber \\
& & \hspace*{1.5in} \left. \cdot \bigotimes_{k=1,2,3,\cdots}
S_{q^k}((TX)^{\mathbb C}) 
\bigotimes_{k=1,2,3,\cdots}
 \wedge_{-q^k}\left((y {\cal E})^{\mathbb C}\right)
\right),
\end{eqnarray}
or equivalently
\begin{eqnarray} \label{eq:32a-y}
\lefteqn{
(-)^{r/2} q^{+(1/12)(r - n)} y^{-r/2}
\int_X \hat{A} (TX)  
\wedge {\rm ch}\Biggl( 
\left( \det {\cal E} \right)^{-1/2}
\wedge_{-1}\left( y {\cal E}\right)
}
\nonumber \\
& & \hspace*{1.5in} \left. \cdot \bigotimes_{k=1,2,3,\cdots}
S_{q^k}((TX)^{\mathbb C}) 
\bigotimes_{k=1,2,3,\cdots}
 \wedge_{-q^k}\left((y {\cal E})^{\mathbb C}\right)
\right),
\end{eqnarray}
This reduces to the earlier expressions in the special case that $y=+1$.
If only a finite subgroup of the left $U(1)$ above is nonanomalous, then one 
can make sense of the expressions above for a finite number of values
of $y$.  We shall see this in examples later.  
(See also \cite{edloopspace}[section 5] 
for a related discussion of constraints on $y$.)

Now, for the moment, let us return to the general, non-Calabi-Yau, case,
to obtain some quick measures of potential (0,2) supersymmetry breaking
from the leading term in the elliptic genus, following the spirit of
\cite{km}.
Let us first compute the index above on the (2,2) locus where
${\cal E} = TX$. 
Using the fact that $S_q({\cal E}) = \wedge_{-q}({\cal E})^{-1}$ for
any vector bundle ${\cal E}$, 
we see that on the (2,2) locus, the Witten genus reduces to
\begin{displaymath}
 \int_X \hat{A} (TX)
\wedge {\rm ch}\Biggl(
 \left( \det TX \right)^{+1/2}
\wedge_{-1}\left(  T^*X \right) \Biggr),
\end{displaymath}
which  
is independent of $q$.  (Since this amounts to a topological field
theory partition function on the (2,2) locus, the $q$-independence
is not surprising.)
Furthermore, for any bundle ${\cal E}$, it is straightforward
to show that
\begin{displaymath}
{\rm ch}\left( (\det {\cal E} )^{+1/2} \otimes \wedge_{-1}( {\cal E}^*)
\right) \: = \: c_{r}({\cal E}) \: + \: \mbox{(higher degree)},
\end{displaymath}
where ${\cal E}$ has rank $r$,
so that on the (2,2) locus,
\begin{displaymath}
{\rm Tr}_{R,R} (-)^{F_R} (-)^{F_L} 
q^{L_0} \overline{q}^{\overline{L}_0}
\: \propto \: \chi(X).
\end{displaymath}
Thus, we recover the standard result that on the (2,2) locus,
the Witten index is given by the Euler characteristic.

Off the (2,2) locus, the $q$ dependence does not drop out.
We can get a quick measure of supersymmetry breaking by examining
the first graded component, namely
\begin{eqnarray}
\lefteqn{
(-)^{r/2} q^{+(1/12)(r - n)} \int_X \hat{A} (TX)
\wedge {\rm ch}\left( \left( \det {\cal E} \right)^{+1/2}
\wedge_{-1}\left(  {\cal E}^{*} \right) \right)
} \nonumber \\
& = &
(-)^{r/2} q^{+(1/12)(r - n)}
\left\{ \begin{array}{cl}
0 & r > n, \\
\int_X c_r({\cal E}) & r=n, \\
\cdots & r < n.
\end{array} \right.
\label{eq:genl-index}
\end{eqnarray}
As this is only one graded component of an infinite series,
it is merely a rather primitive check of supersymmetry breaking.

For later computational purposes, let us rewrite the expression above in 
a few more forms.  In the special case that $\det {\cal E}^* \cong K_X$,
so that the theory admits an A/2 twist, we can use the fact that
\begin{displaymath}
\hat{A}(TX) \: = \: {\rm td}(TX) \exp\left( - \frac{1}{2} c_1(TX) \right)
\end{displaymath}
to write the elliptic genus~(\ref{eq:32}) in the form
\begin{eqnarray*}
\lefteqn{
(-)^{r/2} q^{+(1/12)(r - n)} 
\int_X {\rm td}(TX)
\wedge {\rm ch}\Biggl( 
\wedge_{-1}\left(  {\cal E}^{*} \right)
}
 \\
& & \hspace*{1.5in} \left. \cdot \bigotimes_{k=1,2,3,\cdots}
S_{q^k}((TX)^{\mathbb C}) 
\bigotimes_{k=1,2,3,\cdots}
 \wedge_{-q^k}\left(( {\cal E})^{\mathbb C}\right)
\right),
\end{eqnarray*}
from which we read off the leading term
\begin{equation}
(-)^{r/2} q^{+(1/12)(r - n)} 
\int_X {\rm td}(TX)
\wedge {\rm ch}\left( \wedge_{-1}\left({\cal E}^{*} \right) \right)
\: = \:
(-)^{r/2} q^{+(1/12)(r - n)}
\sum_{s=0}^r (-)^s \chi\left( \wedge^s {\cal E}^* \right),
\end{equation}
which can be used as a crude test for dynamical supersymmetry breaking.

Alternatively, in the special case that $\det {\cal E} \cong K_X$,
so that the theory admits a B/2 twist, we can write the
elliptic genus~(\ref{eq:32a}) in the form
\begin{eqnarray*}
\lefteqn{
(-)^{r/2} q^{+(1/12)(r - n)} 
\int_X {\rm td}(TX)
\wedge {\rm ch}\Biggl( 
\wedge_{-1}\left(  {\cal E}\right)
}
\nonumber \\
& & \hspace*{1.5in} \left. \cdot \bigotimes_{k=1,2,3,\cdots}
S_{q^k}((TX)^{\mathbb C}) 
\bigotimes_{k=1,2,3,\cdots}
 \wedge_{-q^k}\left(( {\cal E})^{\mathbb C}\right)
\right),
\end{eqnarray*}
from which we read off the leading term
\begin{displaymath}
(-)^{r/2} q^{+(1/12)(r - n)} 
\int_X {\rm td}(TX)
\wedge {\rm ch}\left(
\wedge_{-1}\left(  {\cal E}\right)
\right) \: = \:
(-)^{r/2} q^{+(1/12)(r - n)} 
\sum_{s=0}^r (-)^s \chi\left( \wedge^s {\cal E} \right),
\end{displaymath}
which again can be used as a crude test for possible supersymmetry breaking.

As a consistency check, let us apply this in the special case
of a deformation of the (2,2) supersymmetric ${\mathbb C}{\mathbb P}^n$ model
discussed in \cite{tongrev,ggp}.  This deformation involved
decoupling the $\sigma$ field, resulting in a (0,2) theory which dynamically
broke supersymmetry, as could be seen from the one-loop correction to 
the Fayet-Iliopoulos parameter.  (It should be noted that removing the $\sigma$
field from (2,2) GLSMs has long been known to result in
ill-behaved theories \cite{dk3}, so this result is not surprising.)
Geometrically, decoupling the $\sigma$ field
corresponds to replacing the tangent bundle of ${\mathbb P}^n$ with an
extension, specifically the extension given by the Euler sequence
\begin{displaymath}
0 \: \longrightarrow \: {\cal O} \: \longrightarrow \:
\oplus^{n+1} {\cal O}(1) \: \longrightarrow \:
T {\mathbb P}^n \: \longrightarrow \: 0
\end{displaymath}
as the role of the $\sigma$ field is to realize the cokernel above.
Thus, the new gauge bundle is $\oplus^{n+1} {\cal O}(1)$.
Since the
rank is greater than the dimension of the space, our primitive 
supersymmetry index above suggests that supersymmetry may be broken,
which is consistent with the results of \cite{tongrev,ggp}.

In this example, the anomalous axial $U(1)$ is well-known to have
a nonanomalous ${\mathbb Z}_{n+1}$ subgroup, which suggests that
we may be able to form a more refined index by taking $y$ to be an
$(n+1)$th root of unity, not necessarily $+1$.  As this model admits an
A/2 twist, one can repeat earlier analyses to get that for more
general $y$, the elliptic genus should have leading term
\begin{displaymath}
(-)^{r/2} q^{(r-n)/12} y^{+r/2} \sum_{s=0}^r (- y^{-1})^s
\chi\left( \wedge^s {\cal E}^* \right)
\end{displaymath}
for ${\cal E} = \oplus^{n+1} {\cal O}(1)$.
From the Bott formula \cite{okonek}[p. 8]
\begin{displaymath}
h^q\left( {\mathbb P}^n, {\cal O}(k) \right) \: = \:
\left\{ \begin{array}{cl}
\left( \begin{array}{c} n+k \\ k \end{array} \right) & q=0, k \geq 0, \\
\left( \begin{array}{c} -k-1 \\ -k-1-n \end{array} \right) & q=n, k \leq -n-1,\\
0 & \mbox{else},
\end{array} \right.
\end{displaymath}
we have that
\begin{displaymath}
\chi({\cal O}) \: = \: 1, \: \: \:
\chi\left( \wedge^{n+1} {\cal E}^* \right) \: = \: (-)^n,
\end{displaymath}
and $\chi(\wedge^s {\cal E}^*)$ vanishes for $s \neq 0, n+1$.
Thus, the leading term in the elliptic genus is
\begin{displaymath}
(-)^{r/2} q^{(r-n)/12} y^{+r/2}
\left( 1 \: + \: (-y^{-1})^{n+1} (-)^n \right)
\: = \:
(-)^{r/2} q^{(r-n)/12} y^{+r/2}
\left( 1 \: - \: y^{-n-1} \right),
\end{displaymath}
which vanishes for $y$ an $(n+1)$th root of unity.  Thus, even the
refined index is consistent with supersymmetry breaking.

In fact, it is straightforward to show using the methods of \cite{beht}
that the entire elliptic genus for the (0,2) ${\mathbb C}{\mathbb P}^n$
model above, obtained by omitting the $\sigma$ field, vanishes identically,
a stronger sign of supersymmetry breaking.  The point is that since there
is no superpotential and no corresponding analogue of an R-symmetry,
the contribution from each
(0,2) chiral multiplet,
\begin{displaymath}
i \frac{\eta(q)}{\theta_1(q,x)}
\end{displaymath}
cancels the contribution from the corresponding (0,2) Fermi multiplet,
\begin{displaymath}
i \frac{\theta_1(q,x)}{\eta(q)}
\end{displaymath}
leaving one with only the contribution from the $U(1)$ gauge multiplet,
\begin{displaymath}
\frac{2 \pi \eta(q)^2}{i},
\end{displaymath}
which has no pole and hence no residues.

More generally, it will be shown in \cite{dlm} that singular
loci on the (0,2) moduli space, where in the GLSM $E$'s vanish, often
correspond to points where worldsheet supersymmetry is dynamically broken.
Such loci correspond to (singular) rank-changing transitions, and so
in general terms is consistent with our quick-and-dirty computation above.

Now, let us return to Calabi-Yau's.  The nonlinear sigma model for 
a Calabi-Yau has additional symmetries when $\det {\cal E}$ is trivial,
namely both $J_R$ and $J_L$ are separately nonanomalous, so the elliptic
genus admits a finer grading.  Demonstrating supersymmetry breaking,
for example, now requires not only vanishing of the separate coefficients
of powers of $q$, but also the vanishing of the separate coefficients of
powers of $y$.  The leading contribution to the elliptic
genus in this case was computed in \cite{km} (compare also
equation~(\ref{eq:32a-y})) to be proportional to
\begin{displaymath}
q^{+(1/12)(r-n)} y^{-r/2} \int_X {\rm td}(TX) \wedge {\rm ch}
\left( \wedge_{-1} (y {\cal E}) \right).
\end{displaymath}
Reference \cite{km} defined
\begin{eqnarray*}
\chi_y({\cal E}) & \equiv &
\int_X {\rm td}(TX) \wedge {\rm ch}
\left( \wedge_{-1} (y {\cal E}) \right), \\
& = &
\sum_{i=0}^r (-y)^i \chi(\wedge^i {\cal E} ).
\end{eqnarray*}
so that the leading term in the elliptic genus is
\begin{equation}  \label{eq:km-3fold}
q^{+(1/12)(r-n)} y^{-r/2} \chi_y({\cal E}),
\end{equation}
(In passing, an index of this form was independently suggested,
from more abstract considerations of (0,2) analogues of Morse theory
and supersymmetry, in \cite{fln}[section 6.4].)

Results of computations of $\chi_y$ 
can be found in \cite{km}.  One result which we
shall occasionally use, and so repeat here, is that on a Calabi-Yau
3-fold (so that $n=3$),
when the gauge bundle has $c_1({\cal E}) = 0$, 
\begin{displaymath}
\chi_y({\cal E}) \: = \: \left\{ \begin{array}{cl}
0 & r < 3 \\
- \tilde{\chi}({\cal E}) y (1+y) (1-y)^{r-3} & r \geq 3
\end{array} \right.
\end{displaymath}
for
\begin{displaymath}
\tilde{\chi}({\cal E}) \: \equiv \: \frac{1}{2} \int_X c_3({\cal E}).
\end{displaymath}
We shall apply this result explicitly later
to double-check computations of elliptic genera.

In passing, note that in the special case $y=+1$, for $r > 3$ this vanishes,
in agreement with the general result~(\ref{eq:genl-index}).  However,
for Calabi-Yau's, $y$ is not required to be $+1$, and so vanishing of the
elliptic genus is a stronger constraint on Calabi-Yau models than it is
for nonlinear sigma models on other K\"ahler manifolds.

So far, we have only discussed elliptic genera in nonlinear sigma models,
whereas the bulk of this paper is concerned with GLSMs.  However, in
weakly-coupled two dimensional theories, we are not missing any
information.  After all, as gauge fields in two dimensions are not
dynamical, in weak coupling regimes it is physically sensible to
integrate them out and work with the resulting lower-energy nonlinear
sigma model.  Any dynamical supersymmetry breaking in such models should
happen below the scale at which the nonlinear sigma model description
becomes relevant, and necessarily reflects properties of the underlying
geometry and heterotic gauge bundle, and not the GLSM gauge field.

We would like to conclude this section with a few comments on
dynamical supersymmetry breaking in models associated to Calabi-Yau's.
First, note that because the nonlinear sigma model has additional
conserved currents, the elliptic genus admits a finer grading, and so,
just at the level of the index, a vanishing index requires further
constraints than non-Calabi-Yau cases, suggesting that supersymmetry
breaking in (0,2) models associated to Calabi-Yau's may be comparatively
rare relative to supersymmetry breaking in (0,2) nonlinear sigma models
on other K\"ahler manifolds.
This observation is certainly consistent with existing lore in the field.

Furthermore, there is an additional subtlety, namely that even in
cases in which the index vanishes, there are indirect reasons to believe
that supersymmetry might still not be broken.  Specifically, we are thinking
of the old work \cite{dsww1,dsww2}, which argued, essentially by an index
computation, that worldsheet instanton effects should destabilize
(0,2) theories.  A few years after those papers were written, it was
discovered in a succession of papers (see {\it e.g.}
\cite{eva-ed,bcxddh,basu-sethi,beas-ed}) that although index
computations permit it, when one actually sums up all of the worldsheet
instantons in theories derived from GLSMs, the sum vanishes, and the
theory is not destabilized.  Thus, although it was permitted by an
index theory result, no destabilization actually happens.  The mathematical
reasons for this result are, in our opinion, not especially well-understood,
but we mention this as a caution that to convincingly demonstrate 
supersymmetry breaking in Calabi-Yau models built from GLSMs requires
more than just demonstrating that an index vanishes.

\subsection{Overview of bundles on Grassmannians}
\label{sect:review}

Let us briefly define some notation we shall use throughout this paper.
Briefly, all bundles in a (0,2) GLSM, abelian or nonabelian, are ultimately
built from
bundles defined by representations of the gauge group.  In a
GLSM with gauge group $U(1)$, say, all bundles are built as kernels,
cokernels, or cohomologies of monads built from bundles defined by
$U(1)$ charges.  Nonabelian GLSMs are very similar.

In this paper, Grassmannians will form an important prototype for
many constructions, so let us specialize to that case.
A (2,2) GLSM for a Grassmannian $G(k,n)$ of $k$-planes in ${\mathbb C}^n$
is built as a $U(k)$ gauge theory with $n$ fundamentals
\cite{edver}.  

Given a representation
$\rho$ of $U(k)$, we will let ${\cal O}(\rho)$ denote the corresponding
bundle on a Grassmannian.  (We will use the same notation in related
contexts, such as for affine Grassmannians.)
In the special case of a $U(1)$ gauge theory, a representation is defined
by a set of charges, so the description above specializes to give
{\it e.g.} line bundles of the form ${\cal O}(n)$ on projective spaces.

In principle, not every bundle on a Grassmannian is of the form
${\cal O}(\rho)$ for $\rho$ a representation of $U(k)$, just as
not every bundle on a projective spaces is a line bundle.
Instead, bundles of the form ${\cal O}(\rho)$ define a subset of a special
class of bundles, known as homogeneous bundles.  A homogeneous bundle is
defined by a representation of $U(k) \times U(n-k)$; bundles defined solely
by representations of $U(k)$ form what we shall sometimes call
special homogeneous bundles.

Some simple examples are provided by the universal subbundle $S$
and universal quotient bundle $Q$ on $G(k,n)$.  $S$ is rank $k$, $Q$ is
rank $n-k$, and they are related by the short exact sequence
\begin{equation}   \label{eq:sq-defn}
0 \: \longrightarrow \: S \: \longrightarrow \: {\cal O}^n \:
\longrightarrow \: Q \: \longrightarrow \: 0 .
\end{equation}
On a projective space, $S = {\cal O}(-1)$, and $Q = T \otimes {\cal O}(-1)$,
where $T$ denotes the tangent bundle.
In our notation above, $S = {\cal O}({\bf \overline{k}})$, {\it i.e.}
$S$ is a special homogeneous bundle defined by the antifundamental
representation of $U(k)$.  The universal quotient bundle is homogeneous
but not special homogeneous; it is defined by the antifundamental
representation of $U(n-k)$.

Bundles associated to more general representations of $U(k)$
can be built by
expressing the representation as a sum or tensor product of copies of the 
antifundamental and its dual, and then summing or
tensoring together copies of $S$
in the same fashion.  For example:
\begin{displaymath}
{\cal O}\left({\bf \overline{k}} \otimes {\bf \overline{k}}\right) 
\: = \: S \otimes S,
\: \: \:
{\cal O}\left( {\bf \overline{k}} \otimes {\bf k} \right) \: = \:
S \otimes S^*, \: \: \:
{\cal O}\left( {\bf k} \oplus {\bf k} \right) \: = \:
S^* \oplus S^*, \: \: \:
{\cal O}\left( {\rm Sym}^n {\bf k} \right) \: = \:
{\rm Sym}^n S^*,
\end{displaymath}
and so forth.

The prototype for many dualities in $U(k)$ gauge theories in two
dimensions is defined by the relationship $G(k,n) = G(n-k,n)$:
a $U(k)$ gauge theory with $n$ fundamental chirals is in the same
universality class as a $U(n-k)$ gauge theory with $n$ fundamental chirals.
An observation that will be key for many of our later observations is that
under the interchange above,
\begin{displaymath}
\left( S \: \longrightarrow \: G(k,n) \right)
\: = \:
\left( Q^* \: \longrightarrow \: G(n-k,n) \right) ,
\end{displaymath}
{\it i.e.} the interchange $G(k,n) \leftrightarrow G(n-k,n)$ also
exchanges the universal subbundle $S$ with the dual of the universal
quotient bundle $Q$.  Although $Q$ is not special homogeneous, it is
related to $S$ via the three-term exact sequence~(\ref{eq:sq-defn}),
and so $Q$ can be constructed indirectly, as we shall see in examples.

\subsection{Weak coupling limits and spectators}

In two-dimensional theories at low energies, the strength of the coupling
is effectively determined by the Fayet-Iliopoulos parameter, which is additively
renormalized at one-loop, by the sums of the charges of the bosons:
\begin{displaymath}
\Delta r_{{\rm 1-loop}} \: \propto \: \sum_{{\rm bosons}} Q_i.
\end{displaymath}
In a conventional
(2,2) GLSM, it is well-known that
vanishing of this renormalization is equivalent to the Calabi-Yau condition,
and furthermore the signs are such that positively-curved spaces shrink under
RG flow, and negatively-curved spaces expand, precisely as one would expect.

In a (0,2) GLSM describing a Calabi-Yau, it is often the case that the
sums of the charges of the bosonic chiral superfields is nonzero.
However, as observed in \cite{dk}, that does not imply that the
Fayet-Iliopoulos parameter necessarily runs:  one can add `spectators' to the
theory to cancel out charge sums.  Let us briefly review how this works in
abelian GLSMs.
Let $Q_{\alpha}$ denote the sums of the charges of the bosonic chiral
superfields with respect to the $\alpha$th $U(1)$.  Then, we add two
fields to the theory, a bosonic chiral superfield $X$ of $U(1)$ charges
$-Q_{\alpha}$ and a Fermi superfield $\Omega$ of $U(1)$ charges
$+ Q_{\alpha}$, together with a (0,2) superpotential
\begin{displaymath}
W \: = \: m_s \Omega X,
\end{displaymath}
where $m_s$ is a constant, defining the mass of the spectators.
Thanks to the addition of $X$, the sum of the $U(1)$ charges of the
bosonic chiral superfields now vanishes, so that the Fayet-Iliopoulos
parameter is not renormalized.  As we have added matching chiral and
Fermi superfields, anomaly matching is unaffected, and since the superpotential
effectively makes both $X$ and $\Omega$ massive, of mass $m_s$,
they do not contribute
to the IR behavior of the theory.

Thus, after adding spectators, at scales $\Lambda > m_s$ the 
Fayet-Iliopoulos parameter becomes an RG invariant, and so it
can be tuned to any desired value, such as a weak coupling limit in
which geometric descriptions are valid.  Below the scale $m_s$, if the theory
is sufficiently close to a nonlinear sigma model on a Calabi-Yau, the
rest of the RG flow should typically be determined by the mathematical 
properties
of the theory.   

The discussion above was outlined for abelian cases; however,
we can also follow exactly the same procedure in nonabelian (0,2) GLSMs
formally associated to Calabi-Yau geometries.  
For every $U(1)$ factor in the gauge group,
there is a Fayet-Iliopoulos parameter, and one can apply the same
procedure above to add spectators to understand weak coupling limits.

In (0,2) GLSMs formally describing spaces which are not Calabi-Yau,
the sum of the boson charges no longer matches the sum of the
fermion charges.  We can again add spectators to cancel the sum of the
boson charges, which has the effect of cancelling the one-loop
renormalization of Fayet-Iliopoulos
parameters at scales above $m_s$.  However, it is less clear how the
theory behaves at scales below $m_s$, after the spectators have been
integrated out.  Even if one used a small $m_s$ to tune the theory to
a weak coupling regime, below the scale set by $m_s$ the sigma
model coupling would surely begin running again.

In this paper we are primarily concerned with understanding geometric
interpretations in weak-coupling regimes.  Therefore, implicitly we will
add spectators as needed.

\section{Examples on ordinary Grassmannians }
\label{sect:02-grass}

Two-dimensional (2,2) GLSMs for Grassmannians have been discussed in
\cite{edver,horitong}, and for flag manifolds in \cite{meron}.
Briefly, the Grassmannian $G(k,n)$ is constructed via a $U(k)$ gauge
theory with $n$ chiral superfields in the fundamental representation.

Two-dimensional (0,2) theories describing bundles on $G(k,n)$ can be
built from $U(k)$ gauge theories with $n$ (0,2) chiral superfields in the
fundamental and suitable matter to describe the gauge bundle.
These form the prototype for other constructions: understanding 
(0,2) Grassmannian constructions
is essential to understand (0,2) Pfaffian constructions, for example,
and will also be important in our analysis of dualities.

In this section, we will outline some general aspects of (0,2) GLSMs and
their relation to cohomology and bundles on Grassmannians, as simple toy
models to illustrate various phenomena.

\subsection{Anomaly cancellation and Chern classes}

We shall begin by considering anomaly cancellation in nonabelian
(0,2) models, and its relation to cohomology of the underlying space.
In two-dimensional gauge theories, anomaly cancellation requires,
schematically,
\begin{equation}
\sum_{R_{\text{left}}}\text{tr}(T^a T^b) = \sum_{R_{\text{right}}}
\text{tr}(T^a T^b).
\end{equation}
More concretely,
in terms of the Casimirs discussed in appendix~\ref{app:uk-reps}, 
we have the following conditions:
\begin{eqnarray}
\sum_{R_{\text{left}}}\text{dim}(R_{\text{left}})
{\rm Cas}_2(R_{\text{left}}) &
= & \sum_{R_{\text{right}}}\text{dim}(R_{\text{right}})
{\rm Cas}_2(R_{\text{right}}),\\
\sum_{R_{\text{left}}}\text{dim}(R_{\text{left}})
({\rm Cas}_1(R_{\text{left}}))^2 &
= & \sum_{R_{\text{right}}}\text{dim}(R_{\text{right}})
({\rm Cas}_1(R_{\text{right}}))^2 .
\end{eqnarray}
The first condition is the $u(k)^2$ gauge anomaly condition, 
the second the $u(1)^2$
condition; there is no $u(1)-su(k)$ condition, as elements of the Lie
algebra of $su(k)$ are traceless.
Note for $SU(n)$ gauge theories, the second condition is automatically
satisfied, because of the fact that ${\rm Cas}_1(R)=0$
for any representation $R$ of
$SU(n)$.

For example, consider a (0,2) GLSM with right-moving chiral superfields
$\Phi$, $P$, left-moving fermi superfields $\Lambda$, $\Gamma$, and
(left-moving) gauginos. The gauge anomaly cancellation conditions are given by
\begin{equation}
\label{eq:gauge-anomaly}
\begin{split}
\sum_{R_{\Lambda}}\text{dim}(R_{\Lambda}){\rm Cas}_2(R_{\Lambda}) + &
\text{dim}(\text{adj})
{\rm Cas}_2(\text{adj}) \\
&= \sum_{R_{\Phi}}\text{dim}(R_{\Phi}){\rm Cas}_2(R_{\Phi}) + \sum_{R_{P}}
\text{dim}(R_{P}){\rm Cas}_2(R_{P}),\\
\sum_{R_{\Lambda}}\text{dim}(R_{\Lambda})({\rm Cas}_1(R_{\Lambda}))^2 + &
\text{dim}(\text{adj})({\rm Cas}_1(\text{adj}))^2 \\
&=\sum_{R_{\Phi}}\text{dim}(R_{\Phi})({\rm Cas}_1(R_{\Phi}))^2 + 
\sum_{R_{P}}\text{dim}(R_{P})({\rm Cas}_1(R_{P}))^2 .
\end{split}
\end{equation}

In principle, anomaly cancellation in the UV GLSM implies
\begin{equation}
\text{ch}_2(E) = \text{ch}_2(TX)
\end{equation}
in the IR NLSM on the space $X$, and in general is slightly
stronger than the IR condition (see for example \cite{ks}[section 6.5] 
for examples of
anomalous GLSMs associated mathematically to anomaly-free IR geometries).

In appendix~\ref{app:cohomology} we review the cohomology of the
Grassmannian $G(k,n)$ of $k$-planes in ${\mathbb C}^n$.
Briefly, 
\begin{displaymath}
H^2(G(k,n),{\mathbb Z}) \: = \: {\mathbb Z}, \: \: \:
H^4(G(k,n),{\mathbb Z}) \: = \: {\mathbb Z}^2 .
\end{displaymath}
The generators of the cohomology are given by Schubert cycles which are
defined by certain Young diagrams.  For example, we use
$\sigma_{\tiny\yng(1)}$ to denote the generator of $H^2$, and
each of $\sigma_{\tiny\yng(1,1)}$, $\sigma_{\tiny\yng(2)}$,
$\sigma_{\tiny\yng(1)}^2$ describe elements of $H^4$, related by
\begin{displaymath}
\sigma_{\tiny\yng(1)}^2 \: = \:
\sigma_{\tiny\yng(1,1)} \: + \: \sigma_{\tiny\yng(2)} .
\end{displaymath}

Furthermore, as discussed in appendix~\ref{app:uk-reps},
the Chern classes are determined by the Casimirs:
for any given representation $\lambda$, 
\begin{equation}
\label{eq:c1:intro}
c_1({\cal O}(\lambda)) \: = \: \frac{ d_{\lambda} {\rm Cas}_1(\lambda) }{k}
\sigma_{\tiny\yng(1)} ,
\end{equation}
\begin{eqnarray}
{\rm ch}_2({\cal O}(\lambda)) & = &
(1/2) c_1({\cal O}(\lambda))^2 \: - \: c_2({\cal O}(\lambda)), 
\nonumber \\
& = &
d_{\lambda} {\rm Cas}_2(\lambda)\left[ - \frac{1}{k^2-1} 
\sigma_{\tiny\yng(1,1)} \: + \:
\frac{1}{2 k (k+1) } \sigma_{\tiny\yng(1)}^2 \right]
\nonumber \\
& & \hspace*{0.5in} \: + \:
d_{\lambda} {\rm Cas}_1(\lambda)^2
\left[ \frac{1}{k(k^2-1)} \sigma_{\tiny\yng(1,1)}
\: + \:
\frac{1}{2k(k+1)} \sigma_{\tiny\yng(1)}^2 \right] ,
\label{eq:ch2:intro}
\end{eqnarray}
where $d_{\lambda}$ is the dimension of representation $\lambda$.

Let us apply this to heterotic geometries, and check that the
Casimir conditions above imply the mathematical matching of Chern classes
and characters.  Specifically, consider a bundle ${\cal E}$ defined by the
kernel
\begin{equation}   \label{eq:e-kernel}
0 \: \longrightarrow \: {\cal E} \: \longrightarrow \:
\oplus_i {\cal O}(\lambda_i) \: \longrightarrow \: 
\oplus_a {\cal O}(\lambda_a) \: \longrightarrow \: 0 .
\end{equation}
This is defined by a set of Fermi superfields $\Lambda$ in the representations
$\lambda_i$, chiral superfields $P$ in representations dual to $\lambda_a$,
and a (0,2) superpotential encoding the second nontrivial map.
Mathematically, using the additivity properties of Chern characters,
we have
\begin{eqnarray*}
{\rm ch}_2({\cal E}) & = & {\rm ch}_2\left( \oplus_i {\cal O}(\lambda_i) 
\right)
\: - \: {\rm ch}_2\left( \oplus_a {\cal O}(\lambda_a) \right) ,
\\
& = &
\sum_i {\rm ch}_2\left( {\cal O}(\lambda_i) \right) \: - \:
\sum_a {\rm ch}_2\left( {\cal O}(\lambda_a) \right) .
\end{eqnarray*}
The tangent bundle of the Grassmannian $G(k,n)$ is defined as the cokernel
\begin{displaymath}
0 \: \longrightarrow \: S^* \otimes S \: \longrightarrow \:
S^* \otimes {\cal O}^n \: \longrightarrow \: S^* \otimes Q \: = \: 
TG(k,n) \: \longrightarrow \: 0 ,
\end{displaymath}
and so 
\begin{eqnarray*}
{\rm ch}_2( T G(k,n) ) & = &
{\rm ch}_2\left( S^* \otimes {\cal O}^n \right) \: - \:
{\rm ch}_2 \left( S^* \otimes S \right), \\
& = & n \, {\rm ch}_2\left( S^* \right) \: - \: 
{\rm ch}_2 \left( S^* \otimes S \right) .
\end{eqnarray*}
The anomaly-cancellation condition is given by
\begin{displaymath}
{\rm ch}_2(T G(k,n) ) \: = \: {\rm ch}_2({\cal E}) ,
\end{displaymath}
which is equivalent to
\begin{displaymath}
\sum_i {\rm ch}_2\left( {\cal O}(\lambda_i) \right)  \: + \:
{\rm ch}_2 \left( S^* \otimes S \right)
\: = \:
\sum_a {\rm ch}_2\left( {\cal O}(\lambda_a) \right)
\: + \:
n \, {\rm ch}_2\left( S^* \right) .
\end{displaymath}
Writing ${\rm ch}_2$ in terms of Casimirs as in equation~(\ref{eq:ch2:intro})
above, we see that the mathematical anomaly-cancellation condition above
is satisfied if and only if the physical gauge anomaly
constraints~(\ref{eq:gauge-anomaly}) are satisfied, as expected.

Now, let us turn to the A/2 pseudo-topological field theory.
As discussed in \cite{ks,ade}, for a gauge bundle ${\cal E}$ over a space $X$,
in addition to the anomaly-cancellation condition one must also impose
the constraint
\begin{displaymath}
\wedge^{\rm top} {\cal E}^* \: \cong \: K_X ,
\end{displaymath}
which implies $c_1({\cal E}) = c_1(TX)$.
For the gauge bundle defined by~(\ref{eq:e-kernel}) over $X = G(k,n)$,
this constraint becomes
\begin{displaymath}
c_1\left( \oplus_i {\cal O}(\lambda_i) \right) \: - \:
c_1\left( \oplus_a {\cal O}(\lambda)a) \right) \: = \:
c_1\left( S^* \otimes {\cal O}^n \right) \: - \: c_1\left( S^* \otimes S 
\right) ,
\end{displaymath}
which can easily be checked to be equivalent to the statement
\begin{displaymath}
\sum_{R_{\Lambda}} \text{dim}(R_{\Lambda}) {\rm Cas}_1(R_{\Lambda})
\: + \:
\text{dim}(\text{adj}){\rm Cas}_1(\text{adj})
\: = \:
\sum_{R_{\Phi}}\text{dim}(R_{\Phi}){\rm Cas}_1(R_{\Phi})
\: + \:
\sum_{R_{P}}\text{dim}(R_{P}){\rm Cas}_1(R_{P}) ,
\end{displaymath}
or more simply,
\begin{displaymath}
\sum_{R_{\text{left}}} \text{dim}(R_{\text{left}}) {\rm Cas}_1(R_{\text{left}})
\: = \:
\sum_{R_{\text{right}}} \text{dim}(R_{\text{right}}) 
{\rm Cas}_1(R_{\text{right}}) .
\end{displaymath}

\subsection{Examples}
\label{sect:g24-exs}

In table~\ref{table:g24-exs} we list examples of bundles ${\cal E}$ of the form
\begin{displaymath}
0 \: \longrightarrow \: {\cal E} \: \longrightarrow \:
\oplus^{m} {\cal O}(\lambda_{A1}, \lambda_{B1})
\: \longrightarrow \:
\oplus^{n} {\cal O}(\lambda_{A2}, \lambda_{B2})
\: \longrightarrow \: 0
\end{displaymath}
on $G(2,4)$, satisfying anomaly cancellation.  
For simplicity we have chosen to focus on bundles defined by kernels;
however, nonabelian (0,2) GLSMs can also be used to describe cokernels
and cohomologies of monads.  As those constructions are simple generalizations,
we omit their discussion.

\begin{table}[h]
\begin{center}
\begin{tabular}{ccccc}
$m$ & $(\lambda_{A1}, \lambda_{B1})$ & $n$ & $(\lambda_{A2}, \lambda_{B2})$ & rank \\ \hline
5&(-2, -2)&2&(3, 3)&3\\
3&(-1, -1)&1&(1, 1)&2\\
4&(0, -1)&1&(1, -1)&5\\
2&(1, -2)&1&(2, -2)&3\\
5&(2, 2)&2&(3, 3)&3\\
\end{tabular}
\caption{Anomaly-free examples on $G(2,4)$.}
\label{table:g24-exs}
\end{center}
\end{table}

In table~\ref{table:g24-exs}
we have used the notation ${\cal O}(\lambda_1,\lambda_2)$ 
to indicate a vector bundle
on $G(2,4)$ defined by the $(\lambda_1,\lambda_2)$ representation of $U(2)$
($\lambda_1 \geq \lambda_2$).
See appendix~\ref{app:uk-reps} for our conventions.

Looking at the D-term constraints in these theories, we see a potential
issue that may sometimes arise\footnote{
The issue presented here is more subtle for Grassmannians, as the FI
parameter will run, but the same issue can arise in
Calabi-Yau cases, so we present it here as a prototype for
later discussions.
}.  Schematically, if we let $X$'s denote
the chiral superfields defining $G(2,4)$ and $P$'s denote the chiral
superfields corresponding to the third terms in the short exact
sequence defining the gauge bundle ${\cal E}$, then they are
schematically of the form
\begin{displaymath}
X X^{\dag} \: - \: P^{\dag} P \: = \: r_{U(2)} I .
\end{displaymath}
If the $P^{\dag} P$ term is
negative-definite, then for $r \gg 0$, we get that the $X$'s are not all
zero, and so we have a Grassmannian as usual.  If the $P^{\dag} P$ term does not
have that property, then the D term implies a weaker condition, and so
it is no longer clear that the geometry described, semiclassically,
is a Grassmannian.  A closely related issue also arises in abelian
GLSMs:  the total space of ${\cal O}(-1) \rightarrow {\mathbb P}^n$
is easy to describe with a collection of $n+1$ chiral superfields of
charge $+1$ and one of charge $-1$, but the total space of
${\cal O}(+1) \rightarrow {\mathbb P}^n$ cannot be similarly described
in GLSMs, as adding an extra chiral superfield of charge $+1$ would
merely increase the size of the projective space.  We will largely ignore
this potential problem for the time being, but it will crop up occasionally
in our discussion.

Let us describe the maps and superpotentials in these nonabelian
(0,2) models.  In the first entry in table~\ref{table:g24-exs},
we have a map ${\cal O}(-2,-2)^5 \rightarrow {\cal O}(3,3)^2$.
The elements of this map are provided by sections of
\begin{displaymath}
{\cal O}(5,5) \: = \: \left( \det S^* \right)^5 .
\end{displaymath}
A section of $\det S^*$ is a baryon constructed from the chiral superfields
defining the Grassmannian $G(2,4)$, {\it i.e.} an operator of the form
\begin{displaymath}
B_{ij} \: \equiv \: \epsilon_{ab} \phi_i^a \phi_j^b ,
\end{displaymath}
where in this case $i, j \in \{1, \cdots, 4\}$.  Therefore, the maps 
in the bundle in the first entry in table~\ref{table:g24-exs} are provided
by degree five polynomials in the $B^{ij}$, and the (0,2) superpotential
is then of the form
\begin{displaymath}
W \: = \: \Lambda_{\alpha} p_{\gamma} f^{\alpha \gamma}_5(B_{ij}) ,
\end{displaymath}
where $\Lambda$'s are Fermi superfields in representation $(-2,-2)$,
$p$'s are chiral superfields in the representation dual to $(3,3)$,
and $f_5$ is a degree five polynomial.

The second and fifth entries in table~\ref{table:g24-exs} are very similar:
the maps are between powers of $\det S^*$, and so are polynomials in the
$A^{ij}$, of degree 2 in the second entry and of degree 1 in the fifth
entry.

The third entry in the table is more interesting.
The bundles
\begin{displaymath}
{\cal O}(0,-1) \: = \: S \: = \:
(\det S^*)^{-1} \otimes S^*, 
\end{displaymath}
\begin{displaymath}
{\cal O}(1,-1) \: = \: (\det S^*)^{-1} \otimes {\cal O}(2,0) \: = \:
(\det S^*)^{-1} \otimes {\rm Sym}^2 S^* 
\: = \: (\det S^*)^{-1} \otimes K_{\tiny\yng(2)}
S^* ,
\end{displaymath}
(where for any Young diagram $T$ we use $K_T S^*$ to indicate a tensor
product of copies of $S^*$ built in the fashion indicated by $T$),
so we need to describe explicitly maps
\begin{displaymath}
S \: = \:
(\det S^*)^{-1} S^* \: \longrightarrow \:
(\det S^*)^{-1} {\rm Sym}^2 S^* .
\end{displaymath}
In principle, if $\Lambda^a$ couples to $S$, then such maps are of the form
\begin{displaymath}
\Lambda^a \phi^b_i \: + \: \Lambda^b \phi^a_i ,
\end{displaymath}
where the $\phi^a_i$ are sections of $S^*$ corresponding to the 
chiral superfields used to describe the underlying Grassmannian.
If we let $p_{ab}$ denote the chiral superfield in the representation dual
to $(1,-1)$, then the (0,2) superpotential for this case is of the form
\begin{displaymath}
W \: = \: \left( \Lambda_n^a \phi_i^b \: + \: \Lambda_n^b \phi_i^a \right)
f^{in} p_{ab} ,
\end{displaymath}
where the $f^{in}$ are constants.

The fourth entry is also nontrivial.  Here the pertinent bundles are
\begin{displaymath}
{\cal O}(1,-2) \: = \: (\det S^*)^{-2} \otimes {\cal O}(3,0) \: = \:
(\det S^*)^{-2} \otimes {\rm Sym}^3 S^* \: = \:
(\det S^*)^{-2} \otimes K_{\tiny\yng(3)} S^*,
\end{displaymath}
\begin{displaymath}
{\cal O}(2,-2) \: = \: (\det S^*)^{-2} \otimes {\cal O}(4,0) \: = \:
(\det S^*)^{-2} \otimes {\rm Sym}^4 S^*  \: = \:
(\det S^*)^{-2} \otimes K_{\tiny\yng(4)} S^*,
\end{displaymath}
so we need to describe explicitly maps
\begin{displaymath}
(\det S^*)^{-2} \otimes {\rm Sym}^3 S^*
\: \longrightarrow \:
(\det S^*)^{-2} \otimes {\rm Sym}^4 S^* .
\end{displaymath}
We can build such maps in much the same form as for the third entry.
If $\Lambda^{abc}$ couples to $(\det S^*)^{-2} \otimes {\rm Sym}^3 S^*$,
then the needed map is of the form
\begin{displaymath}
\Lambda^{abc} \phi_i^d \: + \: (\mbox{symmetric permutations of }a, b, c, d) ,
\end{displaymath}
and so the (0,2) superpotential for this model is of the form
\begin{displaymath}
W \: = \:
\left( \Lambda^{abc}_n \phi_i^d \: + \: (\mbox{perm's}) \right) f^{ni}
p_{abcd} ,
\end{displaymath}
where $p_{abcd}$ is a chiral superfield in the representation dual to
$(2,-2)$, and $f^{ni}$ are constants as before.

\subsection{Abelian/nonabelian duality to projective space}
\label{sect:g24-duality}

In four dimensions, a Seiberg-like duality between an abelian and a
nonabelian gauge theory seems impossible, as only one of the two could
be asymptotically free.  In two dimensions, however, since the gauge
field does not describe a propagating degree of freedom, more exotic
possibilities exist, including dualities between abelian and nonabelian
gauge theories.

One such example was discussed implicitly in
\cite{beht}[section 4.6], as part of their discussion of
the duality between GLSMs for the Grassmannians $G(k,n)$ and
$G(n-k,n)$.  In the special case $k=1$, this relates
$G(1,n) = {\mathbb P}^{n-1}$, described by an abelian gauge
theory, to $G(n-1,n)$, described by a $U(n-1)$ gauge theory.
Elliptic genera of these two theories were compared in
\cite{beht}[section 4.6], as were elliptic genera for more general
values of $k$, and found to match exactly as one would expect.
(Note that (0,2) dualities built on the equivalence $G(k,n) = G(n-k,n)$
will be described in section~\ref{sect:ggp}.)

We propose that additional dualities of analogous forms should also
exist.  For example,
the Grassmannian $G(2,4)$ has the unusual property\footnote{
We will discuss generalizations of this duality in 
section~\ref{sect:g2n-duals}.
} that it is the same
as a quadric hypersurface in ${\mathbb P}^5$ (see for example
\cite{meron} and references therein), which lends itself to
a natural proposal for another duality between (2,2) supersymmetric
abelian and nonabelian
gauge theories, a duality between the GLSMs for these two presentations of
the same space.  In weak coupling regimes, because the geometries
described are identical, one immediately, trivially, has a matching
between Higgs phases, chiral rings, and global symmetries, which in
four dimensions would typically be sufficient to demonstrate the
existence of the duality.  However, to be thorough, 
in appendix~\ref{app:dual-check}, we also check that global symmetries
and elliptic genera
match, consistent with the proposed duality.

Now, we would also like to use similar mathematical ideas to make
predictions for dualities between (0,2) theories describing gauge
bundles on the spaces above, and to do so, we need to relate bundles
on these dual mathematical descriptions.  For example, 
the universal subbundle and quotient
bundle on $G(2,4)$ correspond to the two spinor bundles on
${\mathbb P}^5[2]$ \cite{ott1,donagipriv}.  To systematically compare
(0,2) GLSMs on $G(2,4)$ to (0,2) GLSMs on ${\mathbb P}^5[2]$,
the essential ingredient is to compare
the restriction of ${\cal O}(1) \rightarrow {\mathbb P}^5$ to the hypersurface,
to a bundle on $G(2,4)$.  Now, sections of the restriction of ${\cal O}(1)$
are just homogeneous coordinates, {\it i.e.}
\begin{displaymath}
B_{ij} \: = \: \phi_i^a \phi_j^b \epsilon_{ab} \mbox{ on } G(2,4) .
\end{displaymath}
The homogeneous coordinates above are sections of $\det S^*$ on $G(2,4)$,
so the restriction of ${\cal O}(1)$ to ${\mathbb P}^5[2]$ is equivalent
to the line bundle $\det S^*$ on $G(2,4)$.  As a consistency check,
note that both have $c_1=1$.  Given this dictionary, from a (0,2) model
on ${\mathbb P}^5[2]$, in principle one could build (0,2) models on $G(2,4)$.
The converse, building (0,2) GLSMs for ${\mathbb P}^5[2]$ from those for
$G(2,4)$, could in principle be done using the fact that the universal
subbundle on $G(2,4)$ maps to a spinor bundle, and the (0,2) GLSM on
$G(2,4)$ will build bundles from tensor products, duals, and so forth of
the universal subbundle.

If two weakly-coupled (0,2) GLSMs describe the same geometry and
gauge bundle, then as before, Higgs phases, chiral rings, and global
symmetries all immediately match, which is the reason we claim a duality.

Let us work through a few specific examples.
Consider the first entry in table~\ref{table:g24-exs}.
This describes the gauge bundle
\begin{displaymath}
0 \: \longrightarrow \: {\cal E} \: \longrightarrow \:
\oplus^5 (\det S^*)^{-2} \: \longrightarrow \: \oplus^2 (\det S^*)^3
\: \longrightarrow \: 0
\end{displaymath}
on $G(2,4)$, which from our analysis above
is the same as the (0,2) GLSM on ${\mathbb P}^5[2]$ with
gauge bundle described as
\begin{displaymath}
0 \: \longrightarrow \: {\cal E} \: \longrightarrow \: \oplus^5 {\cal O}(-2)
\: \longrightarrow \: \oplus^2 {\cal O}(3) \: \longrightarrow \: 0 .
\end{displaymath}
The abelian (0,2) model on ${\mathbb P}^5[2]$ is anomaly-free, just as
its dual on $G(2,4)$.  
The map ${\cal O}(-2)^5 \rightarrow {\cal O}(3)^2$ is defined by
homogeneous polynomials of degree $5$, just as in the analysis of the
bundle on $G(2,4)$.
In a little more detail, we can identify the six
baryons $B_{ij}$ on $G(2,4)$
with homogeneous coordinates $z_{ij}$ on ${\mathbb P}^5$, and
thereby build maps on ${\mathbb P}^5[2]$ from maps on $G(2,4)$.
For example,
\begin{displaymath}
B_{12} B_{13} (B_{24})^3 \: \mapsto \:
z_{12} z_{13} (z_{24})^3. 
\end{displaymath}
That said, the baryons $B^{ij}$ satisfy some additional consistency
conditions, more than just homogeneous coordinates, which are encoded in
the quadric hypersurface condition.  In this fashion, we can construct
a (0,2) GLSM on ${\mathbb P}^5[2]$ from the first entry in 
table~\ref{table:g24-exs}. 

Conversely, given a homogeneous polynomial $p$ on ${\mathbb P}^5$ of degree $n$,
we can construct a section of $(\det S^*)^n$ on $G(2,4)$, by mapping
$z_{ij} \mapsto B_{ij}$.  Some of the terms will drop out after 
making the identification, because the $B_{ij}$'s satisfy an algebraic
equation encoded in the quadric hypersurface.  Put another way, if the
relation between the $B_{ij}$'s is encoded in a quadric $q$,
then to find the remainder after mapping $z_{ij} \mapsto B_{ij}$
we divide:
\begin{displaymath}
p \: = \: m q \: + \: r ,
\end{displaymath}
where $m$ is some degree $n-2$ polynomial and $r$ is a degree $n$ polynomial.
After mapping to $G(2,4)$, the factor $m q$ vanishes automatically,
since by definition $q(B_{ij}) = 0$,
leaving one just with the homogeneous polynomial $r = r(B_{ij})$.

The second and fifth entries in table~\ref{table:g24-exs} are very similar.
The second entry corresponds to the gauge bundle
\begin{displaymath}
0 \: \longrightarrow \: {\cal E} \: \longrightarrow \:
\oplus^3 {\cal O}(-1) \: \longrightarrow \: {\cal O}(1) \:
\longrightarrow \: 0
\end{displaymath}
on ${\mathbb P}^5[2]$, and the fifth entry corresponds to the gauge bundle
\begin{displaymath}
0 \: \longrightarrow \: {\cal E} \: \longrightarrow \:
\oplus^5 {\cal O}(2) \: \longrightarrow \: \oplus^2 {\cal O}(3) \:
\longrightarrow \: 0
\end{displaymath}
on ${\mathbb P}^5[2]$.
Both of these define anomaly-free abelian (0,2) gauge theories.

We have outlined above how to convert (0,2) GLSMs between $G(2,4)$
and ${\mathbb P}^5[2]$, implicitly using the fact that the GLSM for
$G(2,4)$ is built from special homogeneous bundles, {\it i.e.} bundles
defined by $U(2)$ representations, whereas even a general homogeneous
bundle would require a $U(2) \times U(2)$ representation, and analogous
properties of (0,2) GLSMs for complete intersections in projective spaces.
To map a general bundle, one not expressed in terms of a three-term
sequence in which the other terms are of the form above, would in principle
be more complicated.

Examples of this latter form are provided by the third and fourth
entries in table~\ref{table:g24-exs}.  Here, we are not aware of a 
three-term sequence describing symmetric powers of the spinor bundle on
${\mathbb P}^5[2]$, hence we do not understand how to map those
(0,2) GLSMs on $G(2,4)$ to (0,2) GLSMs on ${\mathbb P}^5[2]$.

\subsection{Supersymmetry breaking and checks of dualities}

The purpose of this section has been to give basic toy examples to illustrate
features of the technology of nonabelian (0,2) GLSMs, not to give
viable compactification candidates.  Nevertheless, for completeness,
in this subsection we will check both dualities
and supersymmetry breaking in the examples in table~\ref{table:g24-exs},
by computing elliptic genera.  In particular, we will see the following
interesting results:
\begin{itemize}
\item Although the left $U(1)$ is anomalous, we will formally compute
elliptic genera for all $y$.  (As previously discussed, naively
Jeffrey-Kirwan residue formulas for GLSM elliptic genera can be defined
regardless of whether currents are anomalous.  We leave a proper discussion
of the mathematical interpretation of such genera, if indeed a mathematical
interpretation exists, to future work.)
We will discover that for general $y$,
elliptic genera of duals match.  In principle, as only $y=+1$ is physically
meaningful, such a matching is not necessary.  We are not currently sure
how to interpret this.  Perhaps, for example, the methods we use implicitly
make a gauge choice, and the same gauge choice is being applied to both
genera in each pair.  In any event, it is an intriguing test of duality.
\item At $y=+1$, we will see evidence that
the elliptic genera all vanish, and in particular,
both genera of dual pairs vanish, suggesting that
supersymmetry is broken, and is broken for both elements of the pair.
This is consistent with our expectations:  at weak coupling in two dimensions,
since the gauge field is not dynamical, whether supersymmetry breaks should
be a function of the low-energy nonlinear sigma model, independent of the
details of the presentation of the UV GLSM.
\end{itemize} 

First, a general observation on the entries in that table.
The second, third, and fourth entries obey
\begin{displaymath}
\det {\cal E} \: \cong \: K_X
\end{displaymath}
and so, for example, admit B/2 twists.  The fifth entry obeys
\begin{displaymath}
\det {\cal E}^* \: \cong \: K_X
\end{displaymath}
and so, for example, admits an A/2 twist.
The first entry obeys neither
condition, but does satisfy $c_1({\cal E}) = 
c_1(TX) $ mod 2, hence we can at least define and compute elliptic genera.

Of these examples, the first, second, and fifth entries in 
table~\ref{table:g24-exs}
admit abelian duals, so we will focus on those.

Following the methods in \cite{beht}
and appendix~\ref{app:02-ell-gen}, the elliptic genus
for the first entry in table~\ref{table:g24-exs} is a residue of
\begin{displaymath}
- 2 \pi^2 i \eta(q)^7 
\frac{
\theta_1(q, x_1 x_2^{-1}) \theta_1(q, x_1^{-1} x_2) \theta_1(q, y x_1^{-2}
x_2^{-2})
}{
\theta_1(q,x_1)^4 \theta_1(q,x_2)^4 \theta_1(q,y^{-1} x_1^{-3} x_2^{-3})
}.
\end{displaymath}
The elliptic genus of the abelian dual is a residue of
\begin{displaymath}
2 \pi i \eta(q)^4 
\frac{
\theta_1(q, x^{-2}) \theta_1(q, y x^{-2})^5 
}{
\theta_1(q,x)^6 \theta_1(q, y^{-1} x^{-3})^2
}.
\end{displaymath}
The first few terms in power series in $q$ for both of these elliptic genera
match, and are given by
\begin{eqnarray*}
\lefteqn{
q^{-1/12} \frac{1}{y^{3/2} (y-1)}
\left(
336 + 1559 y + 2460 y^2 + 1559 y^3 + 336 y^4 \right)
} \\
& & \: + \:
q^{11/12} \frac{1}{y^{5/2}(y-1)}
\left(
-4025 - 7137 y + 7157 y^2 + 20510 y^3 + 7157 y^4 - 7137 y^5 - 4025 y^6
\right)
\\
& &
\: + \:
q^{23/12} \frac{1}{y^{7/2}(y-1)} 
\left(
15203 - 23272 y - 91869 y^2 + 31081 y^3 + 168964 y^4 + 31081 y^5
\right. \\
& & \hspace*{2.5in} \left.
- 91869 y^6 - 23272 y^7 + 15203 y^8
\right) 
\: + \: {\cal O}\left( q^{35/12} \right).
\end{eqnarray*} 
Although only the special case $y=+1$ is physically meaningful, it is
at least an intriguing test of dualities that these series match for
more general $y$, something that we will also see in the other
examples in this subsection.  This might reflect a gauge choice implicit in
\cite{beht}, which matches for both computations.  We leave the precise
understanding of this matching for anomalous cases for future work.

Now, let us turn to the physically meaningful case $y=+1$.
Judging from the expressions above, it would appear naively that
the elliptic genus must diverge at $y=+1$; however, one should be careful,
as limits and residues do not commute.  For a simple example, consider
$f(z,u) = 1/(z+u)$.  For this function,
\begin{eqnarray*}
{\rm Res}_{u=0} \left( {\rm Lim}_{z \rightarrow 0} f(z,u) \right)
& = & 1, \\
{\rm Lim}_{z \rightarrow 0} \left( {\rm Res}_{u=0} f(z,u) \right)
& = & 0.
\end{eqnarray*}
In particular, for both of the elliptic genera above, if we first take
$y=+1$ and then compute the residue, we find that the residue
vanishes.  That computation, at $y=+1$, is a bit too naive, as the
pole intersections are, in the language of \cite{beht}, nonprojective,
and so correct version of the Jeffrey-Kirwan residue could be more
complicated.  

That said, we can also independently compute the leading term in the
elliptic genus.  From equation~(\ref{eq:32}), the leading term is
proportional to
\begin{displaymath}
\int_X \hat{A}(TX) \wedge {\rm ch}\left(
( \det {\cal E} )^{+1/2} \wedge_{-1}( {\cal E}^* )
\right)
\end{displaymath}
In general,
\begin{displaymath}
\hat{A}(TX) \: = \: 1 \: - \: \frac{1}{24} \left( c_1(TX)^2 - 2 c_2(TX)
\right) \: + \: (\mbox{degree }8)
\end{displaymath}
and for a rank 3 bundle,
\begin{displaymath}
{\rm ch}\left(
( \det {\cal E} )^{+1/2} \wedge_{-1}( {\cal E}^* )
\right)
\: = \: c_3({\cal E}) \: + \: (\mbox{degree } 5)
\end{displaymath}
hence the leading term vanishes, in agreement with the extremely
naive computation at $y=+1$ above.

If the elliptic genus does in fact vanish at $y=+1$, it suggests that
supersymmetry may be broken dynamically.
It is important to note that
both of the elliptic genera should vanish -- supersymmetry does not break in one
and remain unbroken in the other.  This is because at weak coupling in
two dimensions, since gauge fields have no propagating degrees of freedom,
whether supersymmetry breaks is a function of the low-energy nonlinear
sigma model, independent of the details of the presentation of the UV
GLSM.  The fact that both the elliptic genera vanish is a (weak) check of
the claimed duality.

Later, in discussing Calabi-Yau compactifications, we will see
closely related abelian-nonabelian (0,2) dualities in which supersymmetry
is not broken.

Following the methods in \cite{beht} 
and appendix~\ref{app:02-ell-gen}, the elliptic genus
for the second entry in table~\ref{table:g24-exs} is a residue of
\begin{displaymath}
\frac{ (2\pi)^2 }{2} \eta(q)^8 
\frac{ \theta_1(q,x_1 x_2^{-1}) \theta_1(q,x_2 x_1^{-1}) 
\theta_1(q,y x_1^{-1} x_2^{-1})^3 }{
\theta_1(q, x_1)^4 \theta_1(q,x_2)^4 \theta_1(q,y^{-1}x_1^{-1} x_2^{-1})
} .
\end{displaymath}
The elliptic genus of the abelian dual is a residue of
\begin{displaymath}
- 2 \pi \eta(q)^5 \frac{ \theta_1(q,x^{-2}) \theta_1(q,yx^{-1}) }{
\theta_1(q,x)^6 \theta_1(q,y x^{-1}) } .
\end{displaymath}
The first few terms in power series in $q$ for both of these elliptic
genera match, and are given by
\begin{eqnarray*}
\lefteqn{
q^{-1/6} \frac{ (1+y)^4 }{ y (y-1)^2 } \: + \:
q^{5/6} \frac{1}{2 y^2 (y-1)^2} \left( -36 + 68 y^2 + 64y^3 + 68 y^4 - 
36 y^6 \right)
} \\
& & \: + \:
q^{11/6} \frac{ (1+y)^2 }{y^3 (y-1)^2} \left(
57 - 360 y + 661 y^2 - 660 y^3 + 661 y^4 - 360 y^5 + 57 y^6 \right)
\: + \: {\cal O}\left( q^{17/6} \right).
\end{eqnarray*}
As before, the fact that these expressions match for $y \neq +1$ is
an intriguing test of duality.

Now, let us turn to the physically meaningful special case $y=+1$.
As before, limits and residues do not commute.  For both elliptic genera,
taking the limit $y \rightarrow +1$ and then evaluating the residue, one 
finds that naively, ignoring subtleties due to non-projective
intersections, both of the elliptic genera vanish for $y=+1$, suggesting
that supersymmetry may be broken dynamically.  As before, both of the
genera vanish:  any supersymmetry breaking that occurs, must happen below
the scale at which the nonlinear sigma model becomes a pertinent description.

Now, let us compare the result above to the prediction of
section~\ref{sect:genl-susy-breaking}.  This model satisfies 
$\det {\cal E} \cong K_X$, so the leading term in the elliptic genus
is predicted to be proportional to
\begin{displaymath}
q^{(r-n)/12} \sum_{s=0}^r (-)^s \chi(\wedge^s {\cal E}).
\end{displaymath}
It is straightforward to compute that in this example,
\begin{displaymath}
\chi({\cal O}) \: = \: 1, \: \: \:
\chi({\cal E}) \: = \: 2, \: \: \:
\chi(\wedge^2 {\cal E}) \: = \: 1,
\end{displaymath}
hence
\begin{displaymath}
\sum_{s=0}^r (-)^s \chi\left( \wedge^s {\cal E} \right) \: = \: 0,
\end{displaymath}
in agreement with the naive direct computations.

For the fifth entry in table~\ref{table:g24-exs}, the elliptic genus is
a residue of
\begin{displaymath}
- \frac{i}{2} (2 \pi)^2 \eta(q)^7 \frac{
\theta_1(q,x_1 x_2^{-1}) \theta_1(q,x_2 x_1^{-1}) \theta_1(q,y+x_1^2 x_2^2)^5
}{
\theta_1(q,x_1)^4 \theta_1(q,x_2)^4 \theta_1(q,y^{-1}x_1^{-3}x_2^{-3})^2
}.
\end{displaymath}
The elliptic genus of the abelian dual is a residue of
\begin{displaymath}
+ 2 \pi i \eta(q)^4 \frac{
\theta_1(q,x^{-2}) \theta_1(q,y x^2)^5 
}{
\theta_1(q,x)^6 \theta_1(q,y^{-1} x^{-3})^2
}.
\end{displaymath}
The first few terms in power series in $q$ of these two elliptic genera
match perfectly:
\begin{eqnarray*}
\lefteqn{
\frac{q^{-1/12} }{y^{3/2} (y-1) } \left( 1 - y + 10 y^2 - y^3 + y^4 \right)
\: + \:
\frac{q^{11/12} }{y^{3/2} (y-1) } \left( - 17 + 12 y + 30 y^2 + 12 y^3 
- 17 y^4 \right)
} \\
& & \: + \:
\frac{q^{23/12} }{y^{7/2} (y-1) } \left( 98 - 207 y - 54 y^2 + 216 y^3
- 56 y^4 + 216 y^5 - 54 y^6 - 207 y^7 + 98 y^8 \right)
\\
& & \: + \: {\cal O}\left( q^{35/12} \right).
\end{eqnarray*}
As before, the fact that the two genera match in this form is an interesting
check of duality; however, only the special case $y=+1$ is physically
meaningful.

As before, limits and residues do not commute.  For both elliptic
genera, taking the limit $y\rightarrow
+1$ and then evaluating the residue, one finds that both of the
elliptic genera vanish for $y=+1$.  Ignoring as before subtleties in
non-projective intersections, this suggests that supersymmetry
may be broken dynamically.  As before, both of the genera vanish:
any supersymmetry breaking that occurs, must happen below the scale
at which the nonlinear sigma model becomes a pertinent description.

Now, let us compare to the predictions of 
section~\ref{sect:genl-susy-breaking}.
This model satisfies $\det {\cal E}^* \cong K_X$, so the leading term
in the elliptic genus is predicted to be
\begin{displaymath}
q^{(r-n)/12} \sum_{s=0}^r (-)^s \chi( \wedge^s {\cal E}^* ).
\end{displaymath}
It is straightforward to compute that
\begin{displaymath}
\chi({\cal O}) \: = \: 1, \: \: \:
\chi({\cal E}^*) \: = \: 0, \: \: \:
\chi(\wedge^2 {\cal E}^* ) \: = \: 0, \: \: \:
\chi(\wedge^3 {\cal E}^*) \: = \: 1,
\end{displaymath}
hence the leading term vanishes,
matching our naive computation above.

\section{Calabi-Yau and related examples}
\label{sect:other-cases}

\subsection{Examples on $G(2,4)[4]$}
\label{sect:g24-4}

To build a (0,2) GLSM for a complete intersection, we follow a pattern similar
to that in abelian (0,2) GLSMs:  for each hypersurface $\{ G_a = 0 \}$
(degree $d_a$) in the complete
intersection, we add a Fermi superfield $\Gamma^a$, charged under
$\det U(k)$ with charge $-k d_a$
({\it i.e.} couples to bundle $(\det S^*)^{- d_a} = 
{\cal O}(- d_a,- d_a)$),
and a (0,2) superpotential term
\begin{displaymath}
W \: = \: \Gamma^a G_a(\phi) .
\end{displaymath}
Integrating out the auxiliary field in $\Gamma^a$ forces the vacua to
lie along $\{ G_a = 0 \}$.  The reason for the charge assignments lies in how
the polynomials $G_a$ are defined.  Specifically, these are functions of
baryons in the $U(k)$ theory ({\it i.e.} homogeneous coordinates in the
Pl\"ucker embedding),
\begin{displaymath}
B_{i_1 \cdots i_k} \: = \: \epsilon_{a_1 \cdots a_k} \phi_{i_1}^{a_1} \cdots
\phi_{i_k}^{a_k} ,
\end{displaymath}
which each have $\det U(k)$ charge $k$.

In this language, the Calabi-Yau condition for a complete intersection of
hypersurfaces in $G(k,n)$ is that the sum of the degrees of the hypersurfaces
equals $n$:
\begin{displaymath}
\sum_a d_a \: = \: n .
\end{displaymath}

In table~\ref{table:ci-exs} 
we list anomaly-free examples of bundles ${\cal E}$ of the form
\begin{displaymath}
0 \: \rightarrow \: {\cal E} \: \rightarrow \:
\oplus^{m_1} {\cal O}(\lambda_{A1}, \lambda_{B1}) 
\oplus^{m_2} {\cal O}(\lambda_{A2}, \lambda_{B2})
\: \rightarrow \:
\oplus^{n_1} {\cal O}(\lambda_{A3}, \lambda_{B3}) 
\oplus^{n_2} {\cal O}(\lambda_{A4}, \lambda_{B4}) \: \rightarrow \: 0
\end{displaymath}
on $G(2,4)[4]$ with $c_1({\cal E}) = 0$.
For bundles of the form above,
\begin{eqnarray*}
c_1({\cal E}) & = & m_1 c_1( {\cal O}(\lambda_{A1}, \lambda_{B1}) )
\: + \: m_2 c_1( {\cal O}(\lambda_{A2}, \lambda_{B2}) )
\\
& & \hspace*{0.5in}
\: - \: n_1 c_1( {\cal O}(\lambda_{A3}, \lambda_{B3}) )
\: - \: n_2 c_1( {\cal O}(\lambda_{A4}, \lambda_{B4}) ) , \\
& \propto & d_{(\lambda_{A1}, \lambda_{B1})} {\rm Cas}_1(\lambda_{A1},
\lambda_{B1}) \: + \:
d_{(\lambda_{A2},\lambda_{B2})} {\rm Cas}_1(\lambda_{A2},\lambda_{B2})
\\
& & \hspace*{0.5in} \: - \:
d_{(\lambda_{A3},\lambda_{B3})} {\rm Cas}_1(\lambda_{A3},\lambda_{B3})
\: - \:
d_{(\lambda_{A4},\lambda_{B4})} {\rm Cas}_1(\lambda_{A4},\lambda_{B4}) .
\end{eqnarray*}

\begin{table}[h]
\begin{center}
\begin{tabular}{ccccccccc}
$m_1$ & $(\lambda_{A1}, \lambda_{B1})$ & $m_2$ & $(\lambda_{A2}, \lambda_{B2})$ & $n_1$ & $(\lambda_{A3}, \lambda_{B3})$ & $n_2$ & $(\lambda_{A4}, \lambda_{B4})$ & rank \\ \hline
1&(1, 0)&5&(2, 1)&1&(3, 1)&2&(3, 2)&5 \\
3&(1, 1)&5&(1, 1)&1&(2, 2)&2&(3, 3)&5 \\
5&(1, 1)&5&(2, 0)&4&(2, 2)&2&(3, 0)&8 \\
2&(1, 1)&5&(2, 2)&1&(3, 3)&3&(3, 3)&3 \\
5&(1, 1)&2&(2, 2)&1&(3, 3)&2&(3, 3)&4 \\
2&(2, 1)&5&(2, 2)&2&(3, 2)&2&(3, 3)&3 \\
\end{tabular}
\caption{Anomaly-free examples on $G(2,4)[4]$.}
\label{table:ci-exs}
\end{center}
\end{table}

Let us examine carefully the first entry in table~\ref{table:ci-exs}.
The field content of the (0,2) GLSM pertinent to anomalies is as follows:
\begin{itemize}
\item 1 Fermi superfield in representation (1,0) (for the middle
term defining ${\cal E}$),
\item 5 Fermi superfields in representation (2,1) (for the middle term
defining ${\cal E}$),
\item 1 Fermi superfield $\Gamma$ 
in representation (-4,-4) (for the hypersurface),
\item 1 left-moving gaugino in the adjoint,
\item 4 chiral superfields in the fundamental (1,0) (defining the Grassmannian),
\item 1 chiral superfield in the dual of (3,1) (corresponding to the last
term defining ${\cal E}$),
\item 2 chiral superfields in the dual of (3,2) (corresponding to
the last term defining ${\cal E}$).
\end{itemize}
It is straightforward to check that this field content is anomaly-free,
and defines a theory with $c_1({\cal E}) = 0$.

As another consistency check, let us compute the left and right central
charges of the IR limits of the GLSM, applying c-extremization\footnote{
This is closely analogous to a-maximization in four-dimensional theories
\cite{iw1}.
} as discussed
in \cite{bb1} (see also \cite{ggp} for other recent applications).  
Briefly, the basic idea is that the central charge
can be determined using the fact that the symmetry that becomes the
R-symmetry in the IR SCFT will extremize trial central charges determined
by anomalies.
Consider for example the first entry in table~\ref{table:ci-exs}.
From the matter content listed above and the anomaly
\begin{displaymath}
c_R \: = \: 3 {\rm Tr} \, \gamma^3RR,
\end{displaymath}
one has the trial right-moving
central charge
\begin{displaymath}
c_R \: = \: 3(8(R_{\phi}-1)^2-12R_{\Lambda}^2+7(R_P-1)^2-R_{\Gamma}^2-4),
\end{displaymath}
where the $R$'s denote charges under the left $U(1)$.
We need to find R charges that extremize $c_R$.
Furthermore, from the superpotential terms, there are constraints.
Specifically, terms of the form 
\begin{displaymath}
\int d\theta^+\Gamma G
\end{displaymath}
yield
\begin{displaymath}
-1+R_{\Gamma}+8R_{\Phi} \: = \: 0,
\end{displaymath}
and terms of the form
\begin{displaymath}
\int d\theta^+\Lambda PF
\end{displaymath}
yield constraints
\begin{eqnarray*}
-1+R_\Lambda+R_P+3R_\Phi & = & 0,\\
-1+R_\Lambda+R_P+4R_\Phi & = & 0,\\
-1+R_\Lambda+R_P+R_\Phi & = & 0,\\
-1+R_\Lambda+R_P+2R_\Phi & = & 0.
\end{eqnarray*}
Extremizing the central charge gives
\begin{displaymath}
R_{\Phi}=0, \: \: \: R_{\Gamma}=1, \: \: \:
R_{\Lambda}=0, \: \: \:
R_P=1,
\end{displaymath}
which results in $c_R=9$.  The other central charge, $c_L$, can be
computed from $c_R-c_L={\rm Tr} \gamma^3=8-12+7-1-4=-2$, yielding
altogether $(c_R, c_L)=(9,11)$ for the first entry, exactly right to
describe a (0,2) theory on a 3-fold with a bundle of rank 5.
Proceeding in a similar fashion, the central charges of the other
entries in table~\ref{table:ci-exs} are computed to be
\begin{displaymath}
(c_R, c_L) \: = \:
(9,11), \: \: \: (9,14), \: \: \: (9,9),
\: \: \: (9,10), \: \: \: (9,9),
\end{displaymath}
respectively, exactly correct for the given ranks and dimensions.
We take this as 
evidence for the existence of nontrivial IR
fixed points in these theories.

Maps are given in the same fashion as discussed earlier for 
bundles on $G(2,4)$.  For example,
maps ${\cal O}(1,0) \rightarrow {\cal O}(3,1)$ are of the 
form
\begin{displaymath}
\Lambda^a \: \mapsto \: (\epsilon_{bc} \phi_i^b \phi_j^c)
\left( \Lambda^a \phi_k^b \: + \: \Lambda^b \phi_k^a \right)
\end{displaymath}
maps ${\cal O}(1,0) \rightarrow {\cal O}(3,2)$ are of the form
\begin{displaymath}
\Lambda^a \: \mapsto \: (\epsilon_{bc} \phi_i^b \phi_j^c)^2
\Lambda^a
\end{displaymath}
and so forth, leading to superpotential terms of the form discussed
previously.

Furthermore, just as in section~\ref{sect:g24-duality}, some of the
examples above can be rewritten as examples in the (0,2) GLSM for
${\mathbb P}^5[2]$.  For example, a complete intersection
$G(2,4)[d_1, \cdots, d_n]$ is the same as the complete intersection
\begin{displaymath}
{\mathbb P}^5[2,d_1, \cdots, d_n]
\end{displaymath}
and at least sometimes it is possible to map the bundles, consistent with
the structure of (0,2) GLSMs.  For example, the second entry in
table~\ref{table:ci-exs} corresponds to the bundle
\begin{displaymath}
0 \: \longrightarrow \: {\cal E} \: \longrightarrow \:
\oplus^8 {\cal O}(1) \: \longrightarrow \:
{\cal O}(2) \oplus^2 {\cal O}(3) \: \longrightarrow \: 0
\end{displaymath}
over ${\mathbb P}^5[2,4]$,
which is easily realized as an anomaly-free abelian (0,2) GLSM.
Similarly, the fourth entry in table~\ref{table:ci-exs} corresponds to the
anomaly-free bundle
\begin{displaymath}
0 \: \longrightarrow \: {\cal E} \: \longrightarrow \:
\oplus^2 {\cal O}(1) \oplus^5 {\cal O}(2) \: \longrightarrow \:
\oplus^4 {\cal O}(3) \: \longrightarrow \: 0
\end{displaymath}
on ${\mathbb P}^5[2,4]$, and the fifth entry in table~\ref{table:ci-exs}
corresponds to the anomaly-free bundle
\begin{displaymath}
0 \: \longrightarrow \: {\cal E} \: \oplus^5 {\cal O}(1) \oplus^2 {\cal O}(2)
\: \longrightarrow \: \oplus^3 {\cal O}(3) \: \longrightarrow \: 0
\end{displaymath}
on ${\mathbb P}^5[2,4]$.

In appendix~\ref{app:02-ell-gen} we work through the details of computations
of elliptic genera for the three nonabelian examples above and their
abelian duals.  In each case, the elliptic genera of the proposed duals
match, consistent with geometric expectations.  For the second
entry in table~\ref{table:ci-exs}, the first few terms in the $q$-expansion
of the elliptic genus are shown to be
\begin{eqnarray*}
\lefteqn{
72 \left( - y^{-1/2} + y^{+1/2} \right)^2 
\left( y^{-1/2} + y^{+1/2} \right) q^{1/6} 
} \\
& & 
\: - \: 72 \left( - y^{-1/2} + y^{+1/2} \right)^2
\left( y^{-1/2} + y^{+1/2} \right)^3 \left( y^{-1} - 1 + y \right) q^{7/6}
\\
& &
\: + \:
72 \left( - y^{-1/2} + y^{+1/2} \right)^2 \left( y^{-7/2} - y^{-3/2} +
2 y^{-1/2} + 2 y^{+1/2} - y^{+3/2} + y^{+7/2} \right) q^{13/6} \: + \:
{\cal O}\left(q^{19/6} \right).
\end{eqnarray*}
For the fourth entry in table~\ref{table:ci-exs}, the first few terms
in the $q$-expansion are shown to be
\begin{eqnarray*}
\lefteqn{
88 y^{-1/2} (1 + y) \: - \: 
88 y^{-5/2} \left( 1 - y^2 - y^3 + y^5 \right) q
} \\
& &
\: - \: 88 y^{-7/2} \left( 1 + y \right) \left( -1 + y^3 \right)^2 q^2
\: - \: 88 y^{-7/2} \left( -1 + y\right)^2 \left( 1 + y \right)^3 
\left( 1 + y + y^2 \right) \: + \: {\cal O}\left(q^4\right).
\end{eqnarray*}
For the fifth entry in table~\ref{table:ci-exs}, the first few terms in the
$q$-expansion are shown to be
\begin{eqnarray*}
\lefteqn{
80 \left(y - y^{-1}\right) q^{1/12} \: - \:
80\left(-y^{-3} + y^{-1} - y + y^3\right) q^{13/12}
} \\
& &
 \: - \:
80\left(-y^{-3} + 2 y^{-1} - 2 y + y^3\right) q^{25/12} \: + \:
{\cal O}\left(q^{37/12}\right).
\end{eqnarray*}
In each case, the leading term is independently checked.  Note that in
none of these cases do the elliptic genera vanish, hence we do not expect
supersymmetry breaking in any of these cases.

\subsection{Affine Grassmannians}

Let us next consider some
anomaly-free examples formally associated to the
affine Grassmannian over $G(k,n)$.  This Grassmannian is defined by an $SU(k)$
gauge theory with $n$ chiral multiplets in the fundamental representation.
The ordinary Grassmannian is defined by a $U(k)$ gauge theory with the
same matter.  (See for example \cite{meron}[section 2.5] for more information
on affine and weighted Grassmannians.)

Since the gauge group is $SU(k)$ rather than $U(k)$, there is no
continuously-variable Fayet-Iliopoulos parameter, and hence no way to take
a weak coupling large-radius limit in this theory, 
making any discussion of geometry
rather suspect.  Nevertheless, recently there has been interest in 
{\it e.g.} GLSMs for non-K\"ahler compactifications 
\cite{ael,adl,qss,mqs,mqss} in which the 
overall radius is also fixed.  Thus, with an eye towards applications
of that form, we include here a short discussion of $SU(k)$ GLSMs.
For simplicity, we will characterize them in geometric terms, though as already
noted, geometry should be applied with care here.

In table~\ref{table:affine-exs} we list some anomaly-free examples with
gauge bundle of the form
\begin{displaymath}
0 \: \rightarrow \: {\cal E} \: \rightarrow \:
\oplus^{m_1} {\cal O}(\lambda_{A1}, 0)
\oplus^{m_2} {\cal O}(\lambda_{A2}, 0)
\: \rightarrow \:
\oplus^{n} {\cal O}(\lambda_{A3}, 0)
 \: \rightarrow \: 0
\end{displaymath}
on affine $G(2,4)$.  We compute that for each of the examples in the
table, the elliptic genus vanishes identically, which we take as an
indication of possible dynamical supersymmetry breaking in these toy
models.  As these models have no weak coupling large-radius limit, we are
not surprised, but we list them here regardless as toy examples of the
technology. 

\begin{table}[h]
\begin{center}
\begin{tabular}{ccccccccc}
$m_1$ & $\lambda_{A1}$ & $m_2$ & $\lambda_{A2}$ & $n$ & $\lambda_{A3}$
 & rank \\ \hline
4&3&---&---&2&4&6\\
1&3&3&4&2&5&7\\
5&2&---&---&2&3&7\\
\end{tabular}
\caption{Examples on the affine Grassmannian over $G(2,4)$.}
\label{table:affine-exs}
\end{center}
\end{table}

For example, the first entry in the table involves maps of the form
\begin{displaymath}
\yng(3) \: \longrightarrow \: \yng(4) ,
\end{displaymath}
which are given by, schematically,
\begin{displaymath}
\Lambda^{abcd} \: \mapsto \: \Lambda^{abcd} \phi_i^e \: + \:
(\mbox{symmetric permutations}) .
\end{displaymath}
This would be given physically by a (0,2) superpotential of the form
\begin{displaymath}
W \: = \: f^i \left( \Lambda^{abcd} \phi_i^e \: + \: {\rm perm's} \right)
p_{abcde} ,
\end{displaymath}
where $f^i$'s are constants and
$p_{abcde}$ is a chiral superfield in the representation (dual to)
\begin{displaymath}
\yng(4) .
\end{displaymath}

Note that in general, in $SU(2)$ theories, there will be more possible
maps than in $U(2)$ theories, because one is not constrained by gauge
invariance under the overall $U(1)$.  For example, in an $SU(2)$ gauge
theory with matter in fundamental
representations, we can define a map
\begin{displaymath}
\yng(2) \: \longrightarrow \: \yng(1)
\end{displaymath}
by, schematically,
\begin{displaymath}
\Lambda^{ab} \: \mapsto \: \Lambda^{ab} \Phi^c_i \epsilon_{bc} ,
\end{displaymath}
where $\Lambda^{ab}$ couples to $\tiny\yng(2)$.  
This works because $\epsilon_{ab} \Phi^a$ is the dual of $\Phi^a$ in
an $SU(2)$ theory.  This is not true in a $U(2)$ gauge theory, and
there, a map of the form above would not respect the $\det U(2)$ charges.
Phased another way, our proposed map sends
\begin{displaymath}
\yng(2) \: \longrightarrow \: \yng(2,1) .
\end{displaymath}
As representations of $SU(2)$,
\begin{displaymath}
\yng(2,1) \: \cong \: \yng(1) ,
\end{displaymath}
but in $U(2)$, the representation $(2,1) \neq (1,0)$.

\section{Pfaffian constructions}
\label{sect:pfaff}

\subsection{Review of (2,2) constructions}

The paper \cite{jklmr1} gave two constructions of (2,2) GLSMs associated
to a given Pfaffian variety, denoted the PAX and PAXY models.
Schematically, for an $n \times n$ matrix $A$, each entry a homogeneous
function over some toric variety $V$, each construction defines
a Pfaffian variety given by the locus on $V$ where the 
\begin{displaymath}
{\rm rank}\, A \: \leq \: k
\end{displaymath}
for some $k$.  

In the PAX model, in addition to the gauge-theoretic
data defining the toric variety, one adds a $U(n-k)$ gauge theory
with two chiral superfields $P$, $X$, where $X$ transforms as $n$ copies of the
fundamental\footnote{
To make our (0,2) conventions cleaner, we have made a trivial convention
flip with respect to \cite{jklmr1}, in that $P$ and $X$ are defined in
opposite representations.
} of $U(n-k)$ and $P$ as $n$ copies of the antifundamental
of $U(n-k)$, together with a (2,2) superpotential
\begin{equation}  \label{eq:22-pax-sup}
W \: = \: {\rm tr}\, PAX
\end{equation}
from which the model derives its name.
$P$ and $X$ also have charges under the abelian gauge symmetry defining the
toric variety, so in effect, the model describes a superpotential over
a bundle with fibers that are the total spaces of
\begin{equation}  \label{eq:22pax-fibers}
S^{\oplus n} \: \longrightarrow \: G(n-k,n)
\end{equation}
fibered over the given toric variety.
All charges are required to be such that the 
superpotential~(\ref{eq:22-pax-sup})
is neutral.

The (2,2) GLSM above has two phases, which are closely related.
The D-terms give a constraint of the form
\begin{displaymath}
X X^{\dag}  \: - \: P^{\dag} P \: = \: r I ,
\end{displaymath}
where $r$ is a Fayet-Iliopoulos parameter associated to the overall
$U(1)$.  Without loss of generality, we shall take $r \gg 0$.
The F-terms give constraints of the form
\begin{displaymath}
A X \: = \: 0, \: \: \:
P A \: = \: 0, \: \: \:
P (d A) X \: = \: 0 .
\end{displaymath}
The first constraint defines the variety 
\begin{displaymath}
Z \: \equiv \: \{ (\phi, x) \, | \, A(\phi) x \: = \: 0 \} ,
\end{displaymath}
which is our desired (resolution of a) Pfaffian.  (The constraint forces
$X$ to describe $n-k$ null eigenvectors of $A$, and so only has solutions
when the rank of $A$ is bounded by $k$.)  Under a smoothness 
assumption, the second two F-term constraints imply $P=0$, as discussed
in \cite{jklmr1}[section 3.2].  Thus, we expect that this theory
flows at low energies to a nonlinear sigma model on $Z$.
Nearly an identical analysis applies when $r \ll 0$, except that the roles
of $X$ and $P$ are reversed.

In passing, let us work out the Calabi-Yau condition in a PAX model of
the form above.  First, note that the fibers~(\ref{eq:22pax-fibers}) are
already Calabi-Yau, so we merely need a constraint on charges of the
abelian gauge symmetries defining the underlying toric variety.

Specifically, the space will be Calabi-Yau if the sum of the $U(1)$ charges
vanishes, for each $U(1)$ defining the underlying toric variety.
For example, suppose the underlying toric variety is a projective
space, ${\mathbb P}^m$ for some $m$.  Let $p_i$ denote the $U(1)$ of the
$i$-th fundamental in $P$, and $x_i$ the $U(1)$ charge of the $i$-th
antifundamental in $X$.  Then the Calabi-Yau condition can be
succinctly stated as the condition
\begin{displaymath}
\sum_i (n-k) p_i \: + \: \sum_i (n-k) x_i \: + \: m + 1 \: = \: 0 ,
\end{displaymath}
where we have used the fact that the fundamentals and antifundamentals
both have dimension $n-k$.

In the (2,2) PAXY model, given the GLSM for the underlying
toric variety, one instead adds a $U(k)$ gauge theory
with $n$ fundamentals $\tilde{X}$, $n$ antifundamentals $\tilde{Y}$,
and an $n \times n$ matrix of neutral chiral superfields $\tilde{P}$, together
with a (2,2) superpotential
\begin{equation}  \label{eq:22-paxy-sup}
W \: = \: {\rm tr}\, \tilde{P} \left( A \: - \: \tilde{Y} \tilde{X} \right) .
\end{equation}
Here also, $\tilde{P}$, $\tilde{X}$, $\tilde{Y}$ are charged under the
abelian gauge symmetry defining the underlying toric variety, with
charges such that the superpotential~(\ref{eq:22-paxy-sup}) is gauge invariant.

For a PAXY model over ${\mathbb P}^m$, as before, the Calabi-Yau
condition would be
\begin{displaymath}
k \sum_i x_i \: + \: k \sum_i y_i \: + \: k \sum_i p_i \: + \: m \: + \: 1
\: = \: 0,
\end{displaymath}
where we have used the fact that the fundamentals and antifundamentals
have dimension $k$.

The PAX and PAXY models look different, but for a given Pfaffian,
are equivalent to one another, as we shall review in
section~\ref{sect:pax-vs-paxy}.

\subsection{More general (0,2) examples}

To understand (0,2) models on Pfaffians, let us begin by rewriting the
(2,2) PAX and PAXY models in (0,2) language.

Let us begin with the (0,2) PAX model.  Let $X$, $\Lambda_X$, denote
the (0,2) chiral, Fermi superfields associated to the (2,2) superfield
$X$, all describing $n$ copies of the fundamental, and let $P$, $\Lambda_P$
denote the (0,2) chiral, Fermi superfields associated to the (2,2)
superfield $P$, describing $n$ copies of the antifundamental.
Let $\Phi$, $\Lambda_{\Phi}$ denote the (0,2) chiral, Fermi superfields
associated to the (2,2) $\Phi$ defining the underlying toric variety.
This decomposition of the (2,2) theory also gives rise to an adjoint-valued
(0,2) chiral $\Sigma$, originating in the (2,2) gauge multiplet.

Then, the (0,2) theory is a $U(n-k)$ gauge theory with fields $P$, $\Lambda_P$,
$X$, $\Lambda_X$, $\Phi$, $\Lambda_{\Phi}$, obeying
\begin{displaymath}
\overline{D}_+ \Lambda_P \: \propto \Sigma P
\end{displaymath}
(and similarly for other Fermi superfields),
and with (0,2) superpotential
\begin{displaymath}
W \: = \: {\rm tr}\, \left( \Lambda_P A(\Phi) X \: + \:
P A(\Phi) \Lambda_X \: + \:
P \frac{\partial A(\Phi)}{\partial \Phi^{\alpha}} \Lambda^{\alpha}_{\Phi}
X \right) .
\end{displaymath}
Intuitively, for $r \gg 0$,
one can interpret $\Lambda_P$ as acting as a Lagrange
multiplier, forcing $A X=0$, and $\Lambda_X$, $\Lambda_{\Phi}^{\alpha}$
as describing the fermions in which the gauge bundle lives.

Given the structure above, we can read off the monad whose cohomology
defines the tangent bundle of the Pfaffian:
\begin{eqnarray}
\lefteqn{
0 \: \longrightarrow \: {\cal O}^r \oplus (S^* \otimes S) \: 
\stackrel{*_1}{\longrightarrow}
\: \oplus_{a,\alpha} {\cal O}((0,0),q_{a,\alpha})
 \oplus_i {\cal O}((1,0),x_{a,i})
} \nonumber \\
& \hspace*{2.5in} & 
\: \stackrel{*_2}{\longrightarrow} \:
\oplus_i {\cal O}((1,0),-p_{a,i}) 
\: \longrightarrow \: 0 ,
\label{eq:22-pax-monad}
\end{eqnarray}
where
\begin{displaymath}
*_1 \: = \: \left[ \begin{array}{cc}
q_{a, \alpha} \Phi^{\alpha} & 0 \\
x_{a} X &  X
\end{array} \right], \: \: \:
*_2 \: = \: \left[ 
 \frac{\partial A}{\partial \Phi^{\alpha}} X ,
A  \right] .
\end{displaymath}
As a consistency check, note that the composition of the two maps above
has the form
\begin{displaymath}
*_2 \, *_1 \: = \: \left[ 
q_{a, \alpha} \Phi^{\alpha} \frac{\partial A}{\partial \Phi^{\alpha}} X \: + \:
x_{a} A X ,
A X \right]
\: = \: \left[ 
p_{a} A X , A X  \right],
\end{displaymath}
which vanishes on the Pfaffian, as expected.
The monad above is determined by the field theory, as follows.
The ${\cal O}^r \oplus S^* \otimes S$ is determined by the gauginos;
the other terms are determined by remaining fermions.

Note that the Calabi-Yau condition implied by the monad above is of the
form
\begin{displaymath}
- (n-k) \sum_i p_{a,i} \: = \: (n-k) \sum_i x_{a,i} \: + \: \sum_{\alpha}
q_{a,\alpha}
\end{displaymath}
for each $a$, which specializes to the Calabi-Yau condition discussed
previously in (2,2) models.

A (0,2) deformation of the tangent bundle of the Pfaffian would be
described by a theory with the same matter content, but (0,2)
superpotential
\begin{displaymath}
W \: = \: {\rm tr}\, \left( \Lambda_P A(\Phi) X \: + \:
P A(\Phi) \Lambda_X \: + \:
P \left( \frac{\partial A(\Phi)}{\partial \Phi^{\alpha}} 
\: + \: G_{\alpha}(\Phi)\right) \Lambda^{\alpha}_{\Phi}
X \right) ,
\end{displaymath}
where
\begin{displaymath}
q_{a, \alpha} \Phi^{\alpha} G_{\alpha} \: = \: 0
\end{displaymath}
for each $a$.
This is described by a monad of the same form as in
equation~(\ref{eq:22-pax-monad}), but with maps
\begin{displaymath}
*_1 \: = \: \left[ \begin{array}{cc}
q_{a, \alpha} \Phi^{\alpha} & 0 \\
 x_{a} X & X 
\end{array} \right], \: \: \:
*_2 \: = \: \left[
\left( \frac{\partial A}{\partial \Phi^{\alpha}} X \: + \: G_{\alpha} \right), 
A \right] .
\end{displaymath}

A more general (0,2) model over a Pfaffian, describing a bundle built as
a kernel, based on the PAX model, can be built as follows.
First, to build the Pfaffian itself, we will need a $U(n-k)$
gauge theory, $n$ chiral superfields
in the fundamental, forming an $n\times (n-k)$ matrix denoted $X$,
and $n$ Fermi superfields in the antifundamental, forming an $n\times (n-k)$
matrix of Fermi superfields denoted $\Lambda_0$.  Then, to describe a bundle
${\cal E}$ as a kernel, say,
\begin{displaymath}
0 \: \longrightarrow \: {\cal E} \: \longrightarrow \:
\oplus_{\beta} {\cal O}((\lambda_{\beta 1}, \lambda_{\beta 2}), q_{a, \beta})
\: \stackrel{F^{\gamma}_{\beta} }{\longrightarrow} \:
\oplus_{\gamma} {\cal O}((\lambda_{\gamma 1}, \lambda_{\gamma 2}), 
q_{a, \gamma})
\: \longrightarrow \: 0 ,
\end{displaymath}
we add a set of Fermi superfields $\Lambda^{\beta}$ in the
$(\lambda_{\beta 1}, \lambda_{\beta 2})$ representation of $U(n-k)$
and with charges $q_{a, \beta}$ under the abelian gauge symmetry $U(1)^r$
defining
the toric variety, along with a set of chiral superfields
$P_{\gamma}$ in the $U(n-k)$ representation dual to
$(\lambda_{\gamma 1}, \lambda_{\gamma 2})$ and with charges $- q_{a, \gamma}$
under the abelian gauge symmetry defining the toric variety.
In addition, we have a (0,2) superpotential 
\begin{displaymath}
W \: = \:
{\rm tr}\, \left( \Lambda_0 A(\Phi) X \: + \:
\Lambda^{\beta} F^{\gamma}_{\beta}(\Phi) P_{\gamma}
\right) .
\end{displaymath}
Of course, all representations must be chosen to satisfy
gauge anomaly cancellation for this $U(n-k) \times U(1)^r$ gauge theory.
(Given the kernel construction above,
GLSMs for bundles built as cokernels and as cohomologies of monads
are very straightforward, and so for brevity are omitted.)

Let us briefly check the space of vacua in this theory.
From D-terms for $U(2)$ we have a constraint of the form
\begin{displaymath}
X X^{\dag}  \: + \: \sum_{\gamma} 
 P_{\gamma}^{\dag} P_{\gamma} \: = \: r I
\end{displaymath}
so, for suitable bundle representations, as discussed previously
in section~\ref{sect:g24-exs},
the $X$'s are not all zero.  From the F terms we get
the constraint
\begin{displaymath}
A X \: = \: 0 ,
\end{displaymath}
which describes the underlying Pfaffian variety.  So long as the
nontrivial map determined by $F_{\beta}^{\gamma}$ is surjective,
the $P_{\gamma}$ chiral superfields will all become massive, leaving us
with a gauge bundle contained within the associated Fermi superfields,
as expected.

To get a bundle with $c_1({\cal E}) = 0$, we impose the conditions
\begin{displaymath}
\sum_{\beta} d_{\lambda \beta} {\rm Cas}_1(\lambda_{\beta 1}, \lambda_{\beta 2})\: = \:
\sum_{\gamma} d_{\lambda \gamma} {\rm Cas}_1(\lambda_{\gamma 1},
\lambda_{\gamma 2}) ,
\end{displaymath}
\begin{displaymath}
\sum_{\beta} q_{a, \beta} \: = \: \sum_{\gamma} q_{a, \gamma} .
\end{displaymath}

So far we have outlined (0,2) versions of the PAX model.  Let us now
briefly outline analogues for the PAXY model.
Here, if we start with the (2,2) model and write it in (0,2) language,
following the same convention as previously for the PAX model, we are led
to a $U(k)$ gauge
theory with (0,2) chiral superfields $\tilde{P}$, $\tilde{X}$,
$\tilde{Y}$, $\Phi^{\alpha}$, (0,2) Fermi superfields $\Lambda_{\tilde{P}}$,
$\Lambda_{\tilde{X}}$, $\Lambda_{\tilde{Y}}$, $\Lambda^{\alpha}_{\Phi}$,
and a (0,2)
superpotential of the form
\begin{equation}
W \: = \: {\rm tr}\, \left( \Lambda_{\tilde{P}} \left( A \: - \: 
\tilde{Y} \tilde{X} \right) \: + \:
\tilde{P}\left( \frac{\partial A}{\partial \Phi^{\alpha}}
\Lambda^{\alpha}_{\Phi} \: - \: \Lambda_{\tilde{Y}} \tilde{X}
\: - \: \tilde{Y} \Lambda_{\tilde{X}} \right) \right) .
\end{equation}
A (0,2) theory describing a deformation of the tangent bundle is defined
by the superpotential
\begin{equation}  \label{eq:paxy-02-def}
W \: = \: {\rm tr}\, \left( \Lambda_{\tilde{P}} \left( A \: - \: 
\tilde{Y} \tilde{X} \right) \: + \:
\tilde{P}\left( \left( \frac{\partial A}{\partial \Phi^{\alpha}}
\: + \: G_{\alpha} \right)
\Lambda^{\alpha}_{\Phi} \: - \: \Lambda_{\tilde{Y}} \tilde{X}
\: - \: \tilde{Y} \Lambda_{\tilde{X}} \right) \right) .
\end{equation}

Now, consider a (0,2) theory describing a gauge bundle ${\cal E}$,
given as a kernel
\begin{equation}   \label{eq:02-bundle-paxy}
0 \: \longrightarrow \: {\cal E} \: \longrightarrow \:
\oplus_{\beta} {\cal O}((\lambda_{\beta 1}, \lambda_{\beta 2}), q_{a, \beta})
\: \stackrel{F^{\gamma}_{\beta} }{\longrightarrow} \:
\oplus_{\gamma} {\cal O}((\lambda_{\gamma 1}, \lambda_{\gamma 2}), 
q_{a, \gamma})
\: \longrightarrow \: 0
\end{equation}
over the Pfaffian.  We can describe this following the PAXY pattern as 
follows.  Given the abelian gauge theory for the toric variety,
we add a $U(k)$ gauge theory with 
\begin{itemize}
\item $n$ chiral superfields in the fundamental,
forming a matrix $\tilde{X}$, 
\item $n$ chiral superfields in the antifundamental,
forming a matrix $\tilde{Y}$, 
\item an $n\times n$ matrix of neutral Fermi
superfields $\Lambda_{0}$, 
\item a set of Fermi superfields $\Lambda^{\beta}$
in the $(\lambda_{\beta 1}, \lambda_{\beta 2})$ representation of $U(k)$,
with charges $q_{a, \beta}$ under the abelian gauge symmetry defining
the toric variety, 
\item a set of chiral superfields $P_{\gamma}$ in the $U(k)$
representation dual to $(\lambda_{\gamma 1}, \lambda_{\gamma 2})$
and with charges $-q_{a, \gamma}$ under the abelian gauge symmetry
defining the toric variety, 
\item and finally a (0,2) superpotential
\begin{displaymath}
W \: = \: {\rm tr}\, \left( \Lambda_{0} \left( A(\Phi) \: - \: 
\tilde{Y} \tilde{X} \right) \: + \: \Lambda^{\beta} F^{\gamma}_{\beta}(\Phi)
P_{\gamma} \right) .
\end{displaymath}
\end{itemize}

Note that although the data defining the bundle is formally very similar
to that in the PAX construction, the representations given in
the short exact sequence~(\ref{eq:02-bundle-paxy}) are representations of
$U(k)$, whereas the representations given in the analogue for the PAX
construction are representations of $U(n-k)$.  The relationship
between such representations will be discussed in 
section~\ref{sect:pax-vs-paxy}, 
but is not particularly simple.

\subsection{Examples}

Listed in table~\ref{table:pfaff-exs} 
are some examples of (0,2) models on Pfaffians.
The Pfaffians themselves are all constructed via the (0,2) PAX model for
gauge bundle kernels, as Pfaffians of a $4 \times 4$ matrix $A$,  
defined as the locus where the rank of $A$ is less than or equal to 2.
Hence, we have a $U(4-2) = U(2)$ gauge theory.
The Pfaffians are subvarieties of ${\mathbb P}^7$, so for the PAX
construction we have fibered
\begin{displaymath}
S^{\oplus 4} \: \longrightarrow \: G(2,4)
\end{displaymath}
over ${\mathbb P}^7$, with the fibering defined by the statement that
the $n$ antifundamentals\footnote{
Our conventions in the table are flipped relative to the earlier discussion:
$X$ is here a set of antifundamentals rather than fundamentals,
and $\Lambda_0$ is a set of fundamentals rather than antifundamentals.
The choice is arbitrary.
} $X$ have $U(1)$ charge $0$ and the fundamentals 
$\Lambda_0$ have
$U(1)$ charge $-1$.  The chiral superfields defining ${\mathbb P}^7$ have
charge $1$, and the entries of the matrix $A$ are of degree $1$.
It is straightforward to check that the resulting Pfaffian is 
Calabi-Yau, from the criteria given earlier, and applying the methods of
{\it e.g.} \cite{meron} we see that these are 3-folds.

Table~\ref{table:pfaff-exs} lists data for bundles over the total space of the
$(S^4 \rightarrow G(2,4))$-bundle over ${\mathbb P}^7$.  We have restricted
to bundles built as kernels.  (More general cases are straightforward,
and so are left as exercises.)  
Bundles are kernels of the form
\begin{eqnarray*}
\lefteqn{
0 \: \longrightarrow \: {\cal E} \: \longrightarrow \:
\oplus^{m_1} {\cal O}((\lambda_{A1}, \lambda_{B1}), Q_1)
\oplus^{m_2} {\cal O}((\lambda_{A2}, \lambda_{B2}), Q_2)
} \\
& \hspace*{1in} & 
\: \longrightarrow \:
\oplus^{n_1} {\cal O}((\lambda_{A3}, \lambda_{B3}), Q_3)
\oplus^{n_2} {\cal O}((\lambda_{A4}, \lambda_{B4}), Q_4) 
\: \longrightarrow \: 0 .
\end{eqnarray*}
For each special homogeneous bundle appearing,
we give both a representation of $U(2)$ and also a charge under the $U(1)$
defining the ${\mathbb P}^7$.  Conventions are such that $U(2)$
representation $(\lambda_{Ai}, \lambda_{Bi})$ has ${\mathbb P}^7$ $U(1)$
charge $Q_i$, a fact we have indicated above in subscripts. 
All of the examples in table~\ref{table:pfaff-exs} have $c_1({\cal E}) = 0$.

For completeness, let us describe the first example in
table~\ref{table:pfaff-exs} in detail.
It describes a theory containing charged left-moving fermions as:
\begin{itemize}
\item 5 Fermi superfields in the ((0,0),-1), for part of the gauge bundle,
\item 2 Fermi superfields in the ((2,2),0), for part of the gauge bundle,
\item $\Lambda_0$:  4 Fermi superfields in the ((1,0),-1),
\item 1 $U(2) \times U(1)$ gaugino,
\end{itemize}
and charged right-moving fermions as:
\begin{itemize}
\item $X$:  4 chiral superfields in the ((0,-1),0),
\item 2 chiral superfields in the dual of ((2,2),-1), for part of the
gauge bundle,
\item 1 chiral superfield in the dual of ((1,-1),-1), for part of the gauge
bundle,
\item 8 chiral superfields in the ((0,0),+1), describing homogeneous 
coordinates on ${\mathbb P}^7$.
\end{itemize}
There are several gauge anomaly cancellation conditions that must
be obeyed:  the Cas$_2$ condition and $({\rm Cas}_1)^2$ conditions for
$U(2)$ gauge anomaly cancellation, plus a $q^2$ condition for solely the
extra $U(1)$ for ${\mathbb P}^7$, plus a mixed $U(1)-U(1)$ condition
involving products of the general form  $q {\rm Cas}_1 $.

\begin{table}[h]
\begin{center}
\begin{tabular}{ccccccccccccc}
$Q_1$ & $m_1$ & $(\lambda_{A1}, \lambda_{B1})$ & $Q_2$ & $m_2$ & $(\lambda_{A2},
 \lambda_{B2})$
& $Q_3$ & $n_1$ & $(\lambda_{A3}, \lambda_{B3})$ & $Q_4$ & $n_2$ & $(\lambda_{A4
},
\lambda_{B4})$ & rank \\ \hline
-1&5&(0, 0)&0&2&(2, 2)&-1&2&(2, 2)&-1&1&(1, -1)&2 \\
-1&4&(1, -1)&3&1&(2, 2)&1&1&(2, 2)&-2&1&(2, -2)&7 \\
0&5&(1, 0)&4&2&(2, 2)&4&1&(2, 1)&0&2&(2, 0)&4 \\
-2&2&(0, 0)&1&4&(1, 1)&3&1&(1, 1)&-1&1&(2, 0)&2 \\
-1&4&(0, 0)&0&4&(1, 1)&-1&1&(1, 1)&-1&1&(2, 0)&4 \\
-2&2&(0, 0)&0&5&(1, 1)&-2&2&(1, 1)&0&1&(2, 0)&2 \\
-3&1&(0, -1)&3&5&(1, 0)&2&2&(1, 1)&5&1&(2, -1)&6 \\
-2&5&(0, 0)&0&2&(1, 1)&-2&2&(1, 1)&-2&1&(1, -1)&2 \\
-2&5&(0, 0)&1&1&(1, 1)&-3&1&(1, 1)&-2&1&(1, -1)&2 \\
-2&4&(1, -1)&5&2&(1, 1)&3&2&(1, 1)&-4&1&(2, -2)&7 \\
-1&5&(1, 0)&5&1&(2, 1)&2&2&(2, 0)&-3&2&(1, 0)&2 \\
0&4&(1, 0)&2&1&(2, 1)&0&2&(2, 0)&2&1&(1, 0)&2 \\
0&4&(1, 0)&2&1&(2, 2)&0&2&(2, 0)&2&1&(0, 0)&2 \\
-4&5&(0, 0)&-1&2&(1, 0)&-3&2&(1, 0)&-4&1&(1, -1)&2 \\
-4&5&(0, 0)&0&1&(1, 0)&-4&1&(1, 0)&-4&1&(1, -1)&2 \\
-1&3&(1, -1)&0&4&(1, 0)&-1&1&(0, 0)&-1&2&(2, -1)&8 \\
-4&1&(1, -1)&-1&2&(1, 0)&-4&1&(0, 0)&-3&1&(2, -1)&2 \\
1&2&(0, -1)&4&1&(1, -1)&4&1&(0, 0)&3&1&(1, -2)&2 \\
0&4&(0, -1)&1&3&(1, -1)&1&1&(0, 0)&1&2&(1, -2)&8 \\
-2&1&(-2, -2)&0&4&(0, -1)&-2&1&(0, 0)&0&2&(0, -2)&2 \\
-4&5&(0, 0)&-1&1&(2, -1)&-3&1&(2, -1)&-4&1&(1, -1)&2 \\
0&1&(0, -1)&4&5&(0, 0)&4&1&(1, -1)&4&1&(0, -1)&2 \\
1&2&(0, -1)&4&5&(0, 0)&4&1&(1, -1)&3&2&(0, -1)&2 \\
-1&1&(-1, -1)&2&5&(0, 0)&2&1&(1, -1)&3&1&(-1, -1)&2 \\
0&2&(-1, -1)&2&5&(0, 0)&2&1&(1, -1)&2&2&(-1, -1)&2 \\
\end{tabular}
\caption{Anomaly-free (0,2) models on Pfaffians inside ${\mathbb P}^7$.}
\label{table:pfaff-exs}
\end{center}
\end{table}

As a consistency check, let us work out central charges of the theories
in table~\ref{table:pfaff-exs}, using c-extremization \cite{bb1}
as discussed earlier in section~\ref{sect:g24-4}.
Let us work through the first entry in detail, and summarize results for
the rest of the entries.  Given the field content, it is straightforward
to show that the right-moving central charge ansatz provided by
the identity
\begin{displaymath}
c_R \: = \: 3 {\rm Tr} \, \gamma^3RR
\end{displaymath}
has the form
\begin{displaymath}
c_R \: = \: 3\left(
8(R_{\phi}-1)^2-7R_{\Lambda}^2+8(R_X-1)^2+5(R_P-1)^2-8R_{\Lambda_0}^2-5
\right),
\end{displaymath}
where $R_{\Phi}, R_{\Lambda}, R_X, R_P, R_{\Lambda_0}$ 
denote the R-charge of $\Phi, \Lambda, X, P, \Lambda_0$, respectively.
Furthermore, from the superpotential terms
\begin{displaymath}
\int d\theta^+\Lambda A(\Phi)X,
\end{displaymath}
we have the constraint
\begin{displaymath}
-1+R_{\Lambda_0}+R_{\Phi}+R_X \: = \: 0,
\end{displaymath}
and from the superpotential terms
\begin{displaymath}
\int d\theta^+\Lambda PF,
\end{displaymath}
we have the constraints
\begin{eqnarray*}
-1+R_\Lambda+R_P & = & 0,\\
-1+R_\Lambda+R_P+4R_\Phi & = & 0,\\
-1+R_\Lambda+R_P-4R_\Phi & = & 0.
\end{eqnarray*}
Extremizing the central charge yields
\begin{displaymath}
R_{\Phi}=0, \: \: \:
R_{\Lambda_0}=0, \: \: \:
R_{\Lambda}=0, \: \: \:
R_P=1, \: \: \:
R_X=1,
\end{displaymath}
and the result, $c_R = 9$, is consistent with an IR description as a nonlinear
sigma model on a Calabi-Yau 3-fold, as expected.
Using
\begin{displaymath}
c_R-c_L \: = \: {\rm Tr}\gamma^3 \: = \: 8-7+8+5-8-5 \: = \: 1,
\end{displaymath}
we compute $c_L = 8$, consistent with a rank 2 bundle on a Calabi-Yau 3-fold.
Proceeding in the same fashion, one finds that all other central charges
in the models listed in table~\ref{table:pfaff-exs} are consistent with a 
bundle on a Calabi-Yau 3-fold of the indicated rank.
This supports the conclusion that the PAX models listed do indeed RG
flow to the indicated (0,2) nonlinear sigma models.

In passing, let us comment on the possible existence of a duality to
an abelian description.  Since the nonabelian gauge theory in the PAX model 
describes, in part,
$G(2,4)$, one might hope to use its duality to ${\mathbb P}^5[2]$ to 
find an equivalent abelian model.  Unfortunately, to do so, we would need
a dual description of the universal subbundle on $G(2,4)$.  
On ${\mathbb P}^5[2]$, this is a spinor bundle for which no simple
three-term sequence construction is expected.  Thus, we do not expect
there to exist a dual abelian description of any of the 
theories described in this section.

We will discuss dualities between PAX and PAXY models in 
section~\ref{sect:pax-vs-paxy}.

\section{Mathematics of duality in (2,2) theories}
\label{sect:22-dual}

In the next few sections, we will analyze dualities between two dimensional
(2,2) and (0,2) theories.  We focus on weakly-coupled theories RG flowing to
nonlinear sigma models.  In some cases, we can understand dualities as
relating different presentations of the same mathematical geometry.
In such a case, where we can identify RG endpoints, a duality is immediate
(and as an immediate consequence, one can identify Higgs moduli spaces,
chiral rings, and global symmetries).  Analyses of this form will not
apply to every theory, only to weakly coupled theories with a clear
relationship to geometry, and moreover even in weakly coupled theories
we will later see examples of physical dualities not of this form.  

Although such a mathematical approach does not apply to every theory,
it can be useful for suggesting nonobvious dualities, especially in
theories with no flavor symmetries.  The latter are generic in Calabi-Yau
compactification, where {\it e.g.} superpotentials typically break most if 
not all flavor symmetries.

We shall first discuss the two-dimensional analogue of
Seiberg duality for (2,2) $U(k)$ gauge theories with both fundamentals and
antifundamentals \cite{bc1}.  In particular, although the relation between
$G(k,n)$ and $G(n-k,n)$ is well-known, it is perhaps less well-known that
Seiberg duality itself has an equally simple mathematical description,
only slightly generalizing the $G(k,n)$, $G(n-k,n)$ duality.
We shall discuss the relevant geometry next.

\subsection{$U(k)$ gauge theories with fundamentals and
antifundamentals}
\label{sect:22fundanti}

In this section we will give a geometric understanding of the duality in
$(2,2)$ $U(k)$ gauge theories with both fundamentals and antifundamentals
described in \cite{bc1}.  This is a prototype for many other dualities
we shall discuss in this paper.  It will also serve as a useful caution:
such mathematical dualities are only applicable to weakly-coupled
physical theories.  In particular, in the present case we will
see there is a chain of mathematical equivalences,
but only some of those mathematical equivalences
correspond to relations between weakly coupled theories
and are physically meaningful, as we shall discuss.

Consider a two-dimensional (2,2) GLSM with gauge group $U(k)$,
$n$ multiplets in the fundamental representation, and
$A$ multiplets in the antifundamental representation.
This GLSM has two geometric phases, describing:
\begin{itemize}
\item ${\rm Tot}\, \left( S^A \rightarrow G(k,n) \right)$, and
\item ${\rm Tot}\, \left( S^n \rightarrow G(k,A) \right)$.
\end{itemize}
(In addition, as observed in \cite{mp-c}, there will be discrete Coulomb
vacua in general, but as they will not play an essential role in
our discussion, we omit their details.)

Mathematically, the first phase is equivalent to
\begin{displaymath}
{\rm Tot}\, \left( (Q^*)^A \: \longrightarrow \: G(n-k,n) \right)
\end{displaymath}
as discussed in section~\ref{sect:review}.

Physically, the $Q^*$ must be realized indirectly, from the fact that
\begin{displaymath}
0 \: \longrightarrow \: Q^* \: \longrightarrow \: {\cal O}^n \: \longrightarrow
\: S^* \: \longrightarrow \: 0 .
\end{displaymath}
Specifically, for each $Q^*$ one wishes to implement, one must add chiral
superfields corresponding to ${\cal O}^n$ and the dual of $S^*$, together
with a suitable superpotential.
For example, the phase
\begin{displaymath}
{\rm Tot}\, \left( (Q^*)^A \: \longrightarrow \: G(n-k,n) \right)
\end{displaymath}
above arises in the GLSM with gauge group $U(n-k)$, 
$n$ chiral superfields $\Phi$ in the fundamental representation,
$nA$ neutral chiral superfields $\Gamma$ ($A$ copies of ${\cal O}^n$),
and $A$ chiral superfields $P$ in the antifundamental representation
($A$ copies of the dual of $S^*$), together with the superpotential
\begin{displaymath}
W \: = \: \Gamma \Phi P .
\end{displaymath}

Note in passing that building a (2,2) GLSM to realize
the total space of $Q^{\oplus A}$, rather than $(Q^*)^{\oplus A}$,
would be more problematic.  Formally, one could build each $Q$
as a cokernel, by adding chiral superfields corresponding to ${\cal O}^n$
and the dual of $S$.  However, chiral superfields corresponding to 
$S^*$ are in the fundamental representation, and so physically are 
indistinguishable from the chiral superfields defining the Grassmannian -- 
the result physically would be a larger Grassmannian, rather than a bundle
on the Grassmannian.  A closely related problem exists in abelian GLSMs:
although it is straightforward to build a (2,2) GLSM describing the total
space of the line bundle ${\cal O}(-1)$ on ${\mathbb P}^n$, by adding a 
chiral superfield of opposite charge from the rest, if instead one adds
a chiral superfield of the same charge, the result is a larger projective
space, and not the line bundle ${\cal O}(+1)$ on ${\mathbb P}^n$.

Now, let us return to the case at hand.
In the case $n=A$, which is the case that the space is a noncompact
Calabi-Yau, the data above is the same as the data given by
{\it e.g.} \cite{bstv} to describe the GLSM dual to the $U(k)$ GLSM at top (the
neutral chiral superfields $\Gamma$ being their mesons $M$, for example), 
closely
following the pattern of Seiberg duality in four dimensions.
In effect, we are using mathematics to give a purely geometric understanding
of Seiberg duality, by studying what in four dimensions would be the
classical Higgs branch.

In the case $n \neq A$, this pattern results in a chain of mathematical
dualities between
GLSMs:
\begin{displaymath}
\xymatrix{
S^A \longrightarrow G(k,n) 
\ar@{--}[r] \ar@{<->}[d]_{=} &
S^n \longrightarrow G(k,A) \\
(Q^*)^A \longrightarrow G(n-k,n) 
\ar@{--}[r] &
(Q^*)^n \longrightarrow G(n-k,A)
\ar@{<->}[d]^{=} \\
S^A \longrightarrow G(A-n+k,n)
\ar@{--}[r] \ar@{<->}[d]_{=} &
S^n \longrightarrow G(A-n+k,A)
\\
(Q^*)^A \longrightarrow G(2n-A-k,n)
\ar@{--}[r] &
(Q^*)^n \longrightarrow G(2n-A-k,A) 
\ar@{<->}[d]^{=} \\
S^A \longrightarrow G(2A-2n+k,n)
\ar@{--}[r] \ar@{<->}[d]_{=} &
S^n \longrightarrow G(2A-2n+k,A) \\
(Q^*)^A \longrightarrow G(3n-2A-k,n)
\ar@{--}[r] &
(Q^*)^n \longrightarrow G(3n-2A-k,A)
\ar@{<->}[d]^{=}\\
S^A \longrightarrow G(3A-3n+k,n)
\ar@{--}[r] &
S^n \longrightarrow G(3A-3n+k,A) \\
}
\end{displaymath}
and so forth.
Horizontal rows correspond to the phases of a single GLSM; vertical
arrows indicate mathematical dualities.
We made the arbitrary decision to run the dualities in one direction;
one could also continue in the opposite direction vertically, and it
is straightforward to check that a very similar pattern of dualities
occurs in that direction.
Note that if $A=n$, then the sequence of GLSMs above is 2-periodic.

Now, physics restricts which of the mathematical dualities above
is physically meaningful.  The issue revolves around renormalization
group flow.  In the special case that $A=n$, all the spaces appearing above are
Calabi-Yau, the Fayet-Iliopoulos parameter is a renormalization-group
invariant number.  In other cases, however, the Fayet-Iliopoulos
parameter will flow.
Briefly, the space
\begin{displaymath}
{\rm Tot}\, \left( S^A \: \longrightarrow \: G(k,n) \right)
\end{displaymath}
is positively-curved (and so will shrink) if $A < n$, and is
negatively-curved (and so will expand) if $A > n$.
Note that the two phases of the non-Calabi-Yau
GLSMs have opposite-signed curvature:
since the Fayet-Iliopoulos parameter can only flow in one direction,
if one limit is positively-curved, the other limit must be
negatively-curved, and that is consistent with the mathematics.

Now, in a GLSM, there is a weakly-coupled UV phase in which the
Higgs branches are closely identified with geometry.  As one flows to
the IR in non-Calabi-Yau GLSMs, however, the theory develops isolated
Coulomb vacua \cite{mp-c}.  For example, in the supersymmetric ${\mathbb P}^n$
model, these form the $n+1$ vacua in the asymptotic IR limit of the theory.
Strictly speaking, those Coulomb vacua must be taken into account, and so
a purely geometric description of dualities, one that ignores Coulomb
vacua as we have done, is potentially misleading in the IR.

Thus, the geometric dualities we have outlined need only correspond
to physical dualities in the weakly-coupled UV phases.  If $A < n$,
say, that means one should expect there to be a physical duality
between the UV GLSM phases, of the form
\begin{displaymath}
\xymatrix{
S^A \longrightarrow G(k,n) 
\ar@{--}[r] \ar@{<->}[d]_{=} &
S^n \longrightarrow G(k,A) \\
(Q^*)^A \longrightarrow G(n-k,n) 
\ar@{--}[r] &
(Q^*)^n \longrightarrow G(n-k,A)
}
\end{displaymath}
but the mathematical duality on the other side of the diagram need not
translate to anything physical.
If $A > n$, the opposite mathematical duality should be physical.

This duality is discussed in gauge theories in 
\cite{bc1}[section 7.1].  As each GLSM has the same number of
fundamentals ($n$) and antifundamentals ($A$), checking anomaly
matching is straightforward.  They show $S^2$ partition
functions match for $n > A+1$; their particular expressions for
the cases $n=A, A+1$ do not match, but it is believed \cite{beninipriv}
that the partition functions differ merely by a K\"ahler transformation
in those cases, and so describe equivalent theories.
(The paper \cite{bstv} conjectures differently.)
Later work \cite{beht}[section 4.6.1] shows
elliptic genera match more generally.  Based on the relationship between
the geometries, we conjecture that the theories match in general.

So far we have discussed (2,2) dualities for the total spaces of essentially
two bundles on $G(k,n)$, and Whitney sums thereof:  $S$ and $Q^*$.
It is not clear whether more general bundles can be dualized.  
The problem is to relate a more general representation of $U(k)$ to
representations of $U(n-k)$; as we shall discuss in section~\ref{open-seiberg},
although one can find long exact sequences relating them, and those can
be realized in open strings,
it is not currently known how
to realize those long exact sequences 
in closed-string (2,2) or (0,2) theories, so barring the existence of additional
surprising physical relationships, it is natural to conjecture that
more general bundles cannot be dualized.

\subsection{A proposed duality involving Pfaffians}
\label{sect:g2n-duals}

Proceeding in the same spirit, it is possible to formulate additional
proposals for dualities between GLSMs, motivated by mathematics.
In this subsection we focus on one particular example in (2,2) GLSMs,
relating a Grassmannian $G(2,n)$ of 2-planes in ${\mathbb C}^n$ to
a determinantal variety.

Mathematically (\cite{donagipriv}, \cite{harris}[chapter 9]), 
$G(2,n)$ is the rank 2 locus of the $n \times n$
matrix
\begin{displaymath}
A(z_{ij}) \: = \:
\left[ \begin{array}{cccc}
z_{11}=0 & z_{12} & z_{13} & \cdots \\
z_{21}=-z_{12} & z_{22}=0 & z_{23} & \cdots \\
z_{31}=-z_{13} & z_{32}=-z_{23} & z_{33}=0 & \cdots \\
\cdots & \cdots & \cdots & \cdots
\end{array} \right]
\end{displaymath}
over
\begin{displaymath}
{\mathbb P}^{\tiny{ \left( \begin{array}{c} n \\ 2 \end{array} \right)} - 1}
\end{displaymath}
where the $z_{ij} = - z_{ji}$ are homogeneous coordinates on that projective
space.  In the special case that $n=4$, the rank 2 locus is determined
by the condition that the determinant of the matrix above vanish,
which is checked to be the same as the quadric condition
\begin{displaymath}
z_{12} z_{34} \: - \: z_{13}z_{24} \: + \: z_{14}z_{23} \: = \: 0.
\end{displaymath}
In this special case, one then has
a duality between $G(2,4)$ and ${\mathbb P}^5[2]$ which
we have already discussed.  For more general $n$, the dual cannot
be described as a hypersurface, but instead is a determinantal variety
which can be built using the methods of \cite{jklmr1}.

A PAX model for the dual
is given by a (2,2) $U(n-2) \times U(1)$ gauge theory with matter content  
\begin{itemize}
\item $n! / (2! (n-k)!)$ chiral superfields $\Phi$, neutral under $U(n-2)$ but
charge $+1$ under the $U(1)$, corresponding to homogeneous coordinates
on the projective space,
\item $n$ chiral superfields in the fundamental of $U(n-2)$, neutral under
the $U(1)$, which we label
$X$,
\item $n$ copies of the antifundamental of $U(n-2)$, charge $-1$ under
the $U(1)$, which we label $P$,
\item and a superpotential $W = {\rm tr}\, P A(\Phi) X$.
\end{itemize}

Alternatively, a PAXY model for the dual is given by
a (2,2) $U(2) \times U(1)$ gauge theory with matter content
\begin{itemize}
\item $n! / (2! (n-k)!)$ chiral superfields $\Phi$, neutral under $U(n-2)$ but
charge $+1$ under the $U(1)$, corresponding to homogeneous coordinates
on the projective space,
\item $n$ fundamentals of $U(2)$, neutral under $U(1)$, which we label
$\tilde{X}$,
\item $n$ antifundamentals of $U(2)$, charge $+1$ under $U(1)$, which we
label $\tilde{Y}$,
\item an $n \times n$ matrix of chiral superfields $\tilde{P}$,
neutral under $U(2)$, charge $-1$ under $U(1)$,
\item and a superpotential $W = {\rm tr} \, \tilde{P}\left( A(\Phi) - 
\tilde{Y} \tilde{X} \right)$.
\end{itemize}

As these theories admit weakly-coupled phases describing the same
geometries, we propose that there is a physical Seiberg-like duality
relating them.  The universal subbundle and quotient bundle are realized
as \cite{donagipriv} the image and cokernel, respectively, of the matrix $A$.  
More examples of analogous forms can also be constructed,
and we leave their analyses for future work.

\section{Invariance of (0,2) under gauge bundle dualization}
\label{sect:dualbundle}

In this section we will propose that physical (0,2) theories are
invariant under dualizing the gauge bundle, {\it i.e.}
a (0,2) theory on space $X$ with bundle ${\cal E}$ defines the
same universality class as that for the same space $X$ with
dual bundle ${\cal E}^*$.  We will use this later to help
simplify our description of other (0,2) dualities.

This particular duality
has been discussed previously in pseudo-topological field
theories in \cite{bchir}, as we will review later, and has also been
previously considered by \cite{distpriv,gukovpriv}.  It has also been
used implicitly in \cite{ggp}.  However, we are not aware of published
checks of this duality in physical non-topological theories.

It is extremely straightforward to show that this satisfies some
basic tests, such as leaving massless spectra invariant.  However, to show
that this is true of an entire physical theory, one must also check,
for example, that massive states are also invariant under this operation,
as are worldsheet instanton effects.  We will check such details in the next
several subsections.

\subsection{Initial checks}

Let us begin by considering the worldsheet lagrangian for a two-dimensional
(0,2) theory\footnote{The expression given corrects some minor
typos in the lagrangian written in \cite{gs2}.} \cite{gs2}[equ'n (7)]:
\begin{displaymath}
\begin{split}
\frac{1}{2} \left(g_{\mu \nu} +  i B_{\mu \nu} \right) \partial \phi^{\mu} 
\overline{\partial} \phi^{\nu}                 
&+ \frac{i}{2} g_{\mu \nu} \psi_+^{\mu} D_{\overline{z}} \psi_+^{\nu}
+ \frac{i}{2}h_{\alpha \beta} \lambda_-^{\alpha} D_{z} \lambda_-^{\beta}
+  F_{i \overline{\jmath} a \overline{b}} \psi_+^i \psi_+^{\overline{\jmath}}
 \lambda_-^a \lambda_-^{\overline{b}} \\
&                            
+   h^{a \overline{b}} F_a \overline{F}_{\overline{b}}
+  \psi_+^i \lambda_-^a D_i F_a 
+ \psi_+^{\overline{\imath}} \lambda_-^{\overline{b}} D_{\overline{\imath}} \overline{F}_{\overline{b}} \\
& 
+   h_{a \overline{b}} E^a \overline{E}^{\overline{b}} 
+ \psi_+^i \lambda_-^{\overline{a}} \left( D_i E^b \right) h_{\overline{a} b}
+ \psi_+^{\overline{\imath}} \lambda_-^a \left( D_{\overline{\imath}} 
\overline{E}^{\overline{b}} \right) h_{a \overline{b} } ,
\end{split}
\end{displaymath}
where $\mu, \nu$ are real tangent space indices, $i, j$ holomorphic
tangent space indices, $\alpha, \beta$ real vector bundle indices, and
$a, b$ holomorphic vector bundle indices.
In the expression above, $(E^a) \in \Gamma({\cal E})$ and
$(F_a) \in \Gamma({\cal E}^*)$, and act as the (0,2) analogues of a
superpotential.  They are subject to the constraint
\begin{displaymath}
\sum_a E^a(\phi) F_a(\phi) \: = \: 0.
\end{displaymath}
If we exchange ${\cal E} \rightarrow {\cal E}^*$, simultaneously exchanging
$E^a$ and $F_a$, it is straightforward to check that the lagrangian
above is invariant.  For example, under the bundle interchange
described, $\lambda_-^a$ is exchanged with $h_{a \overline{b}}
\lambda_-^{\overline{b}}$,
which leaves kinetic terms invariant and is needed to 
make sense of the $E^a \leftrightarrow F_a$ exchange.
Under the same interchange, the curvature $F \mapsto -F$; however,
when combined with the $\lambda_-^a \leftrightarrow \lambda_-^{\overline{b}}$ 
exchange, the four-fermi term is left invariant.

Given that the classical action remains invariant, classically the theories
are identical, but there could be (and in fact are) subtleties involving
regularizations, so let us perform additional checks.

As another check, note that anomaly cancellation conditions are invariant
under this dualization:  ${\rm ch}_2({\cal E}) = {\rm ch}_2({\cal E}^*)$.
In UV GLSMs, this is the statement that
gauge anomaly cancellation conditions are invariant under dualizing 
matter representations.

As a further check, consider massless spectra in heterotic Calabi-Yau
compactifications.  As discussed in \cite{dg},
the massless spectra are computed by sheaf cohomology groups of the
form
\begin{displaymath}
H^{\bullet}(X, \wedge^{\bullet} {\cal E}), \: \: \:
H^{\bullet}(X, {\rm End}\, {\cal E}) ,
\end{displaymath}
and it is straightforward to check that these groups are invariant
under ${\cal E} \leftrightarrow {\cal E}^*$ (for bundles of trivial
determinant, as is
typical in Calabi-Yau compactification).  Physical properties are
determined by the gradings; the effect seems to merely be to exchange
particles and antiparticles, a trivial operation.

As another consistency check, dualization of the gauge bundle preserves
stability.  One way to see this is directly in the Donaldson-Uhlenbeck-Yau
equation:
\begin{displaymath}
g^{i \overline{\jmath}} F_{i \overline{\jmath}} \: = \: 0 .
\end{displaymath}
Dualization of the bundle sends $F \mapsto -F$, so the original bundle will
satisfy Donaldson-Uhlenbeck-Yau if and only if the dual bundle also does.
In terms of Mumford stability \cite{tonypriv}, \cite{oss}[lemma II.1.2.4], 
dualization gives a one-to-one
correspondence between saturated subsheaves of ${\cal E}^*$ and quotient
torsion-free sheaves of ${\cal E}$, which preserves slope inequalities.

\subsection{Elliptic genera}

Let us compare elliptic genera for (0,2) theories with complex
vector bundles ${\cal E}$
and ${\cal E}^*$, using the expressions in and notation of \cite{lgeg}.
(In this paper, we only consider complex vector bundles; we make no claims
about invariance under duality for {\it e.g.} real vector bundles.)
For example, the elliptic genera of nonlinear sigma models with left-movers
in an NS sector \cite{lgeg}[equ'n (5)] are of the form
\begin{eqnarray*}
\lefteqn{
\mbox{Tr } (-)^{F_R} \exp(i \gamma (J_L)_0)  q^{L_0} \overline{q}^{
\overline{L}_0}
} \\
& = & q^{-(1/24)(2n + r)}
\int_X {\rm Td}(TX)  
\wedge {\rm ch}\left( \bigotimes_{k=1,2,3,\cdots}
S_{q^k}((TX)^{\bf C}) 
\bigotimes_{k=1/2,3/2,5/2,\cdots}
 \wedge_{q^k}\left((z {\cal E})^{\bf C}\right)
\right) ,
\end{eqnarray*}
where $z = \exp(i \gamma)$, 
\begin{displaymath}
(z{\cal E})^{\bf C} \: = \: z{\cal E} \oplus \overline{z} \overline{{\cal E}},
\end{displaymath}
and other notation follows \cite{lgeg}.
Note from the expression above that
$(z {\cal E})^{\bf C}$ is invariant under the exchange
${\cal E} \leftrightarrow {\cal E}^*$, so long as one simultaneously
exchanges $z \leftrightarrow \overline{z} = z^{-1}$, 
the twist on the left-movers.
As a result, the elliptic genus above is automatically invariant under
the exchange.

For a heterotic nonlinear sigma model with left-moving fermions in an R
sector, the elliptic genus
\begin{displaymath}
\mbox{Tr}_{{\rm R},{\rm R}} (-)^{F_R}  \exp\left( i \gamma (J_L)_0 \right)
q^{L_0} \overline{q}^{\overline{L}_0}
\end{displaymath}
is given by \cite{lgeg}[equ'n (6)]
\begin{eqnarray*} 
\lefteqn{
q^{+(1/12)(r - n)}
} \nonumber \\
& & \cdot \int_X \hat{A} (TX)  
\wedge {\rm ch}\Biggl( 
z^{-r/2} \left( \det {\cal E} \right)^{+1/2}
\wedge_{1}\left( z {\cal E}^* \right)
 \\
& & \hspace*{1.5in} \left. \cdot \bigotimes_{k=1,2,3,\cdots}
S_{q^k}((TX)^{\bf C}) 
\bigotimes_{k=1,2,3,\cdots}
 \wedge_{q^k}\left((z^{-1} {\cal E})^{\bf C}\right)
\right) .
\end{eqnarray*}
Here, invariance under the interchange ${\cal E} \leftrightarrow
{\cal E}^*$, $\gamma \leftrightarrow - \gamma$ is a consequence of the
observations above plus
the fact that   
\begin{equation}  \label{eq:e-id}
z^{-r/2} \left( \det {\cal E} \right)^{+1/2} \wedge_1( z {\cal E}^{*} )
\: = \:
z^{+r/2} \left( \det {\cal E} \right)^{-1/2} \wedge_1( z^{-1} {\cal E} ) .
\end{equation}

Now, let us turn to (0,2) nonlinear sigma models with potential.
The NS sector elliptic genus of a theory describing a cokernel
${\cal E}'$ of an injective map
\begin{displaymath}
0 \: \longrightarrow \: {\cal F}_1 \: \stackrel{ \tilde{E} }{\longrightarrow}
\: {\cal F}_2 \: \longrightarrow \: {\cal E}' \: \longrightarrow \: 0
\end{displaymath}
is given by \cite{lgeg}[equ'n (21)]  
\begin{eqnarray*}   
\lefteqn{
q^{-(1/24)( +2n - r_1  + r_2)}
}  \\
& & \cdot
\int_B {\rm Td}(TB)  
\wedge {\rm ch}\left(
\bigotimes_{k=1,2,3\cdots} S_{q^k}\left((TB)^{\bf C}\right) 
\bigotimes_{k=1/2,3/2,\cdots} S_{-q^k }\left(
( z^{-1} {\cal F}_1)^{\bf C}\right)
\right.  \\
& & \hspace*{2.5in}
\left.
\bigotimes_{k=1/2,3/2,\cdots} \wedge_{q^k }\left( 
( z^{-1} {\cal F}_2)^{\bf C} \right)
\right) . 
\end{eqnarray*}
This should be compared to the NS sector elliptic genus of a theory
describing a kernel of a surjective map
\begin{displaymath}
0 \: \longrightarrow \: {\cal E}' \: \longrightarrow \:
{\cal F}_1 \: \stackrel{F_a}{\longrightarrow} \: {\cal F}_2 \: 
\longrightarrow \: 0 ,
\end{displaymath}
which is given by \cite{lgeg}[equ'n (24)]
\begin{eqnarray*}   
\lefteqn{
q^{-(1/24)(2n - r_2 + r_1)}
} \nonumber \\
& & \cdot
\int_B {\rm Td}(TB)    
\wedge {\rm ch}\left(
\bigotimes_{k=1,2,3,\cdots} S_{q^k}\left( (TB)^{\bf C}
\right) 
\bigotimes_{k=1/2,3/2,\cdots} S_{-q^k }
\left( ( z {\cal F}_2^{*})^{\bf C} \right)
\right.   \\
& & \hspace*{2.5in} \left.
\bigotimes_{k=1/2,3/2,\cdots} \wedge_{q^k }
\left( ( z {\cal F}_1^*)^{\bf C} \right)
\right) .
\end{eqnarray*}

The duality we are checking dualizes the sequences, so we must compare
elliptic genera with
\begin{displaymath}
{\cal F}_2 \: \leftrightarrow \: {\cal F}_1^*
\end{displaymath}
exchanged
at the same time as $z \leftrightarrow z^{-1}$.  It is straightforward to
check that this operations maps the two elliptic genera into one another,
and so these elliptic genera are compatible with the proposed duality.

Now, let us compare the R sector elliptic genera.
For gauge bundles realized as cokernels as above, the R sector elliptic
genus is given by \cite{lgeg}[equ'n (22)]
\begin{eqnarray*}
\lefteqn{ q^{-(1/24)(2n + 2r_1 - 2r_2)} 
} \nonumber \\
& & \cdot \int_B {\rm Td}(TB) \wedge {\rm ch}\Biggl(
z^{+r_2/2} \wedge_{1}(z^{-1} {\cal F}_2) z^{+r_1/2} \wedge_{1}(z^{-1} 
{\cal F}_1)
 \\
& & \hspace*{1.5in} \cdot
\left( \det {\cal F}_2 \right)^{-1/2} 
\left( \det {\cal F}_1 \right)^{-1/2}
\nonumber \\
& & \hspace*{1.5in} \cdot
\bigotimes_{k=1,2,3,\cdots} S_{q^k}((TB)^{\bf C}) 
\bigotimes_{k=0,1,2,\cdots} S_{-q^k}( (z^{-1} {\cal F}_1)^{\bf C} )
 \\
& & \hspace*{1.5in} \left. \cdot
\bigotimes_{k=1,2,3,\cdots} \wedge_{q^k}( (z^{-1} {\cal F}_2)^{\bf C} )
\right) ,
\end{eqnarray*}
and the R sector elliptic genus for a gauge bundle realized as a kernel 
is\footnote{The expression given above corrects a minor typo in
\cite{lgeg}[equ'n (25)], in the first version on the arXiv,
which incorrectly listed a $(\det {\cal F}_2^*)^{1/2}$
which should have been a $(\det {\cal F}_2)^{1/2}$.
}
\cite{lgeg}[equ'n (25)]
\begin{eqnarray*}
\lefteqn{
q^{-(1/24)(2n + 2r_2 - 2r_1)}
} \\
& & \cdot \int_B {\rm Td}(TB) \wedge {\rm ch}\Biggl(
z^{+r_1/2} \wedge_1( z^{-1} {\cal F}_1 )
z^{-r_2/2} \wedge_1( z {\cal F}_2^{*} )
 \\
& & \hspace*{1.5in} \cdot
\left( \det {\cal F}_1 \right)^{-1/2}
\left( \det {\cal F}_2 \right)^{1/2}
 \\
& & \hspace*{1.5in} \cdot
\bigotimes_{k=1,2,3,\cdots} S_{q^k}( (TB)^{\bf C} )
\bigotimes_{k=0,1,2,\cdots} S_{-q^k}( (z {\cal F}_2^{*})^{\bf C} )
 \\
& & \hspace*{1.5in} \left. \cdot
\bigotimes_{k=1,2,3,\cdots} \wedge_{q^k}( (z {\cal F}_1^*)^{\bf C} )
\right) .
\end{eqnarray*}

As before, to compare, we must exchange
\begin{displaymath}
{\cal F}_1 \: \leftrightarrow \: {\cal F}_2^*
\end{displaymath}
as well as $z \leftrightarrow z^{-1}$.
It is straightforward to check that the expressions above are indeed
exchanged under this operation, which implies that the elliptic genus
is invariant under ${\cal E}' \leftrightarrow {\cal E}'^*$.

For completeness, if the gauge bundle is given by the
cohomology of the short complex
\begin{displaymath}
0 \: \longrightarrow \: {\cal F}_0 \: \stackrel{\tilde{E}^a}{\longrightarrow} \:
{\cal F}_1 \: \stackrel{\tilde{F}_a}{\longrightarrow} \: {\cal F}_2 \:
\longrightarrow \: 0 ,
\end{displaymath}
then the NS sector elliptic genus is given by
\begin{eqnarray*}
\lefteqn{
q^{-(1/24)(2n - r_2 - r_0 + r_1) }
} \\
& & \cdot
\int_B {\rm Td}(TB)    
\wedge {\rm ch} \left(
\bigotimes_{k=1,2,3,\cdots} S_{q^k}\left(
(TB)^{\bf C}\right)  
\right.  \\
& & \hspace*{1.5in} \cdot
\bigotimes_{k=1/2,3/2,\cdots} S_{-q^k }
\left( ( z {\cal F}_2^{*})^{\bf C} \right)
\bigotimes_{k=1/2,3/2,\cdots} S_{-q^k }
\left( ( z^{-1} {\cal F}_0)^{\bf C} \right)
\\
& & \hspace*{1.5in} \left. \cdot
\bigotimes_{k=1/2,3/2,\cdots} \wedge_{q^k }
\left( ( z^{-1} {\cal F}_1)^{\bf C} \right)
\right) .
\end{eqnarray*}
In order for the elliptic genus to be invariant under
${\cal E}' \leftrightarrow {\cal E}'^*$ would require invariance of the
expressions above under
\begin{displaymath}
{\cal F}_0 \: \leftrightarrow \: {\cal F}_2^*, \: \: \:
{\cal F}_1 \: \leftrightarrow \: {\cal F}_1^*, \: \: \:
z \: \leftrightarrow \: z^{-1} ,
\end{displaymath}
and it is straightforward to check that the expression above is indeed
so invariant.

The R sector elliptic genus is given by \cite{lgeg}[equ'n (27)]
\begin{eqnarray*}
\lefteqn{
q^{-(1/24)(2n + 2r_0 + 2r_2 - 2r_1)}
} \nonumber \\
& & \cdot \int_B {\rm Td}(TB) \wedge {\rm ch}\Biggl(
z^{+r_1/2} \wedge_1( z^{-1} {\cal F}_1 )
z^{+r_0/2} \wedge_1( z^{-1} {\cal F}_0 )
z^{-r_2/2} \wedge_1( z {\cal F}_2^{*} )
 \\
& & \hspace*{1.5in} \cdot
\left( \det {\cal F}_1 \right)^{-1/2}
\left( \det {\cal F}_0 \right)^{-1/2}
\left( \det {\cal F}_2 \right)^{+1/2}
 \\
& & \hspace*{1.5in} \cdot
\bigotimes_{k=1,2,3,\cdots} S_{q^k}( (TB)^{\bf C} )
\bigotimes_{k=0,1,2,\cdots} S_{-q^k}( (z^{-1} {\cal F}_0)^{\bf C} )
 \\
& & \hspace*{1.5in} \left. \cdot
\bigotimes_{k=0,1,2,\cdots} S_{-q^k}( (z {\cal F}_2^{*} )^{\bf C} )
\bigotimes_{k=1,2,3,\cdots} \wedge_{q^k}( (z^{-1} {\cal F}_1 )^{\bf C} )
\right) .
\end{eqnarray*}
In order for the elliptic genus to be invariant under
${\cal E}' \leftrightarrow {\cal E}'^*$ would require invariance of the
expressions above under
\begin{displaymath}
{\cal F}_0 \: \leftrightarrow \: {\cal F}_2^*, \: \: \:
{\cal F}_1 \: \leftrightarrow \: {\cal F}_1^*, \: \: \:
z \: \leftrightarrow \: z^{-1} ,
\end{displaymath}
and it is straightforward to check that the expression above is indeed
so invariant, using~(\ref{eq:e-id}).

\subsection{Worldsheet instantons}

Worldsheet instanton corrections in this context were discussed in
\cite{bchir}, which argued for a simple relation between the A/2 and
B/2 models:
\begin{displaymath}
{\rm A/2}(X, {\cal E}) \: = \: {\rm B/2}(X, {\cal E}^*) ,
\end{displaymath}
or more precisely, there existed regularizations (compactifications of
the moduli space of worldsheet instantons) compatible with the statements
above.  (For more information on worldsheet instantons in heterotic
strings, see for example \cite{ks,dgks1,dgks2,mm2} and references therein.)

One of the corners specifically explored in \cite{bchir} is the special
case relating the ordinary B model on $X$ (the B/2 model on ($X, {\cal E}=TX$)
to the A/2 model on ($X, {\cal E}^* = T^*X$)).
Specifically, a worldsheet instanton such that $\phi^* TX \cong \phi^* T^*X$,
as arises in genus zero if the normal bundle is ${\cal O} \oplus {\cal O}(-2)$,
seems to provide a potential contradiction:  the B model does not receive
worldsheet instanton corrections, but the A/2 model typically will receive
worldsheet instanton corrections.  It was observed in \cite{bchir} that in
such cases, in simple examples, there were two moduli space compactifications,
one reproducing B model results, the other reproducing A/2 model results.
Thus, so long as the regularization is exchanged consistent with the 
theory, the worldsheet instanton
counting was consistent.

In any event, it is believed that the A/2 and B/2 models are exchanged
when the gauge bundle is dualized, consistent with the interpretation
of flipping the sign of a left $U(1)$ symmetry.

\subsection{Reducible gauge bundles}
\label{sect:gaugedual:factor}

If the gauge bundle is reducible, then 
we conjecture that the (0,2) QFT's remain isomorphic
after dualizing the various factors separately.

Much of our analysis in the rest of this section applies with little
change, for example:
\begin{itemize}
\item Massless spectra in Calabi-Yau compactifications are invariant
under dualizing factors separately.
\item Elliptic genera are invariant (so long as the vector bundle
is complex, which we have assumed throughout).
\item As there are now several left $U(1)$ symmetries, there are
potentially
several analogues of the A/2 and B/2 models, involving different sets of
twists on left-moving fermions, and with different compatibility conditions
generalizing the A/2 condition $\det {\cal E}^* \cong K_X$.
If multiple twists exist, the duality here should exchange them.
\end{itemize}

In the examples we shall encounter in section~\ref{sect:ggp},
there is another way of thinking about this in the UV GLSM.
In those examples, the duality is applied to Fermi superfields which
are not coupled via a superpotential or other supersymmetry transformations
to the other matter fields.  The theory appears invariant under dualizing the
representation of those Fermi superfields, which implies an IR duality of
the form discussed here.

One point that is more subtle, however, involves the role of
stability.  The stability condition shows up in worldsheet beta functions,
and so is necessary to have a nontrivial IR conformal fixed point.
Dualizing one of the factors will flip the sign of the slope of that
factor, likely destabilizing the bundle.  However, because stability
only enters via beta functions, we need only be concerned with its
role in Calabi-Yau compactifications, and
in such compactification, if the gauge
bundle is reducible, each factor will have vanishing slope, hence the slopes
are unaffected by dualizing factors.  Each factor must still be stable,
but as previously discussed, a bundle is stable if and only if its
dual bundle is also stable.
For a more extensive discussion of compactifications on
reducible gauge bundles in the
context of stability, see for example \cite{kcs,aglo1,aglo2,ago1}.

\subsection{Example of (0,2) dual to (2,2)}

For completeness, let us give an example of a nonabelian (0,2) GLSM which,
assuming the conjectured duality is correct, will RG flow to a (2,2)
GLSM, specifically to the (2,2) GLSM for the Grassmannian $G(k,n)$.

Specifically, consider a (0,2) GLSM on $G(k,n)$ for gauge bundle
${\cal E} = T^* G(k,n) = S \otimes Q^*$: 
\begin{displaymath}
0 \: \longrightarrow \: {\cal E} \: \longrightarrow \:
S \otimes {\cal O}^n \: \longrightarrow \: S \otimes S^* \:
\longrightarrow \: 0 .
\end{displaymath}
This is described by the (0,2) $U(k)$ gauge theory with the following
matter content:
\begin{itemize}
\item $n$ chiral superfields $\Phi$ in the fundamental representation,
\item $1$ chiral superfield $P$ in the adjoint representation,
\item $n$ Fermi superfields $\Gamma$ in the antifundamental representation,
\end{itemize}
plus a (0,2) superpotential of the form
\begin{displaymath}
W \: = \:  \Gamma P \Phi .
\end{displaymath}
It is straightforward to check that this nonabelian (0,2) GLSM satisfies
anomaly cancellation.  From our conjectures above, it should be in the
same universality class as the (2,2) GLSM for $G(k,n)$.

\subsection{Relation to (0,2) mirror symmetry}

Depending upon how one defines (0,2) mirror symmetry
(see {\it e.g.} \cite{mss-rev,guffin-rev,mcorist-rev} for some recent reviews),
the duality we have just discussed might be considered an example.
After all,
the duality we have discussed has the properties that it flips the
sign of a left-moving $U(1)$ (in Calabi-Yau examples), it rotates
sheaf cohomology groups, and exchanges the A/2 and B/2 models, in precisely
the same fashion as one would expect of (0,2) mirror symmetry.

On the other hand, when this duality acts on a (2,2) A-twisted theory
on a space $X$, for example it generates the B/2 model on $(X, T^*X)$
rather than a (2,2) B-twisted theory on the ordinary mirror $Y$.
So, it does not specialize to ordinary mirror symmetry, but then again,
we do not expect (0,2) mirror symmetry for most (0,2) theories to be
related easily to ordinary mirror symmetry.  Only when the gauge bundle
is a deformation of the tangent bundle is such a relation possible.

The most conservative description of how this duality relates to
(0,2) mirrors is encapsulated in the following diagram:
\begin{displaymath}
\xymatrix{
{\rm B/2}(X, {\cal E}^*) \ar@{=}[r] \ar@{=}[d] &
{\rm A/2}(Y, {\cal F}^*) \ar@{=}[d] \\
{\rm A/2}(X,{\cal E}) \ar@{=}[r] &
{\rm B/2}(Y, {\cal F}).
}
\end{displaymath}
In this diagram, horizontal lines indicate ordinary (0,2) mirrors, and vertical
lines indicate the duality discussed here.  For example,
the (0,2) theory defined by $(X,{\cal E})$ is
(0,2) mirror -- in the conventional sense -- to $(Y,{\cal F})$.

Another possibility is that the notion of (0,2) mirrors might be much more
general than previously considered.  Much of (0,2) mirror symmetry is motivated
by the example of ordinary mirror symmetry, which is a relation between
single pairs of spaces, hence many workers have long thought of (0,2) mirrors
as also being relations between single pairs of spaces and bundles.
However, it is also possible that a given (0,2) theory might admit a variety
of different (0,2) mirrors -- the family of dualities might be much more
complicated than previously considered.  Perhaps the duality discussed in
this section should be interpreted as an indication of such a more
complicated structure.  We leave this issue for future work.

\section{Mathematics of Gadde-Gukov-Putrov triality}
\label{sect:ggp}

In this section we will describe\footnote{
We have been told this will also be discussed in \cite{ggp-toappear}.
} the Gadde-Gukov-Putrov triality
\cite{ggp} from a mathematical perspective, 
as an example of a nontrivial (0,2) duality.

We begin by working through the mathematical dualities one encounters
in their picture, {\it i.e.} relating $G(k,n)$ to $G(n-k,n)$, with
suitable gauge bundles.  We shall find a twelve-step duality formally;
however, not all of the bundles appearing admit a (0,2) GLSM description.
This can be fixed by applying physical dualities between (0,2) theories
with dual gauge bundles, at which point this will effectively truncate
to a three-step duality, their triality.

To begin, consider the bundle
\begin{displaymath}
S_k^{\oplus A} \oplus (Q_{n-k}^*)^{\oplus B} \: \longrightarrow \:
G(k,n) .
\end{displaymath}
Under the relation $G(k,n) = G(n-k,n)$, the bundles are related
as follows:
\begin{eqnarray*}
S_k & \leftrightarrow & Q_k^* , \\
Q_{n-k}^* & \leftrightarrow & S_{n-k} ,
\end{eqnarray*}
so we see that the bundle above is the same as
\begin{displaymath}
(Q_k^*)^{\oplus A} \oplus S_{n-k}^{\oplus B} \: \longrightarrow \:
G(n-k,n) .
\end{displaymath}

Now, the bundles $Q^*$ above cannot be realized directly in the GLSM,
but they can be realized indirectly, in mathematics as kernels:
\begin{displaymath}
0 \: \longrightarrow \: Q_{n-k}^*  \: \longrightarrow \: {\cal O}^n \: 
\longrightarrow \: S_k^* \: \longrightarrow \: 0 ,
\end{displaymath}
and in physics by adding a set of $n$ neutral Fermi fields and
a chiral superfield transforming in the antifundamental\footnote{
The chiral superfield should couple to the dual of the bundle appearing
in the third term, {\it i.e.} to $S_k$, which means it corresponds to
the antifundamental.
}, together with
a (0,2) superpotential.

For example, ignoring anomalies for the moment, the bundle
\begin{displaymath}
S_k^{\oplus A} \oplus (Q_{n-k}^*)^{\oplus B} \: \longrightarrow \:
G(k,n)
\end{displaymath}
is realized physically by a $U(k)$ gauge theory containing
\begin{itemize}
\item $n$ chiral superfields $\Phi_i$ each in the fundamental representation
of $U(k)$,
\item $B$ chiral superfields $P^i$ each in the antifundamental representation
of $U(k)$,
\item $A$ Fermi superfields in the antifundamental representation of
$U(k)$,
\item $nB$ neutral Fermi superfields $\Gamma$,
\item a (0,2) superpotential $\Gamma \Phi P$.
\end{itemize}

Gauge anomaly cancellation constrains the values of $A$, $B$, $k$, $n$.
For simplicity, we will use the decomposition
$u(k) \cong su(k) \oplus u(1)$ and work out anomaly cancellation in terms
of the constituent summands.
With the benefit of hindsight,
to cancel gauge anomalies, we add Fermi superfields $\Omega$ transforming
only under $\det U(k)$, then the theory above contains the following
matter fields, charged under $su(k) \oplus u(1)$:
\begin{center}
\begin{tabular}{ccccc}
$ $ & type & multiplicity & $su(k)$ & $u(1)$  \\ \hline 
$\Phi$ & chiral & $n$ & ${\bf k}$ & 1  \\
$P$ & chiral & $B$ & ${\bf \overline{k}}$ & -1  \\
$\Gamma$ & Fermi & $nB$ & ${\bf 1}$ & 0  \\
$\Psi$ & Fermi & $A$ & ${\bf \overline{k}}$ & -1  \\
$\lambda$ & fermion & 1 & $ad$ & 0  \\
$\Omega$ & Fermi & 2 & ${\bf 1}$ & $k$  \\
\end{tabular}
\end{center}
Using the indices given in appendix~\ref{app:uk-reps},
the $su(k)^2$ gauge anomaly is
\begin{displaymath}
n k \frac{k^2-1}{k} \: + \: B k \frac{k^2-1}{k}
\: = \: A k \frac{k^2-1}{k} \: + \: (k^2-1)(2k),
\end{displaymath}
and the $u(1)^2$ gauge anomaly is
\begin{displaymath}
nk \: + \: Bk \: = \: Ak \: + \: 2k^2,
\end{displaymath}
which imply the following constraint:
\begin{displaymath}
2k \: = \: n \: + \: B \: - \: A .
\end{displaymath}

The theory describing the same bundle in the dual description, namely
\begin{displaymath}
(Q_k^*)^{\oplus A} \oplus S^{\oplus B}_{n-k} \: \longrightarrow \:
G(n-k,n),
\end{displaymath}
is
a $U(n-k)$ gauge theory containing
\begin{itemize}
\item $n$ chiral superfields $\tilde{\Phi}_i$ 
each in the fundamental representation
of $U(n-k)$,
\item $A$ chiral superfields $\tilde{P}^i$ each in the antifundamental 
representation of $U(n-k)$,
\item $B$ Fermi superfields in the antifundamental representation of
$U(n-k)$,
\item $nA$ neutral Fermi superfields $\tilde{\Gamma}$,
\item a (0,2) superpotential $\tilde{\Gamma} \tilde{\Phi} \tilde{P}$.
\end{itemize}
It is straightforward that adding a pair of Fermi superfields $\Omega$,
each of charge $n-k$, cancels the gauge anomaly so long as the
same constraint from before, namely
\begin{displaymath}
2k \: = \: n \: + \: B \: - \: A
\end{displaymath}
is obeyed.
More generally, it is straightforward to check that in all the duality
frames discussed before, a pair of $\Omega$'s can be added to cancel
anomalies, subject to the same constraint as above, so henceforward
we will omit the $\Omega$'s and take the constraint as given.

Returning to the physical realization of the bundle
\begin{displaymath}
S_k^{\oplus A} \oplus (Q_{n-k}^*)^{\oplus B} \: \longrightarrow \:
G(k,n) ,
\end{displaymath}
it is straightforward to see that
the (0,2) theory describing this phase has a second distinct K\"ahler
phase describing the bundle
\begin{displaymath}
(S_k^*)^{\oplus A} \oplus (Q_{n-k}^*)^{\oplus n} \: \longrightarrow \:
G(k,B) ,
\end{displaymath}
essentially obtained by flipping the interpretation of fundamental and
antifundamental representations.  (Note that the interpretation of the
Fermi superfields describing the $S$ factor also therefore flips, so here
we have $S^*$ rather than $S$ in the gauge bundle.)
This second K\"ahler phase also has
a dual description, and in this fashion we can construct a chain of
dualities.

The first few steps of this chain of dualities are as follows:
\begin{displaymath}
\xymatrix{
 S^A \oplus (Q^*)^{2k+A-n} \rightarrow G(k,n) 
\ar@{--}[r] \ar@{<->}[d]_{=} &
 (S^*)^A \oplus (Q^*)^n \rightarrow G(k,2k+A-n)  
\\
 (Q^*)^A \oplus S^{2k+A-n} \rightarrow G(n-k,n) 
\ar@{--}[r] 
&
 (Q^*)^n \oplus (S^*)^{2k+A-n} \rightarrow G(n-k,A) 
\ar@{<->}[d]^{=} \\
 (S^*)^n \oplus Q^A \rightarrow G(A-n+k,2k+A-n) 
\ar@{--}[r]  
&
 S^n \oplus Q^{2k+A-n} \rightarrow G(A-n+k,A) .  
}
\end{displaymath}
Horizontal (dashed) lines indicate different K\"ahler phases; vertical lines
indicate mathematical dualities between descriptions of the same object.
Formally, if one were to continue for a total of six steps,
one would get to a GLSM with the same Grassmannians as the first line, but 
dual bundles.

We ran into a potential problem in section~\ref{sect:22fundanti} in describing
chains of dualities of the form above, defined by RG flow and the existence
of Coulomb vacua in certain phases.  Although these models have FI parameters
that will certainly RG flow, there is no $\sigma$ field in these models,
hence no Coulomb vacua to obstruct dualities as in
section~\ref{sect:22fundanti}.

A second problem is less trivial.
Specifically, the geometries indicated on the third line above cannot be
realized in (0,2) GLSMs.  The problem is that the gauge bundle on the third
line involves copies of $Q$.  To realize $Q$ as part of the gauge
bundle in a (0,2) GLSM, we would need to realize it as the cokernel in
a short exact sequence of a form previously described, and to do so,
we would need chiral superfields in representations corresponding to the
dual of $S$.  This is a problem -- such chiral superfields would then be
in the same representation as those defining the underlying Grassmannian,
so instead of building a bundle, one would build a larger Grassmannian.
We discussed an analogous difficulty in (2,2) GLSMs in 
section~\ref{sect:22fundanti}; 
as discussed there, the issue here is the analogue of
trying to build a (2,2) GLSM for the total space of ${\cal O}(+1)
\rightarrow {\mathbb P}^n$ -- the chiral superfield for the fibers,
has the same charges as those appearing in the base, so the obvious
GLSM would instead describe ${\mathbb P}^{n+1}$.

Instead, we can dualize the gauge bundle, as in section~\ref{sect:dualbundle}.
Doing so, and using a dashed vertical arrow to indicate a physical
isomorphism which is not also a mathematical equivalence, we are led
to the duality chain
\begin{displaymath}
\xymatrix{
 S^A \oplus (Q^*)^{2k+A-n} \rightarrow G(k,n) 
\ar@{--}[r] \ar@{<->}[d]_{=} &
 (S^*)^A \oplus (Q^*)^n \rightarrow G(k,2k+A-n)  
\\
 (Q^*)^A \oplus S^{2k+A-n} \rightarrow G(n-k,n) 
\ar@{--}[r] 
&
 (Q^*)^n \oplus (S^*)^{2k+A-n} \rightarrow G(n-k,A) 
\ar@{<-->}[d]^{\cong} \\
 S^n \oplus (Q^*)^A \rightarrow G(A-n+k,2k+A-n) 
\ar@{--}[r] \ar@{<->}[d]_{=} 
&
 (S^*)^n \oplus (Q^*)^{2k+A-n} \rightarrow G(A-n+k,A)  \\
 (Q^*)^n \oplus S^A \rightarrow G(k,2k+A-n) 
\ar@{--}[r] 
& 
(Q^*)^{2k+A-n} \oplus (S^*)^A \rightarrow G(k,n) 
\ar@{<-->}[d]^{\cong} \\
 S^{2k+A-n} \oplus (Q^*)^n \rightarrow G(n-k,A) 
\ar@{--}[r] \ar@{<->}[d]_{=} &
 (S^*)^{2k+A-n} \oplus (Q^*)^A \rightarrow G(n-k,n)  \\
 (Q^*)^{2k+A-n} \oplus S^n \rightarrow G(k+A-n,A) 
\ar@{--}[r] &
 (Q^*)^A \oplus (S^*)^n \rightarrow G(k+A-n,2k+A-n) 
\ar@{<-->}[d]^{\cong} \\
 S^A \oplus (Q^*)^{2k+A-n} \rightarrow G(k,n) 
\ar@{--}[r] &
 (S^*)^A \oplus (Q^*)^n \rightarrow G(k,2k+A-n) . 
}
\end{displaymath}

After six steps we have returned to our starting point, but in fact
one can do better.  If we were to dualize the $S$ factors in the gauge
bundle in the fourth line, applying the duality discussed in
section~\ref{sect:gaugedual:factor}, then the diagram above would reduce to
\begin{displaymath}
\xymatrix{
 S^A \oplus (Q^*)^{2k+A-n} \rightarrow G(k,n) 
\ar@{--}[r] \ar@{<->}[d]_{=} &
 (S^*)^A \oplus (Q^*)^n \rightarrow G(k,2k+A-n)  
\\
 (Q^*)^A \oplus S^{2k+A-n} \rightarrow G(n-k,n) 
\ar@{--}[r] 
&
 (Q^*)^n \oplus (S^*)^{2k+A-n} \rightarrow G(n-k,A) 
\ar@{<-->}[d]^{\cong} \\
 S^n \oplus (Q^*)^A \rightarrow G(A-n+k,2k+A-n) 
\ar@{--}[r] \ar@{<-->}[d]_{\cong} 
&
 (S^*)^n \oplus (Q^*)^{2k+A-n} \rightarrow G(A-n+k,A)  \\
 (Q^*)^n \oplus (S^*)^A \rightarrow G(k,2k+A-n) 
\ar@{--}[r] 
& 
(Q^*)^{2k+A-n} \oplus S^A \rightarrow G(k,n) .
}
\end{displaymath}
The fourth line is now identical to the first, except that the
Fayet-Iliopoulos parameter has been reversed.  In this fashion we can
understand this as a triality symmetry, as described in \cite{ggp}.

As in section~\ref{sect:22fundanti}, we have only described the duality
chain moving in one direction.  One could also move in the opposite direction,
yielding equivalent results.

More degenerate examples exist with shorter periodicities.
For example, if $n=2k$, $A=2k$, then we have
\begin{displaymath}
\xymatrix{
S^{2k} \oplus (Q^*)^{2k} \rightarrow G(k,2k)
\ar@{--}[r] \ar@{<->}[d]_{=} &
(S^*)^{2k} \oplus (Q^*)^{2k} \rightarrow G(k,2k) \\
(Q^*)^{2k} \oplus S^{2k} \rightarrow G(k,2k) 
\ar@{--}[r] &
(Q^*)^{2k} \oplus (S^*)^{2k} \rightarrow G(k,2k) .
}
\end{displaymath}
In effect, the (0,2) GLSM defined by
\begin{displaymath}
S^{2k} \oplus (Q^*)^{2k} \rightarrow G(k,2k)
\end{displaymath}
is self-dual.

\section{Relation between models of Pfaffians}
\label{sect:pax-vs-paxy}

In section~\ref{sect:pfaff} we reviewed the construction of (2,2) GLSMs for 
Pfaffian varieties, and also extended those constructions to (0,2) GLSMs.
For any given Pfaffian and bundle, there were a pair of constructions,
known as the PAX and PAXY models.
Reference \cite{jklmr1} described how the (2,2) PAX and PAXY models were
related.

In this section, we will use (2,2) and (0,2) nonabelian gauge theory dualities
to update the discussion of \cite{jklmr1}[section 3.4],
and also extend to (0,2) cases.

\subsection{(2,2) GLSMs}

Let us begin by rewriting the analysis of \cite{jklmr1}[section 3.4]
utilizing the two-dimensional analogue of Seiberg duality introduced
in \cite{bc1} and reviewed in section~\ref{sect:22fundanti}.

Briefly, begin with the PAX model.  Here one has, in addition to the
data defining a toric variety and a matrix $A$ defined over that toric
variety, a $U(n-k)$ gauge theory, a set of $n$ fundamentals encoded in an
$n \times (n-k)$ matrix $P$, a set of $n$ antifundamentals encoded in an
$n \times (n-k)$ matrix $X$, and a superpotential of the form
\begin{displaymath}
W \: = \: {\rm tr}\, PAX .
\end{displaymath}

Now, let us apply the duality of \cite{bc1}.  The dual theory will be
a $U(k)$ gauge theory, with an $n \times n$ matrix of neutral mesons
$\tilde{P}$, related to the charged matter of the original theory by
\begin{displaymath}
\tilde{P} \: = \: X P ,
\end{displaymath}
as well as a new set of $n$ fundamentals, encoded in a $n \times k$ matrix
$\tilde{X}$, a new set of $n$ antifundamentals, encoded in an $n \times k$
matrix $\tilde{Y}$, and, just from the duality, a superpotential
\begin{displaymath}
W' \: = \: {\rm tr}\, \tilde{X} \tilde{P} \tilde{Y} ,
\end{displaymath}
closely following the pattern of four-dimensional Seiberg duality
\cite{seibdual}.
If we combine the duality contribution with the original superpotential
written in dual variables, we find that the complete superpotential
for the theory dual to the PAX model is
\begin{displaymath}
W \: = \: {\rm tr}\, \tilde{P}\left( A(\Phi) \: + \: 
\tilde{Y} \tilde{X} \right) .
\end{displaymath}
After a trivial field redefinition, this becomes
\begin{displaymath}
W \: = \: {\rm tr}\, \tilde{P}\left( A(\Phi) \: - \: 
\tilde{Y} \tilde{X} \right) ,
\end{displaymath}
which exactly matches the PAXY model.  

Thus, we see that the (2,2) PAX and PAXY models are related by a simple
application of the duality discussed in \cite{bc1} and 
section~\ref{sect:22fundanti}.

\subsection{(0,2) generalizations}

Now, let us apply analogous ideas to compute the dual of a more general
(0,2) PAX model.  
We shall begin by studying how deformations of the tangent bundle in the
PAX model map to analogous deformations in the PAXY model.
Recall that deformations of the tangent bundle are described by a (0,2)
PAX model fields as described in section~\ref{sect:pfaff} 
and with superpotential
\begin{displaymath}
W \: = \: {\rm tr}\, \left( \Lambda_P A(\Phi) X \: + \:
P A(\Phi) \Lambda_X \: + \:
P \left( \frac{\partial A(\Phi)}{\partial \Phi^{\alpha}} 
\: + \: G_{\alpha}(\Phi)\right) \Lambda^{\alpha}_{\Phi}
X \right) .
\end{displaymath}
Since the deformation is encoded in the superpotential, the fields themselves
are the same as in the (2,2) GLSM, so we can apply essentially the same
duality as in the (2,2) case, albeit re-expressed in terms of (0,2)
superfields.  Thus, the dual gauge theory will be a $U(k)$ gauge theory
(plus another abelian factor, which will go along for the ride),
with
\begin{enumerate}
\item an $n\times n$ matrix of neutral (meson) chiral superfields
$\tilde{P}$ and
Fermi superfields $\Lambda_{\tilde{P}}$, related to fields of the
original theory by
\begin{displaymath}
\tilde{P} \: = \: XP, \: \: \:
\Lambda_{\tilde{P}} \: = \: \Lambda_X P \: + \: X \Lambda_P ,
\end{displaymath}
\item a new set of $n$ fundamentals , encoded in a $n \times k$ matrix
$\tilde{X}$,
\item a new set of $n$ antifundamentals, encoded in a $n \times k$ matrix
$\tilde{Y}$,
\item a superpotential term
\begin{displaymath}
W' \: = \: {\rm tr}\, \left(
\Lambda_{\tilde{P}} \tilde{X} \tilde{Y} \: + \:
\tilde{P} \Lambda_{\tilde{X}} \tilde{Y} \: + \:
\tilde{P} \tilde{X} \Lambda_{\tilde{Y}}
\right) .
\end{displaymath}
\end{enumerate}
When the new superpotential term is
added to the previous superpotential expressed in terms of the
dual variables, namely
\begin{displaymath}
{\rm tr}\, \left( \Lambda_{\tilde{P}} A(\Phi) \: + \: \tilde{P}\left(
\frac{\partial A(\Phi) }{\partial \Phi^{\alpha} } \: + \:
G_{\alpha}(\Phi) \right) \Lambda^{\alpha}_{\Phi}
\right) ,
\end{displaymath}
we get the full (0,2) superpotential of the dual theory:
\begin{displaymath}
W \: = \:
 \left( \Lambda_{\tilde{P}} A(\Phi) \: + \: \tilde{P}\left(
\frac{\partial A(\Phi) }{\partial \Phi^{\alpha} } \: + \:
G_{\alpha}(\Phi) \right) \Lambda^{\alpha}_{\Phi}
\: + \: 
\Lambda_{\tilde{P}} \tilde{X} \tilde{Y} \: + \:
\tilde{P} \Lambda_{\tilde{X}} \tilde{Y} \: + \:
\tilde{P} \tilde{X} \Lambda_{\tilde{Y}}
\right) .
\end{displaymath}
Modulo absorbing signs into trivial field redefinitions, this is
the same as the PAXY theory for the deformation off the (2,2) locus
given in equation~(\ref{eq:paxy-02-def}).

Thus, we see the duality between PAX and PAXY models extends to
deformations off the (2,2) locus.

Now, let us consider an example of a more general case,
a gauge bundle given as a kernel.
We follow the same conventions as in section~\ref{sect:pfaff}.
In other words, to build the Pfaffian itself, we will need a $U(n-k)$
gauge theory, $n$ chiral superfields
in the fundamental, forming an $n\times (n-k)$ matrix denoted $X$,
and $n$ Fermi superfields in the antifundamental, forming an $n\times (n-k)$
matrix of Fermi superfields denoted $\Lambda_0$. 
If the gauge bundle ${\cal E}$ is given as a kernel of the form
\begin{displaymath}
0 \: \longrightarrow \: {\cal E} \: \longrightarrow \:
\oplus_{\beta} {\cal O}((\lambda_{\beta 1}, \lambda_{\beta 2}), q_{a, \beta})
\: \stackrel{F^{\gamma}_{\beta} }{\longrightarrow} \:
\oplus_{\gamma} {\cal O}((\lambda_{\gamma 1}, \lambda_{\gamma 2}, q_{a, \gamma})
\: \longrightarrow \: 0 ,
\end{displaymath}
then we add a set of Fermi superfields $\Lambda^{\beta}$ in the
$(\lambda_{\beta 1}, \lambda_{\beta 2})$ representation of $U(n-k)$
and with charges $q_{a, \beta}$ under the abelian gauge symmetry $U(1)^r$
defining
the toric variety, along with a set of chiral superfields
$P_{\gamma}$ in the $U(n-k)$ representation dual to
$(\lambda_{\gamma 1}, \lambda_{\gamma 2})$ and with charges $- q_{a, \gamma}$
under the abelian gauge symmetry defining the toric variety.
In addition, we have a (0,2) superpotential 
\begin{displaymath}
W \: = \:
{\rm tr}\, \left( \Lambda_0 A(\Phi) X \: + \:
\Lambda^{\beta} F^{\gamma}_{\beta}(\Phi) P_{\gamma}
\right) .
\end{displaymath}

However, we quickly run into a problem.  The gauge bundle ${\cal E}$
above is defined in terms of representations of $U(n-k)$, in the PAX
model.  However, in the PAXY model, the gauge bundle is defined in 
terms of representations of $U(k)$.  Now, it is possible to write down
long exact sequences relating representations of one to the other,
as we shall discuss in section~\ref{open-seiberg}, 
but as we shall discuss there,
to be relevant for (0,2) constructions, we must restrict to duals involving
three-term sequences, which are comparatively rare.  Thus, we do not
expect to be able to construct PAXY duals of most (0,2) PAX models, and
also conversely.  This is a special case of a more general obstruction
we shall discuss in section~\ref{sect:02-obs}.

\section{More general bundles and obstructions to duality}
\label{open-seiberg}

So far, we have discussed dualities for closed string (2,2) and (0,2)
$U(k)$ gauge theories with matter in fundamental and antifundamental 
representations.  In this section, we will discuss more general matter
representations.  We will discuss how arbitrary matter representations can
be dualized in open strings, and also discuss obstructions to duality for
more general matter representations in closed string (2,2) and (0,2) theories.

\subsection{Duality for $U(k)$ gauge theories in open strings}

The key to our deliberations so far has been that the bundles $S_k$, $Q_{n-k}$
over the Grassmannian $G(k,n)$ are the same as the bundles
$Q^*_k$, $S^*_{n-k}$ over the Grassmannian $G(n-k,n)$.
In each case, the universal subbundle is defined by the antifundamental
representation (in conventions in which the matter defining the Grassmannian
itself is in the fundamental representation), and the universal quotient
bundle is built as the cokernel in a short exact sequence, which can be
realized physically.

In open strings, the Chan-Paton factors couple to complexes of
bundles defined by representations of the gauge group, so in open strings
on a GLSM for $G(k,n)$, the Chan-Paton factors are defined by complexes
of bundles defined by $U(k)$ representations.   

Suppose we start with Chan-Paton factors coupling to a single bundle
${\cal O}(\rho)$, defined by some representation $\rho$ of $U(k)$.
In the notation of appendix~\ref{app:uk-reps}, if the representation is
defined by
\begin{displaymath}
\rho \: \equiv \: (\lambda_1, \lambda_2, \cdots, \lambda_k) ,
\end{displaymath}
where each $\lambda_i \geq \lambda_{i+1}$, then we can construct
${\cal O}(\rho)$ from suitable tensor products of powers of $S$ and
$S^*$.  Schematically,
\begin{displaymath}
{\cal O}(\rho) \: = \: K_{\rho}(S) \otimes (\det S^*)^{\lambda_k}
\end{displaymath}
where $K_{\rho}$ is the tensor product defined by the $SU(k)$ Young
diagram associated to $\rho$.

In the dual $U(n-k)$ gauge theory, the Chan-Paton factors in principle
should couple to
\begin{displaymath}
K_{\rho}(Q^*) \otimes (\det Q)^{\lambda_k} .
\end{displaymath}
However, the bundle $Q \rightarrow G(n-k,n)$ 
is not given directly by a representation of
$U(n-k)$.  Instead, it is always possible to find a long exact
sequence of bundles defined by representations of $U(n-k)$ that
`resolves' the bundle above, and so we can replace the bundle above
by its resolution.  The resolution then gives well-defined Chan-Paton
factors in the dual gauge theory, which in principle must result in
an open string in the same universality class as the original open
string.

As a consistency check, note that the tangent bundle of the Grassmannian
is the cokernel of
\begin{displaymath}
\left\{
0 \: \longrightarrow \: S^{\vee} \otimes S \: \longrightarrow \:
S^{\vee} \otimes {\cal O}^n
\right\}
\: = \:
S^{\vee} \otimes \left\{
0 \: \longrightarrow \: S \: \longrightarrow \: {\cal O}^n \right\} ,
\end{displaymath}
which is precisely $S^{\vee}$ tensored with the dual of the complex
representing the dual $S$, and hence is manifestly symmetric
under the duality $G(k,n) \leftrightarrow G(n-k,n)$,
which is very satisfying.

Let us consider a less trivial example, namely the bundle
${\cal O}({\bf k}
\otimes {\bf k})$.  This is dual to the tensor product of two
copies of the complex $\{ S_{n-k} \rightarrow {\cal O}^n \}$ on
$G(n-k,n)$.  In general, give two chain complexes
$P_{\cdot}$, $Q_{\cdot}$, we can define a complex
$P \otimes Q$ by taking \cite{weibel}[chapter 2.7]
\begin{displaymath}
( P \otimes Q )_n \: = \: \bigoplus_{p+q=n} P_p \otimes Q_q ,
\end{displaymath}
with differential $d \otimes 1 + (-)^p \otimes d$.
In the present case, this yields the complex
\begin{displaymath}
{\cal O}^{n^2} \: \longrightarrow \:
\oplus_1^2 (S^{*}_{n-k})^{\oplus n} \: \longrightarrow \:
S^{*}_{n-k} \otimes S^{*}_{n-k},
\end{displaymath}
which we claim is the open string dual to the Chan-Paton bundle
${\cal O}({\bf k}
\otimes {\bf k})$ in the $U(k)$ gauge theory corresponding to
$G(k,n)$ (in the bulk of the open string).
(As a check, note that the rank is $n^2 - 2n(n-k) + (n-k)^2 = k^2$,
as expected.)

For another example, suppose instead the Chan-Paton bundle in the
$U(k)$ gauge theory corresponding to $G(k,n)$ was given by
the bundle
$\wedge^p S \rightarrow G(k,n)$
for some $p > 1$.  Under the duality, $\wedge^p S \mapsto \wedge^p Q^*$.
However, $\wedge^p Q^*$ can not be resolved by a three-term sequence involving
only bundles defined by representations of $U(k)$.  Instead, it can be
resolved as
\begin{displaymath}
0 \: \rightarrow \: \wedge^p Q^* \: \rightarrow \:
\wedge^p {\cal O}^n \: \rightarrow \: S^* \otimes \wedge^{p-1} {\cal O}^n
\: \rightarrow \: \cdots \: \rightarrow \: 
{\rm Sym}^{p-1} S^* \otimes {\cal O}^n \: \rightarrow \:
{\rm Sym}^p S^* \: \rightarrow \: 0 .
\end{displaymath}
Thus, in the dual gauge theory, Chan-Paton factors describing the
complex
\begin{displaymath}
\wedge^p {\cal O}^n \: \longrightarrow \: S^* \otimes \wedge^{p-1} {\cal O}^n
\: \longrightarrow \: \cdots \: \longrightarrow \: 
{\rm Sym}^{p-1} S^* \otimes {\cal O}^n \: \longrightarrow \:
{\rm Sym}^p S^* 
\end{displaymath}
over $G(n-k,n)$
should be in the same universality class as the original Chan-Paton bundle
\begin{displaymath}
\wedge^p S
\end{displaymath}
over $G(k,n)$
in the original gauge theory.

\subsection{Obstructions to duality in (0,2) theories}
\label{sect:02-obs}

Now, let us apply the same ideas to (0,2) and closed-string (2,2) theories.
In a (0,2) theory, we can talk about dualizing the gauge bundle;
in a closed-string (2,2) theory describing the total space of the
bundle, we can talk about dualizing to another closed-string theory
describing the same bundle.

In both cases, there is a potential obstruction to making sense of the
duality, lying in the fact that in each case, it is not currently known
how to realize longer than three-term complexes.

For example, in a (0,2) GLSM, the gauge bundle can be realized as a 
kernel, cokernel, or as the cohomology of a three-term monad, but it is
not currently known how to physically realize
sequences longer than three terms.

As a result, for example, although we can certainly write down a
(0,2) GLSM describing the bundle $\wedge^p S \rightarrow G(k,n)$ for
$p>1$, it is not known at present how to realize
its dual over $G(n-k,n)$ in a (0,2) GLSM, because it
involves a complex of length greater than three, barring the use
of a physical duality that does not correspond to a mathematical one.

Similarly in (2,2) nonabelian GLSMs, 
we can describe target spaces that are total
spaces of any bundle defined by a $U(k)$ representation over $G(k,n)$,
for example, but unless the dual is defined in terms of a three-term
sequence, we do not currently know how to describe it with superpotentials 
and so forth,
and so we cannot currently describe it.

For this reason, we conjecture\footnote{
We hesitate to formulate this as a no-go theorem, 
as we are reminded of the old saying, ``Never trust a no-go
theorem until after some counterexamples are known.''
} that (0,2) and closed string (2,2) GLSMs
describing bundles over $G(k,n)$ corresponding to
representations of $U(k)$ other than fundamentals, antifundamentals,
and adjoints, do not have Seiberg-like duals.  Of course, new 
physical relationships, unmotivated by mathematics, could easily
modify that conclusion.

By contrast, in open string theories there is no such restriction,
and we expect all Chan-Paton factors in corresponding open string
GLSMs to have duals.

\section{Conclusions}

In this paper we have discussed a variety of basic aspects of nonabelian
(0,2) GLSMs in two dimensions. 
We began with a general discussion of dynamical supersymmetry breaking and
the role of spectators in understanding weak coupling limits. 
We then worked through the details of some
toy examples of nonabelian (0,2) theories on Grassmannians, studying
dualities and supersymmetry breaking.  We then turned to Calabi-Yau
and related examples, such as complete
intersections in Grassmannians, Pfaffians, and other related spaces,
studying basic properties and dualities.  We then turned to a study of
dualities in (2,2) and (0,2) theories.  We observed that two-dimensional
analogues of four-dimensional Seiberg duality have a purely mathematical
understanding, as a simple generalization of the relationship
$G(k,n) = G(n-k,n)$, which in GLSMs relates $U(k)$ and $U(n-k)$ gauge
theories, and also used existing mathematical relationships to observe
the existence of additional dualities in (2,2) GLSMs between Grassmannians
and certain Pfaffians.  
We then worked through the details of another proposed
duality relating theories with dual gauge bundles, and applied these
two dualities 
to give a mathematical picture of 
the recent Gadde-Gukov-Putrov triality.  We also
reviewed the relation between
PAX and PAXY constructions of GLSMs for Pfaffians, and concluded with
a description of dualities in open strings and a demonstration that analogous
dualities for more general nonabelian (0,2) theories are unlikely.

One simple possibility for future work would be to better understand
Landau-Ginzburg points.  We have focused almost exclusively on
large-radius geometries in this paper.

Another possibility for future work is to consider variations on the
gauge groups given here.  We have discussed, albeit briefly, how
a $SU(k)$ gauge theory with chiral superfields in the fundamental
can result in an affine Grassmannian instead of
an ordinary Grassmannian, and more could be done to follow that up.
For another example, if we take the gauge group to be $U(1) \times SU(k)$
rather than $U(k)$, then instead of a Grassmannian $G(k,n)$ one gets
a ${\mathbb Z}_k$ gerbe on the Grassmannian, in fact a gerbe that generates
other other ${\mathbb Z}_k$ gerbes \cite{tonypriv}.  On a related point,
we would also like to better understand the role of 
two-dimensional discrete theta angles (recently discussed in \cite{horiknapp}),
related to four-dimensional discrete theta angles recently discussed
in \cite{ast1,gmn1}.  We hope to return to these questions in
future work.

Another possibility left for future work is to find D-brane realizations of the
(2,2) and (0,2) dualities and duality chains discussed here,
following \cite{hh1}.

Yet another possibility is to apply the geometries discussed in this paper
to four-dimensional Seiberg duality, to better understand existing
gauge theory dualities and perhaps extract a few more.

In terms of Pfaffian realizations in GLSMs, another direction to pursue
would be to realize extremal transition for Pfaffians (see {\it e.g.}
\cite{kapustka})
in terms of the construction of \cite{jklmr1}.

\section{Acknowledgements}

We would like to thank M.~Ando, F.~Benini,
J.~Distler, R.~Donagi, R.~Eager, S.~Gukov, S.~Katz, A.~Knutson, J.~Lapan,
I.~Melnikov, L.~Mihalcea,
and T.~Pantev for
useful discussions.
E.S. was partially supported by NSF grant PHY-1068725.

\appendix

\section{GLSMs and cohomology}
\label{app:cohomology}

In this appendix we relate GLSM operators to cohomology, focusing
in particular on GLSMs for Grassmannians.

Consider a GLSM that is described by gauging the action of
some Lie group $G$ on a vector space $V$, supersymmetrically.
We claim that the cohomology ring seen by the GLSM is
the $G$-equivariant cohomology of $V$, {\it i.e.},
\begin{displaymath}
H^*_G(V) \: = \: H^*_G( {\rm pt} ) \: = \: H^*(BG),
\end{displaymath}
and that this cohomology ring is realized by operators build from the
adjoint-valued scalars $\sigma$ in the two-dimensional gauge multiplet.

Let us work through the details a little more.
Suppose we have an abelian GLSM (assumed without a superpotential),
which
describes, in one K\"ahler phase, a toric variety
\begin{displaymath}
\frac{ V - E }{ ( {\bf C}^{\times} )^k } ,
\end{displaymath}
where $V$ is a vector space, and $E$ the exceptional set for that
phase.

The cohomology seen by the GLSM is the $({\bf C}^{\times})^k$-equivariant
cohomology of the vector space $V$ -- equivalently, the
$U(1)^k$-equivariant cohomology, as $G_{\bf C}$-equivariant
cohomology is the same as $G$-equivariant cohomology.
Using the inclusion $V - E \hookrightarrow V$, the equivariant
cohomology of $V$ can be pulled back to the equivariant cohomology
of $V-E$, which (assuming there are no fixed points) descends to the
ordinary cohomology of the toric variety.
(This is a special case of the Kirwan surjectivity
theorem, valid for rational coefficients.)
This can be done for every exceptional set, and so the equivariant
cohomology of $V$ defines something universal for all phases of the
GLSM.

We can compute the $({\bf C}^{\times})^k$-equivariant cohomology of $V$
by using the fact that $V$ is contractible;
the result is just $H^*(B U(1) \times \cdots \times BU(1))$,
($k$ copies of $BU(1)$), which is the polynomial ring in k variables:
\begin{displaymath}
H^*_{ ( {\bf C}^{\times})^k }(V) \: = \:
{\bf C}[x_1, \cdots, x_k] ,
\end{displaymath}
independent of the dimension of the vector space $V$.
Physically, each $x_i$ corresponds to a $\sigma_i$ in the
vector supermultiplet.

In principle something closely analogous should happen in nonabelian
GLSMs.  All of the analysis above applies, except that the equivariant
cohomology itself now has different values.
To this end, recall
\begin{itemize}
\item $H^*(BSU(n), {\bf Z}) = {\bf Z}[c_2, c_3, \cdots, c_n], $
\item $H^*(BU(n), {\bf Z}) = {\bf Z}[c_1, c_2, \cdots, c_n], $
\end{itemize}
where $c_i$ has degree $2i$, and corresponds to a Chern class.

In more detail, the cohomology ring of $B GL(k) = BU(k) = G(k,\infty)$ is
discussed in \cite{ms}[section 16]:  it is the
ring of symmetric polynomials in $k$ indeterminates.

We can relate the formal structures above to physics as follows.
Recall the Cartan model of equivariant cohomology \cite{cmr}[section 10.7]
is the multiplet
\begin{eqnarray*}
d A & = & \psi , \\
d \psi & = & - D_A \sigma , \\
d \sigma & = & 0 .
\end{eqnarray*}
The field $\sigma$ is a Lie-algebra-valued scalar; see also
\cite{cmr}[sections 10.9-10.10].  As discussed in
\cite{mp}[section 3.6], this structure is realized in GLSMs.
The generators of the equivariant cohomology rings above correspond to
operators of the form ${\rm Tr}\, \sigma^k$ for various $k$.

It is a standard result (see {\it e.g.} \cite{gh}[chapter 1.5],
\cite{chern}[chapter 8])
that the integral homology of the Grassmannian
$G(k,n)$ has no torsion and is freely generated by cycles in one-to-one
correspondence with Young diagrams (unlabelled Young tableaux),
specified by a sequence of $k$
positive integers $d_1, \cdots, d_k$, where
\begin{displaymath}
n-k \: \geq \: d_1 \: \geq \: d_2 \: \geq \: \cdots \: \geq \: d_k
\: \geq \: 0
\end{displaymath}
({\it i.e.} $d_i$ is the number of boxes on row $i$)
and where the Young diagram above corresponds to a cycle of
real codimension
\begin{displaymath}
2 \sum_i d_i .
\end{displaymath}
For example,
\begin{displaymath}
H^2(G(k,n),{\bf Z}) \: = \: {\bf Z}
\end{displaymath}
corresponds to
\begin{displaymath}
\yng(1) . 
\end{displaymath}
Similarly,
\begin{displaymath}
H^4(G(k,n),{\bf Z}) \: = \: {\bf Z}^2
\end{displaymath}
corresponds to
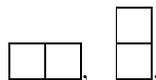
\begin{displaymath}
\yng(2), \: \: \: \yng(1,1) .
\end{displaymath}
Intersection theory on these (Schubert) cycles, cup products on the
cohomology, are determined in the same way as representations of
$GL(k)$.  Schur polynomials provide the link between Young diagrams and
symmetrized polynomials that were used earlier to describe the cohomology
of the Grassmannian, in terms of equivariant cohomology.

Let us work through some examples in more detail.
In general terms, $H^2$ is generated by
${\rm Tr}\, \sigma$;
$H^4$ is generated by ${\rm Tr}\, \sigma^2$ and $( {\rm Tr}\, \sigma)^2$,
and so forth.  For example, for three indeterminates, we have the
Schur polynomials\footnote{
See appendix~\ref{app:Schur} for a short introduction to Schur polynomials.}
\begin{eqnarray*}
\left( s_{\tiny\yng(1)}(x_1, x_2, x_3) \right)^2 & = &
x_1^2 \: + \: x_2^2 \: + \: x_3^2 \: + \: 2 x_1 x_2  \: + \:
2 x_1 x_3 \: + \: 2 x_2 x_3, \\
s_{\tiny\yng(1,1)}(x_1, x_2, x_3) & = &
x_1 x_2 \: + \: x_1 x_3 \: + \: x_2 x_3, \\
s_{\tiny\yng(2)}(x_1,x_2,x_3) & = &
x_1^2 \: + \: x_2^2 \: + \: x_3^2 \: + \: x_1 x_2 \: + \:
x_2 x_3 \: + \: x_1 x_3 ,
\end{eqnarray*}
from which we see that
\begin{displaymath}
\left( s_{\tiny\yng(1)} \right)^2 \: = \:
s_{\tiny\yng(1,1)} \: + \:
s_{\tiny\yng(2)} ,
\end{displaymath}
so that there are only two independent quantities.
Clearly if we associate ${\rm Tr}\, \sigma$ to $\tiny\yng(1)$,
then $( {\rm Tr}\, \sigma)^2$ is associated to $\tiny\yng(1,1) + \tiny\yng(2)$.
This is the sum of Sym$^2$ and Alt$^2$, so it makes perfect sense that the
result is just the square.
Similarly, looking at the indeterminates as eigenvalues,
\begin{eqnarray*}
{\rm Tr}\, \sigma^2 & = & x_1^2 \: + \: x_2^2 \: + \: x_3^2, \\
& = & s_{\tiny\yng(2)}(x_1,x_2,x_3) \: - \:
s_{\tiny\yng(1,1)}(x_1, x_2, x_3),
\end{eqnarray*}
so we see that ${\rm Tr}\, \sigma^2$ is associated to
$\tiny\yng(2) - \tiny\yng(1,1)$.

We will argue next that all cohomology of $G(k,n)$ can be constructed from
operators of the form 
\begin{displaymath}
{\rm Tr}\, \sigma^k \: = \: \sum_i x_i^k .
\end{displaymath}

In general, the dimension of $H^{2\bullet}(G(k,n),
{\bf Z}$ is the same as the number of gauge-invariant
independent $\sigma$ polynomials
of degree $\bullet$ in $\sigma$.  To see this, we define an isomorphism
between Young diagrams and $\sigma$ polynomials as follows.
First,
to a Young diagram $(d_1, \cdots, d_k)$,
one can associate
\begin{displaymath}
\prod_i  {\rm Tr}\, \sigma^{d_i}  .
\end{displaymath}
(Note that this association is not intended to relate representations
of elements of cohomology, but rather is merely meant to be used in
a set-theoretic counting.)
Conversely,
given a gauge-invariant $\sigma$ polynomial
\begin{displaymath}
\prod_i \left( {\rm Tr} \, \sigma^i \right)^{a_i} ,
\end{displaymath}
where $j$ is the highest power of $\sigma$ appearing in a trace,
the largest $j$ such that $a_j \neq 0$,
we associate a Young diagram defined by
\begin{displaymath}
(d_1, \cdots, d_k) \: = \: \left( 
\underbrace{j, \cdots, j,}_{a_j}
\underbrace{j-1, \cdots, j-1,}_{a_{j-1}} \cdots
\underbrace{1, \cdots, 1}_{a_1}
\right) .
\end{displaymath}
As a consistency check, the $\sigma$ polynomial should contribute to
cohomology in degree
\begin{displaymath}
2 \sum_i i a_i
\end{displaymath}
and the Young diagram indicated should contribute to cohomology in
the same degree.  It is straightforward to check that these two
maps are inverses of one another, and so we see that the dimension
of the cohomology of $G(k,n)$ is the same as the number of
gauge-invariant $\sigma$ polynomials of the same degree.

\section{Schur polynomials}
\label{app:Schur}

Since Schur polynomials are not often encountered in the physics literature,
in this appendix we briefly review some of their pertinent properties.

Briefly, Schur polynomials are polynomials in $k$ variables
associated to Young diagrams (unlabelled Young tableaux) describing
representations of $SL(k)$ or $SU(k)$.
Such a Young diagram can be characterized by a sequence of $k$
positive integers $d_1, \cdots, d_k$, where
\begin{displaymath}
d_1 \: \geq \: d_2 \: \geq \: \cdots \: \geq \: d_k
\end{displaymath}
and $d_i$ gives the number of boxes in row $i$ of the Young diagram.

Define
\begin{displaymath}
a_{(d_1, \cdots, d_k)}(x_1, \cdots, x_k) \: = \:
\det \left[ \begin{array}{cccc}
x_1^{d_1} & x_2^{d_1} & \cdots & x_k^{d_1} \\
x_1^{d_2} & x_2^{d_2} & \cdots & x_k^{d_2} \\
\vdots & \vdots & & \vdots \\
x_1^{d_k} & x_2^{d_k} & \cdots & x_k^{d_k}
\end{array} \right] ,
\end{displaymath}
then the Schur polynomial corresponding to the Young diagram defined
by $(d_1, \cdots, d_k)$ is
\begin{displaymath}
s_{(d_1, \cdots, d_k)}(x_1, \cdots, x_k) \: = \:
\frac{
a_{(d_1 + k - 1, d_2 + k - 2, \cdots, d_k + 0)}(x_1, \cdots, x_k)
}{
a_{(k-1, k-2, \cdots, 0)}(x_1, \cdots, x_k)
} .
\end{displaymath}
(For a different perspective on the Schur polynomials, compare the
characters given in \cite{cmr}[equ'n (4.5)].)

For example, it is straightforward to compute that
\begin{displaymath}
s_{\tiny\yng(1)}(x_1, x_2, x_3) \: = \:
s_{(1,0,0)}(x_1, x_2, x_3) \: = \:
x_1 \: + \: x_2 \: + \: x_3 ,
\end{displaymath}
\begin{displaymath}
s_{\tiny\yng(1,1)}(x_1, x_2, x_3) \: = \:
s_{(1,1,0)}(x_1, x_2, x_3) \: = \:
x_1 x_2 \: + \: x_1 x_3 \: + \: x_2 x_3 ,
\end{displaymath}
\begin{displaymath}
s_{\tiny\yng(2)}(x_1,x_2,x_3) \: = \:
s_{(2,0,0)}(x_1, x_2, x_3) \: = \:
x_1^2 \: + \: x_2^2 \: + \: x_3^2 \: + \: x_1 x_2 \: + \:
x_2 x_3 \: + \: x_1 x_3 ,
\end{displaymath}
\begin{displaymath}
s_{\tiny\yng(2,1)}(x_1, x_2, x_3) \: = \:
s_{(2,1,0)}(x_1, x_2, x_3) \: = \:
x_1^2( x_2 + x_3) \: + \: x_1( x_2 + x_3)^2 \: + \:
x_2 x_3(x_2 + x_3) ,
\end{displaymath}
\begin{displaymath}
s_{\tiny\yng(2,2)}(x_1, x_2, x_3) \: = \:
s_{(2,2,0)}(x_1, x_2, x_3) \: = \:
x_1^2 x_2^2 \: + \: x_1^2 x_2 x_3 \: + \: x_1^2 x_3^2 \: + \:
x_2^2 x_1 x_3 \: + \: x_2^2 x_3^2 \: + \: x_3^2 x_1 x_2 .
\end{displaymath}

\section{Representations of $U(k)$}
\label{app:uk-reps}

The representation theory of $SU(k)$ is certainly well-known;
however, representations of $U(k)$ can be more complicated, because of the
possibility of tensoring in powers of the determinant.
In this appendix, we give our conventions for describing
representations of $U(k)$.

Any irreducible unitary representation of $U(k)$ is given by a $k$-tuple
of ordered integers \cite{u(n)rep}[sections 19-22]
\begin{equation}
\lambda = (\lambda_1\geq\lambda_2\geq...\geq\lambda_k), \ \lambda_i \in
\mathbb{Z}, \ \forall i .
\end{equation}
This is the highest weight of the corresponding representation.
For completeness, here are a few examples \cite{u(n)rep}[sections 19-22]:
\begin{itemize}
\item The defining fundamental representation of $U(k)$
has highest weight $(1,0,\cdots,0)$, while its conjugate, the
antifundamental representation, has highest weight $(0,\cdots,0,-1)$.
\item The exterior product representation on $\wedge^{\ell} 
{\mathbb C}^k$ has highest weight $(1,1,\cdots,1,0,0,\cdots,0)$
($\ell$ 1's).  In particular, the determinant representation has highest
weight $(1,1,\cdots,1)$.
\item The  adjoint representation of $U(k)$ is reducible:  
$ad = (1,0,\cdots,0) \otimes (0,0,\cdots,-1) = (1,0,\cdots,0,-1)\oplus
(0,0,\cdots,0)$.
\end{itemize}

Below are some frequently used formulas for $U(k)$ representations
\cite{casimir}[chapter 5]:
\begin{itemize}
\item The dimension of $\lambda$ is given by \cite{casimir}[equ'n (4.56)]
      \begin{equation}
      d_{\lambda} = \prod_{i<j}\frac{l_i - l_j}{l^0_i - l^0_j} = 
\prod_{i<j}\frac{\lambda_i - \lambda_j+j-i}{j-i} ,
      \end{equation}
      where $l^0_i=k-i$, and $l_i=\lambda_i+k-i$, with $i,j=1,\cdots,k$.
\item The eigenvalue of the first Casimir operator on $\lambda$ is
\cite{casimir}[equ'n (5.24), table 5.1]
      \begin{equation}
      {\rm Cas}_1(\lambda)= \sum_i \lambda_i ,
      \end{equation}
\item The eigenvalue of the second Casimir operator on $\lambda$ is
\cite{casimir}[equ'n (5.24), table 5.1]
      \begin{equation}
      {\rm Cas}_2(\lambda)= \sum_i \lambda_i(\lambda_i+k+1-2i) .
      \end{equation}
\end{itemize}

In terms of bundles on $G(k,n)$ of the form ${\cal O}(\lambda)$ for
some representation $\lambda$, it is straightforward to show that
\begin{equation}
c_1({\cal O}(\lambda)) \: = \: \frac{ d_{\lambda} {\rm Cas}_1(\lambda) }{k}
\sigma_{\tiny\yng(1)} ,
\end{equation}
where $\sigma_{\tiny\yng(1)}$ denotes the Schubert cycle
generating
$H^2(G(k,n),{\mathbb Z})$, which is one-dimensional, and
\begin{eqnarray}
{\rm ch}_2({\cal O}(\lambda)) & = &
(1/2) c_1({\cal O}(\lambda))^2 \: - \: c_2({\cal O}(\lambda)), 
\nonumber \\
& = &
d_{\lambda} {\rm Cas}_2(\lambda)\left[ - \frac{1}{k^2-1} 
\sigma_{\tiny\yng(1,1)} \: + \:
\frac{1}{2 k (k+1) } \sigma_{\tiny\yng(1)}^2 \right]
\nonumber \\
& & \hspace*{0.5in} \: + \:
d_{\lambda} {\rm Cas}_1(\lambda)^2
\left[ \frac{1}{k(k^2-1)} \sigma_{\tiny\yng(1,1)}
\: + \:
\frac{1}{2k(k+1)} \sigma_{\tiny\yng(1)}^2 \right] ,
\end{eqnarray}
where $\sigma_{\tiny\yng(1,1)}$ and $\sigma_{\tiny\yng(2)}$ generate
\begin{displaymath}
H^4(G(k,n),{\mathbb Z}) \: = \: {\mathbb Z}^2
\end{displaymath}
and 
\begin{displaymath}
\sigma_{\tiny\yng(1)}^2 \: = \: \sigma_{\tiny\yng(1,1)} \: + \:
\sigma_{\tiny\yng(2)} ,
\end{displaymath}
as we demonstrated in appendix~\ref{app:Schur}.

As a consistency check, recall that the bundle $\wedge^p S^* \rightarrow
G(k,n)$ has rank
\begin{displaymath}
\left( \begin{array}{c} k \\ p \end{array} \right)
\end{displaymath}
and 
\begin{displaymath}
c_1\left( \wedge^p S^* \right) \: = \:
\left( \begin{array}{c} k-1 \\ p-1 \end{array} \right) 
\sigma_{\tiny\yng(1)},
\end{displaymath}
and these are both consistent with the formulas above for the
representation 
\begin{displaymath}
(1, 1, \cdots, 1, 0, \cdots, 0)
\end{displaymath}
($p$ $1$'s) of $U(k)$, which defines the bundle $\wedge^p S^*$.
We list here results for a few other cases, which can also be used
to check the general formulas above.
For $p=1$ \cite{leopriv}, \cite{manivel}[prop. 3.5.5],
\begin{displaymath}
c_2(S^*) \: = \: \sigma_{\tiny\yng(1,1)}, \: \: \:
{\rm ch}_2(S^*) \: = \: (1/2) \sigma_{\tiny\yng(1)}^2 \: - \:
\sigma_{\tiny\yng(1,1)},
\end{displaymath}
and in fact $c_i(S^*)$ is given by the Schubert cycle associated to the
Young diagram with $i$ vertical boxes.
In the special case $p=2$,
\begin{displaymath}
c_2(\wedge^2 S^*) \: = \: \left( \begin{array}{c} k-1 \\ 2 \end{array} \right)
\sigma_{\tiny\yng(1)}^2 \: + \: (k-2) \sigma_{\tiny\yng(1,1)},
\end{displaymath}
which one can use to show that for the representation $(2,0,\cdots,0)$,
\begin{displaymath}
{\rm rk} \, {\rm Sym}^2 S^* \: = \: \frac{ k (k+1) }{2}, \: \: \:
c_1({\rm Sym}^2 S^*) \: = \: (k+1) \sigma_{\tiny\yng(1)}, 
\end{displaymath}
\begin{displaymath}
{\rm ch}_2({\rm Sym}^2 S^*) \: = \: 
\frac{k+3}{2} \sigma_{\tiny\yng(1)}^2 \: - \:
(k+2) \sigma_{\tiny\yng(1,1)},
\end{displaymath}
and similarly
\begin{displaymath}
c_2(\wedge^3 S^*)
\: = \:
\frac{k (k-1) (k-2) (k-3) }{8} \sigma_{\tiny\yng(1)}^2 \: + \:
\frac{(k-2) (k-3)}{2} \sigma_{\tiny\yng(1,1)},
\end{displaymath}
\begin{displaymath}
{\rm ch}_2(\wedge^3 S^*) \: = \: \frac{(k-1) (k-2)}{4} 
\sigma_{\tiny\yng(1)}^2 \: - \: 
\frac{(k-2) (k-3)}{2} \sigma_{\tiny\yng(1,1)} .
\end{displaymath}

At the level of Lie algebras, $u(k) \cong su(k) \oplus u(1)$.  
Therefore, given a representation $\lambda$ of $u(k)$, we can get
an irreducible representation of $su(k)\oplus u(1)$: the representation of
$su(k)$ is given by the Young diagram
$(\lambda_1-\lambda_k\geq\lambda_2-\lambda_k\geq\cdots\geq 0)$,
and the representation of $u(1)$ is given by the integer
${\rm Cas}_1(\lambda)$.

For completeness, the eigenvalue of an $su(k)$ second Casimir operator
on the $su(k)$ representation
$\lambda= (\lambda_1\geq\lambda_2\geq\cdots\geq\lambda_k\geq 0)$
is given by
\cite{casimir}[equ'n (5.24), table 5.1]
\begin{equation}
{\rm Cas}_2(\lambda)= \sum_i \left(\lambda_i - \frac{\sum_i \lambda_i}{k}
\right)\left(\lambda_i-
 \frac{\sum_i \lambda_i}{k} +2k -2i\right) .
\end{equation}
For example:
\begin{eqnarray*}
{\rm Cas}_2(ad) & = & 2k, \\
{\rm Cas}_2(1,0,\cdots,0) & = & (k^2-1)/k .
\end{eqnarray*}
As a consistency check, \cite{mckay-patera}[equ'n (2.18)] 
lists an index for $su(2)$ representations defined by Young diagrams with
$n$ boxes:
\begin{displaymath}
I_2(n) \: = \: \frac{1}{6} n (n+1)(n+2) ,
\end{displaymath}
where
\begin{displaymath}
{\rm Tr}\left( T_R^a T_R^b \right) \: = \:  
I_2(R) \delta^{ab} .
\end{displaymath}

It is straightforward to check that
\begin{displaymath}
I_2(n) \: = \: \frac{ d_{(n,0)} {\rm Cas}_2(n,0) }{ {\rm dim}\, su(2) } ,
\end{displaymath}
where $d_{(n,0)} = n+1$ and
\begin{displaymath}
{\rm Cas}_2(n,0) \: = \: (n/2)(n/2+4-2) + (-n/2)(-n/2 +4-4)
\: = \: (1/2) n(n+2) .
\end{displaymath}

\section{Checks of (2,2) abelian/nonabelian duality}
\label{app:dual-check}

In this section, we shall use compare elliptic genera as a check of
the duality between the (2,2) GLSMs for $G(2,4)$ and ${\mathbb P}^5[2]$
proposed in the text.  As discussed earlier, as the two GLSMs have
weak-coupling limits describing the same geometry, they have, by construction, 
the same IR limit, making checks of elliptic genera somewhat unnecessary.
Nevertheless, to be thorough, in this appendix we will verify that
elliptic genera match.

To fix notation, 
for a (2,2) supersymmetric gauge theory with global symmetry $K$,  
the elliptic genus is the quantity
\begin{equation}
Z_{T^2}(\tau,u) := \text{Tr}_{\text{RR}} (-1)^{F} q^{L_0} 
\overline{q}^{\overline{L}_0} y^J \prod_a x^{K_a}_a
\end{equation}
where $F$ is the fermion number operator, and $q=e^{2\pi i\tau}$ on our $T^2$ 
defined by $\tau$. 
In addition, we define $x_a = e^{2\pi i u_a}$ coming from fugacities $u_a$ 
of global and gauge symmetries, and $y=e^{2\pi iz}$ coming from the 
fugacity of the left-moving $U(1)$ R-symmetry $J$.

In \cite{beht1,beht}, this index was computed for general
(2,2) gauge theories in two dimensions.  In particular, they derived
\begin{equation}
Z_{T^2}(\tau,u,\xi) = \frac{1}{|W|} 
\sum_{u_* \in \mathfrak{M}^*_{\text{sing}}} 
\text{JK-Res}(Q(u_*,\eta))Z_{\text{1-loop}}.
\end{equation}
(See \cite{beht1,beht} for notation.)

Note in passing that since these GLSMs are not Calabi-Yau, 
the left-moving R-symmetry
$J$ is anomalous, so in principle we can only expect a physically
unambiguous result for special values of $y$.  Nevertheless, we will
compute for general values of $y$ and find matching, 
a strong check.  (For related analyses in different contexts see
for example \cite{murthy1,adt1}).

Let's use this index to test the abelian/nonabelian duality between the GLSM 
on $G(2,4)$ and the GLSM on $\mathbb{P}^5[2]$. 
The elliptic genus of the GLSM on $G(2,4)$ was computed in \cite{beht1,beht}, 
so let's compute the elliptic genus of the GLSM on $\mathbb{P}^5[2]$. 
The GLSM is a $U(1)$ gauge theory with 6 chiral superfields $\Phi^i$ with 
charge 1, a chiral superfield $P$ with charge -2, 
and a superpotential $W=PG(\Phi)$ 
where $G(\Phi)$ is a generic polynomial of degree 2.

The 1-loop determinant coming from the $\Phi$'s is
\begin{equation}
Z_{\Phi} = \left( \frac{\theta_1(q,y^{-1}x)}{\theta_1(q,x)} \right)^6 ,
\end{equation}
since $\Phi^i$ has R-charge 0. The 1-loop determinant coming from $P$ is
\begin{equation}
Z_{P} = \frac{\theta_1(q,x^{-2})}{\theta_1(q,yx^{-2})} ,
\end{equation}
since $P$ has R-charge 2. Finally, the 1-loop determinant coming from the 
vector multiplet is
\begin{equation}
Z_{V} = \frac{2\pi\eta(q)^3}{\theta_1(q,y^{-1})} du .
\end{equation}

Then, applying the methods of \cite{beht1,beht}, we recover the elliptic
genus (in the geometric phase):
\begin{equation}
Z_{T^2}(q,z) = \frac{\eta(q)^3}{i \theta_1(q,y^{-1})}\oint_{u=0} du 
\left( \frac{\theta_1(q,y^{-1}x)}{\theta_1(q,x)} \right)^6 
\frac{\theta_1(q,x^{-2})}{\theta_1(q,yx^{-2})} .
\end{equation}
One can use Mathematica to evaluate this integral. 
In the limit $z\rightarrow 0$, one finds
$Z_{T^2}(q,z\rightarrow 0) = 6$, independent of the value of $q$, 
which matches precisely the corresponding computation for
the GLSM describing $G(2,4)$ 
(or the Euler characteristic of $G(2,4)$), given in \cite{beht}[equ'n (4.43)].

Now, to properly compare elliptic genera, let us take into account
the action of the $G(2,4)$ symmetries on ${\mathbb P}^5[2]$.  
Let $z_{ij}$ denote homogeneous coordinates on ${\mathbb P}^5$,
which are related to the fundamentals $\phi^a_i$ defining $G(2,4)$ as
the baryons
\begin{displaymath}
z_{ij} \: = \: \epsilon_{ab} \phi^a_i \phi^b_j.
\end{displaymath}
Now, one of symmetries of $G(2,4)$ used in \cite{beht} in computing the
elliptic genus is the rescaling symmetry
\begin{displaymath}
\phi^a_i \: \mapsto \: e^{2 \pi i \xi_i} \phi^a_i,
\end{displaymath}
from which we read off that on ${\mathbb P}^5[2]$, we should have
the symmetry
\begin{displaymath}
z_{ij} \: \mapsto \: e^{2 \pi i (\xi_i + \xi_j) } z_{ij}.
\end{displaymath}
A generic quadric would break rescaling symmetries of this form,
but in the present case, we are interested in a quadric which is
a linear combination of $z_{12} z_{34}$, $z_{13} z_{24}$, $z_{14} z_{23}$,
and so it is preserved by the symmetry.
With this in mind, we can now read off the flavored elliptic genus
of ${\mathbb P}^5[2]$, taking into account this symmetry:
\begin{eqnarray*}
Z_{T^2}(q,z,\xi_i) & = &
\frac{2\pi\eta(q)^3}{\theta_1(q,y^{-1})}\oint du 
\left( \prod_{i,j} 
\frac{ \theta_1\left(q, y^{-1} x e^{2 \pi i (\xi_i + \xi_j) }\right) }{
\theta_1\left( q, x e^{2 \pi i (\xi_i + \xi_j) }\right) }
\right)
\frac{ \theta_1\left(q, x^{-2} e^{2 \pi i (- \xi_1 - \xi_2 - \xi_3 - \xi_4)}
\right) }{
\theta_1\left( q, y x^{-2} e^{2 \pi i (- \xi_1 - \xi_2 - \xi_3 - \xi_4) }
\right) }.
\end{eqnarray*}
The residues are computed at six poles, at the locations
\begin{displaymath}
u \: = \: - \xi_i \: - \: \xi_j
\end{displaymath}
for $i \neq j$.  For example, the residue at $u = - \xi_1 - \xi_2$ is given by
\begin{displaymath}
\frac{ \theta_1\left(q, y^{-1} e^{2 \pi i (\xi_1 - \xi_3)} \right) }{
\theta_1\left(q, e^{2 \pi i (\xi_1 - \xi_3)} \right) }
\frac{ \theta_1 \left(q, y^{-1} e^{2 \pi i (\xi_1 - \xi_4)} \right) }{
\theta_1\left(q, e^{2 \pi i (\xi_1 - \xi_4)} \right) }
\frac{ \theta_1\left(q, y^{-1} e^{2 \pi i (\xi_2 - \xi_3)} \right) }{
\theta_1\left(q, e^{2 \pi i (\xi_2 - \xi_3)} \right) }
\frac{ \theta_1\left(q, y^{-1} e^{2 \pi i (\xi_2 - \xi_4)} \right) }{
\theta_1\left(q, e^{2 \pi i (\xi_2 - \xi_4)} \right) }.
\end{displaymath}
Each residue precisely corresponds to a term in the expression for
the flavored elliptic genus for $G(2,4)$ given in
\cite{beht}[equ'n (4.42)].
Thus, we see that the flavored elliptic genus of the abelian GLSM for
${\mathbb P}^5[2]$ precisely matches that of the nonabelian GLSM for
$G(2,4)$ computed in \cite{beht}, as expected from the proposed duality.

So far we have used a $({\mathbb C}^{\times})^4$ symmetry group
common to both $G(2,4)$ and ${\mathbb P}^5[2]$.
More generally, there is a global $GL(4,{\mathbb C})$ symmetry acting
linearly on the four fundamentals defining $G(2,4)$.  Under this symmetry,
\begin{displaymath}
\phi^a_i \: \mapsto \: V_i^j \phi^a_j
\end{displaymath}
and so
\begin{displaymath}
z_{ij} \: \mapsto V_i^{i'} V_j^{j'} z_{i'j'}
\end{displaymath}
(transforming in the $\wedge^2 {\bf 4}$ representation, in other words).
Furthermore, 
the quadric hypersurface in ${\mathbb P}^5$ is invariant.  Specifically,
the hypersurface polynomial
\begin{displaymath}
z_{12} z_{34} \: - \: z_{13} z_{24} \: + \: z_{14} z_{23}
\end{displaymath}
transforms to
\begin{displaymath}
V_1^i V_2^j V_3^k V_4^m \left( z_{ij} z_{km} \: - \:
z_{ik} z_{jm} \: + \: z_{im} z_{jk} \right) 
\: = \: (\det V) (z_{12} z_{34} \: - \: z_{13} z_{24} \: + \: z_{14} z_{23}
),
\end{displaymath}
where we have used the fact that
\begin{displaymath}
 z_{ij} z_{km} \: - \:
z_{ik} z_{jm} \: + \: z_{im} z_{jk}
\end{displaymath}
is completely antisymmetric in all its indices.

\section{(0,2) elliptic genera in Calabi-Yau duals}
\label{app:02-ell-gen}

In this appendix we will outline the computation of some (0,2) elliptic
genera, to check for dynamical supersymmetry breaking and as evidence
of dualities.  We will follow the conventions of \cite{beht}.

\subsection{Second entry}

We will begin with the second entry in table~\ref{table:ci-exs}.
This describes a bundle ${\cal E}$ on the Calabi-Yau hypersurface
$G(2,4)[4]$, given by
\begin{displaymath}
0 \: \longrightarrow \: {\cal E} \: \longrightarrow \:
\oplus^8 {\cal O}(1,1) \: \longrightarrow \:
{\cal O}(2,2) \oplus^2 {\cal O}(3,3) \: \longrightarrow \: 0 .
\end{displaymath}

The field content and corresponding contributions to the elliptic genus
are as follows:
\begin{itemize}
\item 4 chiral multiplets each in the fundamental of $U(2)$:
\begin{displaymath}
\left( i \frac{ \eta(q) }{ \theta_1(q,x_1) } \right)^4
\left( i \frac{ \eta(q) }{ \theta_1(q,x_2) } \right)^4,
\end{displaymath}
\item 1 Fermi multiplet in the $(-4,-4)$ representation of $U(2)$,
enforcing the hypersurface condition:
\begin{displaymath}
i \frac{ \theta_1(q,  x_1^{-4} x_2^{-4} ) }{\eta(q)},
\end{displaymath}
\item 8 Fermi multiplets in the $(1,1)$ representation of $U(2)$,
partially defining the gauge bundle:
\begin{displaymath}
\left( i \frac{ \theta_1(q, y x_1 x_2 ) }{ \eta(q) } \right)^8,
\end{displaymath}
\item 1 chiral multiplet in the $(-2,-2)$ representation of $U(2)$,
partially defining the gauge bundle:
\begin{displaymath}
i \frac{ \eta(q) }{ \theta_1(q, y^{-1} x_1^{-2} x_2^{-2} ) },
\end{displaymath}
\item 2 chiral multiplets in the $(-3,-3)$ representation of $U(2)$,
partially defining the gauge bundle:
\begin{displaymath}
\left( i \frac{ \eta(q) }{ \theta_1(q, y^{-1} x_1^{-3} x_2^{-3} ) } \right)^2,
\end{displaymath}
\item and finally the $U(2)$ gauge field contributes
\begin{displaymath}
\left( \frac{2 \pi \eta(q)^2 }{i} \right)^2 
i \frac{\theta_1(q, x_1 x_2^{-1} )}{\eta(q)}
i \frac{\theta_1(q, x_2 x_1^{-1} )}{\eta(q)}
du_1 du_2.
\end{displaymath}
\end{itemize}

In this particular example, 
the sum of the charges of the chiral superfields vanishes
without any spectators.  The dual, on the other hand, will contain
spectators, but as we shall argue there, spectators cancel out of elliptic
genus computations.

In the expressions above we have implicitly used a left-moving $U(1)$ symmetry,
under which the Fermi multiplets defining the gauge bundle have charge $+1$
and the chiral multiplets defining the gauge bundle have charge $-1$.

Assembling these components gives an elliptic genus of the form
\begin{displaymath}
\frac{1}{2} \frac{(2 \pi)^2 \eta(q)^4}{(2\pi i)^2}
\oint du_1 du_2 \frac{
\theta_1(q, x_1 x_2^{-1} ) \theta_1(q, x_2 x_1^{-1} )  
\theta_1(q,  x_1^{-4} x_2^{-4} ) 
\theta_1(q, y x_1 x_2 )^8
}{
\theta_1(q,x_1)^4
\theta_1(q,x_2)^4
 \theta_1(q, y^{-1} x_1^{-2} x_2^{-2} )
\theta_1(q, y^{-1} x_1^{-3} x_2^{-3})^2
}.
\end{displaymath}
(The overall factor of 1/2 is from the Weyl group of $SU(2)$.) 
Poles lie along the hypersurfaces $\{u_1=0\}$, $\{u_2=0\}$,
$\{z + 2(u_1+u_2) = 0\}$, $\{z + 3(u_1 + u_2) = 0\}$.  
The intersection of these hypersurfaces is projective\footnote{
The multiplicity of the $\theta_1^2$ in the denominator does not count.
}.  Let us work in a geometric phase, specified by $\eta = (1,1)$.
The only pole in the corresponding chamber is at the origin, so we compute
the repeated residue there.

Expanding the genus above in a power series in $q$, the first few terms are
\begin{eqnarray*}
\lefteqn{
72 \left( - y^{-1/2} + y^{+1/2} \right)^2 
\left( y^{-1/2} + y^{+1/2} \right) q^{1/6} 
} \\
& & 
\: - \: 72 \left( - y^{-1/2} + y^{+1/2} \right)^2
\left( y^{-1/2} + y^{+1/2} \right)^3 \left( y^{-1} - 1 + y \right) q^{7/6}
\\
& &
\: + \:
72 \left( - y^{-1/2} + y^{+1/2} \right)^2 \left( y^{-7/2} - y^{-3/2} +
2 y^{-1/2} + 2 y^{+1/2} - y^{+3/2} + y^{+7/2} \right) q^{13/6} \: + \:
{\cal O}\left(q^{19/6} \right).
\end{eqnarray*}

As described in section~\ref{sect:g24-4},
the example above is mathematically equivalent to an abelian GLSM
describing a bundle ${\cal E}$ on ${\mathbb P}^5[2,4]$, given by
\begin{displaymath}
0 \: \longrightarrow \: {\cal E} \: \longrightarrow \:
{\cal O}(1)^8 \: \longrightarrow \: {\cal O}(2) \oplus {\cal O}(3)^2
\: \longrightarrow \: 0.
\end{displaymath}

The field content and corresponding contributions to the elliptic genus
are as follows:
\begin{itemize}
\item 6 chiral multiplets each of charge $+1$:
\begin{displaymath}
\left( i \frac{ \eta(q) }{ \theta_1(q, x) } \right)^6,
\end{displaymath}
\item 8 Fermi multiplets $\Lambda_{\alpha}$ each of charge $+1$:
\begin{displaymath}
\left( i \frac{ \theta_1(q,y x) }{\eta(q)} \right)^8,
\end{displaymath}
\item one chiral multiplet $p_1$ 
of charge $-2$, helping to form the gauge bundle:
\begin{displaymath}
 i \frac{ \eta(q) }{ \theta_1(q,y^{-1} x^{-2}) },
\end{displaymath}
\item two chiral multiplets $p_{2,3}$
of charge $-3$, helping to form the gauge bundle:
\begin{displaymath}
\left(  i \frac{ \eta(q) }{ \theta_1(q,y^{-1} x^{-3}) } \right)^2,
\end{displaymath}
\item one Fermi multiplet $\Gamma_1$
of charge $-2$, enforcing a hypersurface condition:
\begin{displaymath}
 i \frac{ \theta_1(q,  x^{-2}) }{\eta(q)},
\end{displaymath}
\item one Fermi multiplet $\Gamma_2$
of charge $-4$, enforcing a hypersurface condition:
\begin{displaymath}
 i \frac{ \theta_1(q,  x^{-4}) }{\eta(q)},
\end{displaymath}
\item one chiral multiplet of charge $+2$, one of the spectators:
\begin{displaymath}
 i \frac{ \eta(q) }{ \theta_1(q,x^{+2}) },
\end{displaymath}
\item one Fermi multiplet of charge $-2$, one of the spectators:
\begin{displaymath}
i \frac{ \theta_1(q, x^{-2}) }{\eta(q)},
\end{displaymath}
\item and finally the $U(1)$ gauge field contributes
\begin{displaymath}
\frac{2 \pi \eta(q)^2}{i} du.
\end{displaymath}
\end{itemize}

This theory has a (0,2) superpotential of the form
\begin{displaymath}
W \: = \: \Lambda_{\alpha} p_a F^{\alpha a}(\phi) \: + \:
\Gamma_1 G_2(\phi) \: + \: \Gamma_2 G_4(\phi)
\end{displaymath}
(plus a term for spectators).
This theory has a nonanomalous global symmetry acting on the fermions,
under which the left-moving fermions $\lambda_{\alpha}$ 
have charge $+1$ and the chiral multiplets $p_a$
have charge $-1$.  We implicitly used this global symmetry to flavor the
elliptic genus contributions above, as this is the symmetry defining the
variable $y$.

Using the identity \cite{beht}[equ'n (A.5)],
\begin{displaymath}
\theta_1(q,x) \: = \: - \theta_1(q, x^{-1}),
\end{displaymath}
it is straightforward to see that the contribution from the spectators
cancel out.  This is a (0,2) analogue of an observation in
\cite{beht}[section 2.1], that in (2,2) supersymmetry, a pair of
chiral multiplets in conjugate representations of the gauge group 
and with R-charges obeying $R_1 + R_2 = 2$ will cancel out of the
elliptic genus, reflecting the fact that with those R-charges,
there can be a superpotential term pairing them up to become massive.

Putting this together, we get the elliptic genus
\begin{eqnarray*}
\lefteqn{
\frac{1}{2\pi i}\frac{2 \pi \eta(q)^2}{i} \oint_{u=0} du
\left( i \frac{ \eta(q) }{ \theta_1(q, x) } \right)^6
\left( i \frac{ \theta_1(q,y x) }{\eta(q)} \right)^8
 i \frac{ \eta(q) }{ \theta_1(q,y^{-1} x^{-2}) }
\left(  i \frac{ \eta(q) }{ \theta_1(q,y^{-1} x^{-3}) } \right)^2
} \\
& & \hspace*{2.5in} \cdot
 i \frac{ \theta_1(q, x^{-2}) }{\eta(q)}
 i \frac{ \theta_1(q, x^{-4}) }{\eta(q)}
\\
& \hspace*{0.5in} = &
- 
\frac{\eta(q)}{i} \oint_{u=0} du \frac{ \theta_1(q, y x)^8 \theta_1(q, x^{-2} )
\theta_1(q, x^{-4} ) }{
\theta_1(q,x)^6 \theta_1(q,y^{-1} x^{-2}) \theta_1(q, y^{-1} x^{-3})^2
}.
\end{eqnarray*}

Expanding this genus in a power series in $q$, we compute
the same first few terms as in the proposed dual:
\begin{eqnarray*} 
\lefteqn{
72 \left( - y^{-1/2} + y^{+1/2} \right)^2 
\left( y^{-1/2} + y^{+1/2} \right) q^{1/6} 
} \\
& & 
\: - \: 72 \left( - y^{-1/2} + y^{+1/2} \right)^2
\left( y^{-1/2} + y^{+1/2} \right)^3 \left( y^{-1} - 1 + y \right) q^{7/6}
\\
& &
\: + \:
72 \left( - y^{-1/2} + y^{+1/2} \right)^2 \left( y^{-7/2} - y^{-3/2} +
2 y^{-1/2} + 2 y^{+1/2} - y^{+3/2} + y^{+7/2} \right) q^{13/6} \: + \: 
{\cal O}\left(q^{19/6} \right).
\end{eqnarray*}
Thus, we have good evidence that the proposed (0,2) duals are, in fact,
dual, consistent with the fact that weakly-coupled limits describe
the same geometry and gauge bundle.

For completeness, let us also compare the leading term above to what
one would expect from the general analysis of \cite{km}.  Recall
equation~(\ref{eq:km-3fold}) says the leading term in the elliptic genus
on a Calabi-Yau 3-fold, for a rank 5 bundle, is given by
\begin{displaymath}
q^{(5-3)/12} y^{-5/2} (-) \tilde{\chi}({\cal E}) y (1+y)(1-y)^{5-3}
\: = \:
- \tilde{\chi}({\cal E}) q^{+1/6} y^{-5/2} y (1 - y - y^2 + y^3).
\end{displaymath}
It is straightforward to compute in this case that 
$\tilde{\chi}({\cal E}) = -72$,
and a bit of algebra suffices to demonstrate that the leading term above
matches the prediction of \cite{km}.

\subsection{Fourth entry}

We now compute the elliptic genus of the fourth entry in
table~\ref{table:ci-exs} and compare to the elliptic genus of the
proposed abelian dual.

The fourth entry describes the bundle
\begin{displaymath}
0 \: \longrightarrow \: {\cal E} \: \longrightarrow \:
{\cal O}(1,1)^2 \oplus {\cal O}(2,2)^5 \:
\longrightarrow \: {\cal O}(3,3)^4 \: \longrightarrow \: 0
\end{displaymath}
on the Calabi-Yau threefold $G(2,4)[4]$.

The field content and corresponding contributions to the elliptic
genus are as follows:
\begin{itemize}
\item 4 chiral multiplets each in the fundamental of $U(2)$
\begin{displaymath}
\left( i \frac{ \eta(q) }{ \theta_1(q,x_1) } \right)^4
\left( i \frac{ \eta(q) }{ \theta_1(q,x_2) } \right)^4,
\end{displaymath}
\item 1 Fermi multiplet in the $(-4,-4)$ representation of $U(2)$,
enforcing the hypersurface condition:
\begin{displaymath}
i \frac{ \theta_1(q,  x_1^{-4} x_2^{-4} ) }{\eta(q)},
\end{displaymath}
\item 2 Fermi multiplets in the $(1,1)$ representation of $U(2)$,
forming part of the gauge bundle:
\begin{displaymath}
\left( i \frac{ \theta_1(q, y x_1 x_2) }{ \eta(q) } \right)^2,
\end{displaymath}
\item 5 Fermi multiplets in the $(2,2)$ representation of $U(2)$,
forming part of the gauge bundle:
\begin{displaymath}
\left( i \frac{ \theta_1(q, y x_1^2 x_2^2 ) }{\eta(q)} \right)^5,
\end{displaymath}
\item 4 chiral multiplets in the $(-3,-3)$ representation of $U(2)$,
forming part of the gauge bundle:
\begin{displaymath}
\left( i \frac{ \eta(q) }{\theta_1(q, y^{-1} x_1^{-3} x_2^{-3} ) } 
\right)^4,
\end{displaymath}
\item and finally the $U(2)$ gauge field contributes
\begin{displaymath}
\left( \frac{2 \pi \eta(q)^2 }{i} \right)^2 
i \frac{\theta_1(q, x_1 x_2^{-1} )}{\eta(q)}
i \frac{\theta_1(q, x_2 x_1^{-1} )}{\eta(q)}
du_1 du_2.
\end{displaymath}
\end{itemize}
(We omit spectators, as they do not contribute.)

Putting this together, we get the elliptic genus
\begin{displaymath}
\frac{1}{2} \frac{( 2 \pi)^2}{(2\pi i)^2} \eta(q)^6 \oint du_1 du_2
\frac{
\theta_1(q, x_1^{-4} x_2^{-4} ) \theta_1(q, y x_1 x_2)^2 
\theta_1(q, y x_1^2 x_2^2)^5 \theta_1(q, x_1 x_2^{-1})
\theta_1(q, x_1^{-1} x_2)
}{
\theta_1(q, x_1)^4 \theta_1(q, x_2)^4 
\theta_1(q, y^{-1} x_1^{-3} x_2^{-3} )^4
}.
\end{displaymath}
This has poles along the hypersurfaces $\{u_1=0\}$, $\{u_2=0\}$,
$\{-z - 3 u_1 - 3 u_2 = 0 \}$, which have projective intersections.  
Proceeding as before, we compute the residue at $u_1=u_2=0$.  Expanding
in a power series in $q$, the first few terms are
\begin{eqnarray*}
\lefteqn{
88 y^{-1/2} (1 + y) \: - \:
88 y^{-5/2} \left( 1 - y^2 - y^3 + y^5 \right) q
} \\
& &
\: - \: 88 y^{-7/2} \left( 1 + y \right) \left( -1 + y^3 \right)^2 q^2
\: - \: 88 y^{-7/2} \left( -1 + y\right)^2 \left( 1 + y \right)^3 
\left( 1 + y + y^2 \right) \: + \: {\cal O}\left(q^4\right).
\end{eqnarray*}

The abelian dual to this GLSM describes the bundle
\begin{displaymath}
0 \: \longrightarrow \: {\cal O}(1)^2 \oplus {\cal O}(2)^5 \:
\longrightarrow \: {\cal O}(3)^4 \: \longrightarrow \: 0
\end{displaymath}
on ${\mathbb P}^5[2,4]$.

The field content and corresponding contributions to the elliptic genus
are as follows:
\begin{itemize}
\item 6 chiral multiplets each of charge $+1$:
\begin{displaymath}
\left( i \frac{ \eta(q) }{ \theta_1(q, x) } \right)^6,
\end{displaymath}
\item 1 Fermi multiplet of charge $-2$, enforcing a hypersurface condition:
\begin{displaymath}
 i \frac{ \theta_1(q, x^{-2} ) }{\eta(q)},
\end{displaymath}
\item 1 Fermi multiplet of charge $-4$, enforcing a hypersurface condition:
\begin{displaymath}
 i \frac{ \theta_1(q, x^{-4} ) }{\eta(q)},
\end{displaymath}
\item 2 Fermi multiplets of charge $+1$, forming part of the gauge bundle:
\begin{displaymath}
\left( i \frac{ \theta_1(q,y x) }{\eta(q)} \right)^2,
\end{displaymath}
\item 5 Fermi multiplets of charge $+2$, forming part of the gauge bundle:
\begin{displaymath}
\left( i \frac{ \theta_1(q,y x^2) }{\eta(q)} \right)^5,
\end{displaymath}
\item 4 chiral multiplets of charge $-3$, forming part of the gauge bundle:
\begin{displaymath}
\left( i \frac{ \eta(q) }{ \theta_1(q, y^{-1} x^{-3}) } \right)^4,
\end{displaymath}
\item and finally the $U(1)$ gauge field contributes
\begin{displaymath}
\frac{2 \pi \eta(q)^2}{i} du.
\end{displaymath}
\end{itemize}

Putting this together, we get the elliptic genus
\begin{displaymath}
- \frac{2 \pi}{2\pi i} \eta(q)^3 \oint du \frac{
\theta_1(q, x^{-2}) \theta_1(q, x^{-4}) \theta_1(q, yx)^2 
\theta_1(q, y x^2)^5 }{
\theta_1(q,x)^6 \theta_1(q, y^{-1} x^{-3})^4
}.
\end{displaymath}
We compute the residue at $u=0$ and, expanding in a power series in $q$,
get the same result as for the dual:
\begin{eqnarray*}
\lefteqn{
88 y^{-1/2} (1 + y) \: - \:
88 y^{-5/2} \left( 1 - y^2 - y^3 + y^5 \right) q
} \\
& &
\: - \: 88 y^{-7/2} \left( 1 + y \right) \left( -1 + y^3 \right)^2 q^2
\: - \: 88 y^{-7/2} \left( -1 + y\right)^2 \left( 1 + y \right)^3 
\left( 1 + y + y^2 \right) \: + \: {\cal O}\left(q^4\right).
\end{eqnarray*}
This is a good check that the proposed (0,2) duals are, in fact, dual,
consistent with the fact that weakly-coupled limits describe the same
geometry and gauge bundle.

For completeness, let us also compare the leading term above to what one would
expect from the general analysis of \cite{km}.  Recall
equation~(\ref{eq:km-3fold}) says that the leading term in the elliptic genus
on a Calabi-Yau 3-fold, for a rank 3 bundle, is given by
\begin{displaymath}
q^{(3-3)/12} y^{-3/2} (-) \tilde{\chi}({\cal E}) y (1+y) \: = \:
- \tilde{\chi}({\cal E}) y^{-1/2} (1+y).
\end{displaymath}
It is straightforward to compute in this case that
$\tilde{\chi}({\cal E}) = -88$, 
and so the leading term computed above is consistent
with the predictions of \cite{km}.

\subsection{Fifth entry}

Next, we shall compare elliptic genera for the example given in the
fifth entry in table~\ref{table:ci-exs}), and that of its abelian dual.

The fifth entry is the GLSM for the bundle
\begin{displaymath}
0 \: \longrightarrow \: {\cal E} \: \longrightarrow \:
{\cal O}(1,1)^5 \oplus {\cal O}(2,2)^2 \: \longrightarrow \:
{\cal O}(3,3)^3 \: \longrightarrow \: 0
\end{displaymath}
on the Calabi-Yau $G(2,4)[4]$.

The field content and corresponding contributions to the elliptic genus
are as follows:
\begin{itemize}
\item 4 chiral multiplets each in the fundamental of $U(2)$
\begin{displaymath}
\left( i \frac{ \eta(q) }{ \theta_1(q,x_1) } \right)^4
\left( i \frac{ \eta(q) }{ \theta_1(q,x_2) } \right)^4,
\end{displaymath}
\item 1 Fermi multiplet in the $(-4,-4)$ representation of $U(2)$,
enforcing the hypersurface condition:
\begin{displaymath}
i \frac{ \theta_1(q,  x_1^{-4} x_2^{-4} ) }{\eta(q)},
\end{displaymath}
\item 5 Fermi multiplets in the $(1,1)$ representation of $U(2)$,
partially defining the gauge bundle:
\begin{displaymath}
\left(
i \frac{ \theta_1(q, y x_1 x_2 ) }{\eta(q)}
\right)^5,
\end{displaymath}
\item 2 Fermi multiplets in the $(2,2)$ representation of $U(2)$,
partially defining the gauge bundle:
\begin{displaymath}
\left(
i \frac{ \theta_1(q, y x_1^2 x_2^2 ) }{\eta(q)}
\right)^2,
\end{displaymath}
\item 3 chiral multiplets in the $(-3,-3)$ representation of $U(2)$,
partially defining the gauge bundle:
\begin{displaymath}
\left(  i \frac{ \eta(q) }{ \theta_1(q, y^{-1} x_1^{-3} x_2^{-3}) } \right)^3,
\end{displaymath}
\item and finally the $U(2)$ gauge field contributes
\begin{displaymath}
\left( \frac{2 \pi \eta(q)^2 }{i} \right)^2 
i \frac{\theta_1(q, x_1 x_2^{-1} )}{\eta(q)}
i \frac{\theta_1(q, x_2 x_1^{-1} )}{\eta(q)}
du_1 du_2.
\end{displaymath}
\end{itemize}

Putting this together, we get the elliptic genus
\begin{displaymath}
-\frac{i}{2} \frac{(2 \pi)^2}{(2\pi i)^2} \eta(q)^5 \oint du_1 du_2
\frac{
\theta_1(q, x_1^{-4} x_2^{-4} ) \theta_1(q, y x_1 x_2)^5
\theta_1(q, y x_1^2 x_2^2)^2 \theta_1(q, x_1 x_2^{-1})
\theta_1(q, x_1^{-1} x_2)
}{
\theta_1(q,x_1)^4 \theta_1(q,x_2)^4 \theta_1(q, y^{-1} x_1^{-3} x_2^{-3})^3
}.
\end{displaymath}
Expanding as before in a series in $q$, the first few terms of the elliptic
genus above are given by
\begin{eqnarray*}
\lefteqn{
80 \left(y - y^{-1}\right) q^{1/12} \: - \:
80\left(-y^{-3} + y^{-1} - y + y^3\right) q^{13/12}
} \\
& &
 \: - \:
80\left(-y^{-3} + 2 y^{-1} - 2 y + y^3\right) q^{25/12} \: + \:
{\cal O}\left(q^{37/12}\right).
\end{eqnarray*}

The proposed abelian dual describes the bundle
\begin{displaymath}
0 \: \longrightarrow \: {\cal E} \: \longrightarrow \:
{\cal O}(1)^5 \oplus {\cal O}(2)^2 \: \longrightarrow \:
{\cal O}(3)^3 \: \longrightarrow \: 0
\end{displaymath}
on the Calabi-Yau ${\mathbb P}^5[2,4]$.

The field content and corresponding contributions to the elliptic genus
are as follows:
\begin{itemize}
\item 6 chiral multiplets each of charge $+1$:
\begin{displaymath}
\left( i \frac{ \eta(q) }{ \theta_1(q, x) } \right)^6,
\end{displaymath}
\item one Fermi multiplet $\Gamma_1$
of charge $-2$, enforcing a hypersurface condition:
\begin{displaymath}
 i \frac{ \theta_1(q,  x^{-2}) }{\eta(q)},
\end{displaymath}
\item one Fermi multiplet $\Gamma_2$
of charge $-4$, enforcing a hypersurface condition:
\begin{displaymath}
 i \frac{ \theta_1(q,  x^{-4}) }{\eta(q)},
\end{displaymath}
\item 5 Fermi multiplets of charge $+1$, partially defining the gauge bundle:
\begin{displaymath}
\left( 
 i \frac{ \theta_1(q, y x) }{\eta(q)}
\right)^5,
\end{displaymath}
\item 2 Fermi multiplets of charge $+2$, partially defining the gauge bundle:
\begin{displaymath}
\left( 
 i \frac{ \theta_1(q, y x^2) }{\eta(q)}
\right)^2,
\end{displaymath}
\item 3 chiral multiplets of charge $-3$, partially defining the gauge bundle:
\begin{displaymath}
\left( i \frac{ \eta(q) }{ \theta_1(q, y^{-1} x^{-3}) } \right)^3,
\end{displaymath}
\item and finally the $U(1)$ gauge field contributes
\begin{displaymath}
\frac{2 \pi \eta(q)^2}{i} du.
\end{displaymath}
\end{itemize}

Assembling these pieces, we find that the elliptic genus is given by
\begin{displaymath}
\eta(q)^2 \oint du
\frac{
\theta_1(q, x^{-2} ) \theta_1(q, x^{-4}) \theta_1(q, y x)^5
\theta_1(q, y x^2)^2
}{
\theta_1(q,x)^6 \theta_1(q, y^{-1} x^{-3})^3
},
\end{displaymath}
and expanding in a power series in $q$, we find the same expression
as in the dual theory:
\begin{eqnarray*}
\lefteqn{
80 \left(y - y^{-1}\right) q^{1/12} \: - \:
80\left(-y^{-3} + y^{-1} - y + y^3\right) q^{13/12}
} \\
& &
 \: - \:
80\left(-y^{-3} + 2 y^{-1} - 2 y + y^3\right) q^{25/12} \: + \:
{\cal O}\left(q^{37/12}\right).
\end{eqnarray*}

For completeness, let us also compare the leading term above to what one would
expect from the general analysis of \cite{km}.  Recall
equation~(\ref{eq:km-3fold}) says that the leading term in the elliptic genus
on a Calabi-Yau 3-fold, for a rank 4 bundle, is given by
\begin{displaymath}
q^{(4-3)/12} y^{-4/2} (-) \tilde{\chi}({\cal E}) y (1+y)(1-y) \: = \:
- \tilde{\chi}({\cal E}) q^{+1/12} y^{-1} \left( 1 - y^2 \right).
\end{displaymath}
It is straightforward to compute in this case that 
$\tilde{\chi}({\cal E}) = -80$,
and so the leading term computed above is consistent with the predictions
of \cite{km}.


\begin{thebibliography}{199}

\addcontentsline{toc}{section}{References}

\bibitem{edphases} E. Witten, ``Phases of $N=2$ theories in two dimensions,''
Nucl. Phys. {\bf B403} (1993) 159-222,
{\tt hep-th/9301042}.

\bibitem{ael} A. Adams, M. Ernebjerg, J. Lapan, ``Linear models for
flux vacua,'' Adv. Theor. Math. Phys. {\bf 12} (2008) 817-851,
{\tt hep-th/0611084}.

\bibitem{adl} A. Adams, E. Dyer, J. Lee, ``GLSM's for non-K\"ahler
geometries,'' JHEP 1301 (2013) 044, {\tt arXiv:  1206.5815}.

\bibitem{qss} C. Quigley, S. Sethi, M. Stern, ``Novel branches of (0,2)
theories,'' JHEP 1209 (2012) 064, {\tt arXiv:  1206.3228}.

\bibitem{mqs} I. Melnikov, C. Quigley, S. Sethi, ``Target spaces from
chiral gauge theories,'' JHEP 1302 (2013) 111, {\tt arXiv:  1212.1212}.

\bibitem{mqss} I. Melnikov, C. Quigley, S. Sethi, M. Stern,
``Target spaces from chiral gauge theories,''
JHEP 1302 (2013) 111,
{\tt arXiv:  1212.1212}.

\bibitem{jklmr1} H. Jockers, V. Kumar, J. Lapan, D. Morrison,
M. Romo, ``Nonabelian 2d gauge theories for determinantal
Calabi-Yau varieties,'' JHEP 1211 (2012) 166, 
{\tt arXiv:  1205.3192}.

\bibitem{horitong} K. Hori, D. Tong, ``Aspects of non-abelian gauge dynamics
in two dimensional $N=(2,2)$ theories, JHEP 0705 (2007) 079,
{\tt hep-th/0609032}.

\bibitem{hori2} K. Hori, ``Duality in two-dimensional (2,2) supersymmetric
non-abelian gauge theories,'' {\tt arXiv:  1104.2853}.

\bibitem{meron} R. Donagi, E. Sharpe, ``GLSM's for partial flag manifolds,''
J. Geom. Phys. {\bf 58} (2008) 1662-1692,
{\tt arXiv:  0704.1761}.

\bibitem{horiknapp} K. Hori, J. Knapp, ``Linear sigma models with
strongly coupled phases---one parameter models,''
{\tt arXiv:  1308.6265}.

\bibitem{cdhps} A. Caldararu, J. Distler, S. Hellerman, T. Pantev,
E. Sharpe, ``Non-birational twisted derived equivalences in abelian
GLSMs,'' Comm. Math. Phys. {\bf 294} (2010) 605-645,
{\tt arXiv:  0709.3855}.

\bibitem{as2} N. Addington, E. Segal, E. Sharpe, ``D-brane probes, 
branched double covers, and noncommutative resolutions,'' {\tt arXiv:
1211.2446}.

\bibitem{me-rflat} E. Sharpe, ``A few Ricci-flat stacks as phases of
exotic GLSMs,'' Phys. Lett. {\bf B726} (2013) 390-395,
{\tt arXiv:  1306.5440}.

\bibitem{hkm} J. Halverson, V. Kumar, D. Morrison, ``New methods for
characterizing phases of 2d supersymmetric gauge theories,''
JHEP 1309 (2013) 143,
{\tt arXiv:  1305.3278}.

\bibitem{bdfik} M. Ballard, D. Deliu, D. Favero, M. Umut Isik, L. Katzarkov,
``Homological projective duality via variation of geometric invariant
theory quotients,'' {\tt arXiv:  1306.3957}.

\bibitem{bc1} F. Benini, S. Cremonesi, ``Partition functions of $N=(2,2)$
gauge theories on $S^2$ and vortices,'' {\tt arXiv:  1206.2356}.

\bibitem{dgfl} N. Doroud, J. Gomis, B. Le Floch, S. Lee,
``Exact results in $D=2$ supersymmetric gauge theories,''
{\tt arXiv:  1206.2606}.

\bibitem{jklmr2} H. Jockers, V. Kumar, J. Lapan, D. Morrison,
M. Romo, ``Two-sphere partition functions and Gromov-Witten invariants,''
{\tt arXiv:  1208.6244}.

\bibitem{ncgw} E. Sharpe, ``Predictions for Gromov-Witten invariants
of noncommutative resolutions,'' J. Geom. Phys. {\bf 74} (2013) 256-265,
{\tt arXiv:  1212.5322}.

\bibitem{bstv} G. Bonelli, A. Sciarappa, A. Tanzini, P. Vasko,
``Vortex partition functions, wall crossing, and equivariant
Gromov-Witten invariants,'' {\tt arXiv:  1307.5997}.

\bibitem{gg0} A. Gadde, S. Gukov, ``2d index and surface operators,''
{\tt arXiv:  1305.0266}.

\bibitem{beht1} F. Benini, R. Eager, K. Hori, Y. Tachikawa, ``Elliptic
genera of two-dimensional $N=2$ gauge theories with rank one gauge
groups,'' {\tt arXiv:  1305.0533}.

\bibitem{beht} F. Benini, R. Eager, K. Hori, Y. Tachikawa,
``Elliptic genera of 2d $N=2$ gauge theories,''
{\tt arXiv:  1308.4896}.

\bibitem{murthy} S. Murthy, ``A holomorphic anomaly in the elliptic genus,''
{\tt arXiv:  1311.0918}.

\bibitem{adt} S. Ashok, N. Doroud, J. Troost, ``Localization and real
Jacobi forms,'' {\tt arXiv:  1311.1110}.

\bibitem{ggp} A. Gadde, S. Gukov, P. Putrov, ``(0,2) trialities,''
{\tt arXiv:  1310.0818}.

\bibitem{kutasovlin} D. Kutasov, J. Lin, ``(0,2) dynamics from four
dimensions,'' {\tt arXiv:  1310.6032}. 

\bibitem{hpt} K. Hori, C. Park, Y. Tachikawa, ``2d SCFT's from M2-branes,''
{\tt arXiv:  1309.3036}.

\bibitem{am1} P. Aspinwall, D. Morrison, ``Chiral rings do not suffice:
$N=(2,2)$ theories with nonzero fundamental group,''
Phys. Lett. {\bf B334} (1994) 79-86,
{\tt hep-th/9406032}.



\bibitem{hetstx} L. Anderson, B. Jia, R. Manion, B. Ovrut, E. Sharpe,
``General aspects of heterotic string compactifications on stacks
and gerbes,'' {\tt arXiv:  1307.2269}.

\bibitem{edoldeg1} E. Witten, ``Elliptic genera and quantum field theory,''
Comm. Math. Phys. {\bf 109} (1987) 525-536.

\bibitem{murthy1} S. Murthy, ``A holomorphic anomaly in the elliptic
genus,'' {\tt arXiv:  1311.0918}.

\bibitem{adt1} S. Ashok, N. Doroud, J. Troost, ``Localization and
real Jacobi forms,'' {\tt arXiv:  1311.1110}.

\bibitem{km} T. Kawai, K. Mohri, ``Geometry of (0,2) Landau-Ginzburg
orbifolds,'' {\tt hep-th/9402148}.

\bibitem{lgeg} M. Ando, E. Sharpe, ``Elliptic genera of Landau-Ginzburg
models over nontrivial spaces,'' Adv. Theor. Math. Phys. {\bf 16} (2012)
1087-1144, {\tt arXiv:  0905.1285}.

\bibitem{edloopspace} E. Witten, ``The index of the Dirac operator in
loop space,'' pp. 161-181 in {\it Elliptic curves and modular forms in 
algebraic topology}, ed. P. Landweber, 
Lecture notes in math. 1326, Springer-Verlag, Berlin, 1988. 


\bibitem{tongrev} D. Tong, ``Quantum vortex strings: a review,''
{\tt arXiv:  0809.5060}. 

\bibitem{dk3} J. Distler, S. Kachru, ``Quantum symmetries and
stringy instantons,'' Phys. Lett. {\bf B336} (1994) 368-375,
{\tt hep-th/9406091}.

\bibitem{okonek} C. Okonek, M. Schneider, H. Spindler,
{\it Vector bundles on complex projective spaces}, Birkh\"auser, Boston,
1980.

\bibitem{dlm} R. Donagi, V. Lu, I. Melnikov, to appear.

\bibitem{fln} E. Frenkel, A. Losev, N. Nekrasov, ``Instantons beyond
topological theory I,'' {\tt hep-th/0610149}.

\bibitem{dsww1} M. Dine, N. Seiberg, X. G. Wen, E. Witten,
``Nonperturbative effects on the string worldsheet,''
Nucl. Phys. {\bf B278} (1986) 769-789.

\bibitem{dsww2} M. Dine, N. Seiberg, X. G. Wen, E. Witten,
``Nonperturbative effects on the string worldsheet 2,''
Nucl. Phys. {\bf B289} (1987) 319-363.

\bibitem{eva-ed} E. Silverstein, E. Witten, ``Criteria for
conformal invariance of (0,2) models,'' Nucl. Phys. {\bf B444} (1995)
161-190, {\tt hep-th/9503212}.

\bibitem{bcxddh} P. Berglund, P. Candelas, X. de la Ossa, E. Derrick,
J. Distler, T. Hubsch, ``On the instanton contributions to the masses
and couplings of $E_6$ singlets,''
Nucl. Phys. {\bf B454} (1995) 127-163,
{\tt hep-th/9505164}.

\bibitem{basu-sethi} A. Basu, S. Sethi, ``Worldsheet stability of
(0,2) linear sigma models,'' Phys. Rev. {\bf D68} (2003) 025003,
{\tt hep-th/0303066}.

\bibitem{beas-ed} C. Beasley, E. Witten, ``Residues and worldsheet
instantons,'' JHEP {\bf 0310} (2003) 065, {\tt hep-th/0304115}.



\bibitem{edver} E. Witten, ``The Verlinde algebra and the
cohomology of the Grassmannian,'' {\tt hep-th/9312104}.

\bibitem{dk} J. Distler, S. Kachru, ``(0,2) Landau-Ginzburg theory,''
Nucl. Phys. {\bf B413} (1994) 213-243, {\tt hep-th/9309110}.




\bibitem{ks} S. Katz, E. Sharpe, ``Notes on certain (0,2) correlation
functions,'' Comm. Math. Phys. {\bf 262} (2006) 611-644,
{\tt hep-th/0406226}.

\bibitem{ade} A. Adams, J. Distler, M. Ernebjerg, ``Topological
heterotic rings,'' Adv. Theor. Math. Phys. {\bf 10} (2006) 657-682,
{\tt hep-th/0506263}.


\bibitem{iw1} K. Intriligator, B. Wecht, ``The exact superconformal R symmetry
maximizes a,'' Nucl. Phys. {\bf B667} (2003) 183-200,
{\tt hep-th/0304128}.

\bibitem{bb1} F. Benini, N. Bobev, ``Two-dimensional SCFT's from wrapped
branes and c-extremization,'' JHEP 1306 (2013) 005, 
{\tt arXiv:  1302.4451}.

\bibitem{donagipriv} R. Donagi, private communication.

\bibitem{ott1} G. Ottaviani, ``Spinor bundles on quadrics,''
Trans. Amer. Math. Soc. {\bf 307} (1988) 301-316.

\bibitem{mp-c} I. Melnikov, R. Plesser, ``The Coulomb branch in gauged
linear sigma models,'' JHEP 0506 (2005) 013,
{\tt hep-th/0501238}.

\bibitem{dk2} J. Distler, S. Kachru, ``Duality of (0,2) string vacua,''
Nucl. Phys. {\bf B442} (1995) 64-74,
{\tt hep-th/9501111}.

\bibitem{br1} R. Blumenhagen, T. Rahn, ``Landscape study of target space
duality of (0,2) heterotic string models,'' JHEP 1109 (2011) 098,
{\tt arXiv:  1106.4998}.


\bibitem{beninipriv} F. Benini, private communication.

\bibitem{harris} J. Harris, {\it Algebraic geometry}, Graduate texts in
mathematics 133, Springer-Verlag, New York, 1992.


\bibitem{bchir} E. Sharpe, ``Notes on certain other (0,2) correlation
functions,'' Adv. Theor. Math. Phys. {\bf 13} (2009) 33-70,
{\tt hep-th/0605005}.

\bibitem{distpriv} J. Distler, private communication. 

\bibitem{gukovpriv} S. Gukov, private communication.

\bibitem{ilarionpriv} I. Melnikov, private communication.

\bibitem{katzpriv} S. Katz, private communication.

\bibitem{gs2} J. Guffin, E. Sharpe, ``A-twisted heterotic Landau-Ginzburg
models,'' J. Geom. Phys. {\bf 59} (2009) 1581-1596,
{\tt arXiv:  0801.3955}.

\bibitem{dg} J. Distler, B. Greene, ``Aspects of (2,0) string
compactifications,'' Nucl. Phys. {\bf B304} (1988) 1-62.

\bibitem{tonypriv} T. Pantev, private communication.

\bibitem{oss} C. Okonek, M. Schneider, H. Spindler, {\it Vector bundles
on complex projective spaces}, Birkh\"auser, Boston, 1980.

\bibitem{dgks1} R. Donagi, J. Guffin, S. Katz, E. Sharpe, ``A mathematical
theory of quantum sheaf cohomology,'' {\tt arXiv:  1110.3751}.

\bibitem{dgks2} R. Donagi, J. Guffin, S. Katz, E. Sharpe, ``Physical aspects
of quantum sheaf cohomology for deformations of tangent bundles of toric
varieties,'' {\tt arXiv:  1110.3752}.

\bibitem{mm2} J. McOrist, I. Melnikov, ``Summing the instantons in
half-twisted linear sigma models,'' JHEP 0902 (2009) 026,
{\tt arXiv:  0810.0012}.

\bibitem{kcs} E. Sharpe, ``K\"ahler cone substructure,''
Adv. Theor. Math. Phys. {\bf 2} (1999) 1441-1462,
{\tt hep-th/9810064}.

\bibitem{aglo1} L. Anderson, J. Gray, A. Lukas, B. Ovrut, ``The edge of
supersymmetry:  stability walls in heterotic theory,''
Phys. Lett. {\bf B677} (2009) 190-194,
{\tt arXiv:  0903.5088}.

\bibitem{aglo2} L. Anderson, J. Gray, A. Lukas, B. Ovrut,
``Stability walls in heterotic theories,''
JHEP 0909 (2009) 026,
{\tt arXiv:  0905.1748}.

\bibitem{ago1} L. Anderson, J. Gray, B. Ovrut, ``Yukawa textures from
heterotic stability walls,'' JHEP 1005 (2010) 086,
{\tt arXiv:  1001.2317}.

\bibitem{mss-rev} I. Melnikov, S. Sethi, E. Sharpe, ``Recent developments
in (0,2) mirror symmetry,'' SIGMA {\bf 8} (2012) 068,
{\tt arXiv:  1209.1134}.

\bibitem{guffin-rev} J. Guffin, ``Quantum sheaf cohomology, a precis,''
{\tt arXiv:  1101.1305}.

\bibitem{mcorist-rev} J. McOrist, ``The revival of (0,2) linear sigma
models,'' Int. J. Mod. Phys. {\bf A26} (2011) 1-41,
{\tt arXiv:  1010.4667}.

\bibitem{ggp-toappear} A. Gadde, S. Gukov, P. Putrov, to appear.

\bibitem{seibdual} N. Seiberg, ``Electric-magnetic duality in
supersymmetric non-abelian gauge theories,''
Nucl. Phys. {\bf B435} (1995) 129-146,
{\tt hep-th/9411149}.

\bibitem{weibel}   C. Weibel, {\it An introduction to homological
algebra}, Cambridge University Press, Cambridge, 1994.

\bibitem{ast1} O. Aharony, N. Seiberg, Y. Tachikawa, ``Reading between the
lines of four-dimensional gauge theories,'' JHEP 1308 (2013) 115,
{\tt arXiv:  1305.0318}.

\bibitem{gmn1} D. Gaiotto, G. Moore, A. Neitzke, ``Framed BPS states,''
{\tt arXiv:  1006.0146}.

\bibitem{hh1} A. Hanany, K. Hori, ``Branes and $N=2$ theories in two
dimensions,'' {\tt hep-th/9707192}.

\bibitem{kapustka} M. Kapustka, ``Mirror symmetry for Pfaffian Calabi-Yau
3-folds via conifold transitions,'' {\tt arXiv:  1310.2304}.

\bibitem{ms} J. Milnor, J. Stasheff, {\it Characteristic
classes}, Annals of Math. Studies 76, Princeton University Press,
Princeton, New Jersey, 1974.

\bibitem{cmr} S. Cordes, G. Moore, S. Ramgoolam, ``Lectures on 2d
Yang-Mills theory, equivariant cohomology, and topological field
theories,'' Nucl. Phys. Proc. Suppl. {\bf 41} (1995) 184-244,
{\tt hep-th/9411210}.

\bibitem{mp} D. Morrison, R. Plesser, ``Summing the instantons:  quantum
cohomology and mirror symmetry in toric varieties,'' Nucl. Phys.
{\bf B440} (1995) 279-354, {\tt hep-th/9412236}.

\bibitem{gh} P. Griffiths, J. Harris, {\it Principles of
algebraic geometry}, John Wiley \& Sons, New York, 1978.

\bibitem{chern} S.-S. Chern, {\it Complex manifolds without potential
theory}, second edition, Springer-Verlag, New York, 1968, 1995.

\bibitem{u(n)rep} M. Taylor, {\it Lectures on Lie groups}, available at
{\tt http://www.unc.edu/math/Faculty/met/m273.pdf}

\bibitem{casimir} F. Iachello,  {\it Lie algebras and applications},
Lecture Notes in Physics, Volume 708, Springer, 2006.

\bibitem{leopriv} L. Mihalcea, private communication.

\bibitem{manivel} L. Manivel, {\it Symmetric functions, Schubert polynomials,
and degeneracy loci}, Amer. Math. Soc., Providence, Rhode Island, 2001.

\bibitem{mckay-patera} W. G. McKay, J. Patera, {\it Tables of dimensions,
indices, and branching rules for representations of simple Lie algebras},
Marcel Dekker, New York, 1981.








\end{thebibliography}
\end{document}